\documentclass[10pt, twoside, british]{article}

\usepackage[T1]{fontenc}
\usepackage[utf8]{inputenc}
\usepackage[english]{babel}

\usepackage[
    useregional,
]{datetime2}

\usepackage{amsmath}
\usepackage{amsfonts}
\usepackage{amssymb}
\usepackage{bm}

\usepackage[
    centering,
    left=2.8cm,
    right=2.8cm,
    top=2.6cm,
    bottom=2.8cm,
    a4paper,
]{geometry}

\usepackage[dvipsnames]{xcolor}
\usepackage{caption}
\usepackage{subcaption}
\usepackage{float}
\usepackage{ragged2e}
\usepackage{enumitem}
\usepackage{centernot}
\usepackage{dsfont}

\usepackage{mlmodern} 
\usepackage{ClearSans} 
\usepackage[scaled=0.85]{sourcecodepro} 

\DeclareFontSeriesDefault[tt]{md}{sb}
\DeclareFontSeriesDefault[rm]{bf}{bx}

\definecolor{colorbackground}{HTML}{F5F5F5}
\definecolor{colorcomment}{HTML}{699ADA}
\definecolor{colorprimary}{HTML}{224499}
\definecolor{colorlogo}{HTML}{003989}

\definecolor{darkerblue}{HTML}{003989}
\definecolor{darkblue}{HTML}{224499}
\definecolor{lightblue}{HTML}{D8E9FF}

\usepackage{dcolumn}

\usepackage{mfirstuc}
\usepackage{graphicx} 
\usepackage{caption}
\usepackage{booktabs}
\usepackage{makecell}
\usepackage{multirow}
\usepackage{nicematrix}
\NiceMatrixOptions{cell-space-top-limit=3pt, cell-space-bottom-limit=3pt}
\usepackage{authblk}

\usepackage{float}
\usepackage{ragged2e}
\usepackage{tensor}
\usepackage{centernot}
\usepackage{dsfont}
\usepackage{titlesec}
\usepackage[subfigure]{tocloft}
\usepackage{csquotes}
\usepackage{tikz}
\usetikzlibrary{shapes}
\usetikzlibrary{quantikz2}
\usetikzlibrary{arrows}
\usetikzlibrary{arrows.meta}
\usetikzlibrary{calc}
\usetikzlibrary{decorations.markings}
\usetikzlibrary{decorations.text}
\usepackage{pgfplots}
\pgfplotsset{compat=1.17}
\usepackage[outline]{contour}
\usepackage[compat=1.1.0]{tikz-feynman}
\usepackage[edges]{forest}
\usepackage{etoolbox}
\PassOptionsToPackage{hyphens}{url}
\usepackage[%
colorlinks,
linktoc=all,
pdfencoding=unicode,
pdftitle={Qhronology: A Python package for studying quantum models of closed timelike curves},
pdfauthor={Lachlan G. Bishop}
]{hyperref}
\usepackage{xparse}
\ExplSyntaxOn
\DeclareExpandableDocumentCommand{\eval}{m}{\fp_eval:n {#1}}
\ExplSyntaxOff

\usepackage[skins,breakable]{tcolorbox}

\usepackage[frozencache,cachedir=.,newfloat]{minted}
\setminted{%
ignorelexererrors=true,
bgcolorpadding=0cm,
frame=none,
python3=true,
escapeinside=$$,
breaklines=true,
breakautoindent=true,
breaksymbolright=\textcolor{colorcomment}{\tiny\ensuremath{\hookleftarrow}},
breaksymbolleft=\textcolor{colorcomment}{\tiny\ensuremath{\hookrightarrow}},
breaksymbolindentrightnchars=3,
breaksymbolseprightnchars=1,
breakafter=,
listparameters=\setlength{\topsep}{0pt},
}

\setmintedinline{
bgcolorpadding=0.15em,
framesep=0.15em,
bgcolor=colorbackground,
}

\newcommand{\py}[1]{\mintinline{py}{#1}}

\newtcolorbox[auto counter]{code}[1][]{
enhanced,
breakable,
segmentation at break=true,
before skip balanced=0.35\baselineskip + 2pt,
after skip balanced=0.35\baselineskip + 2pt,
enlarge top by=0.05cm,
enlarge bottom by=0.05cm,
arc=0mm,
outer arc=0mm,
boxrule=0.25mm,
boxsep=0cm,
toptitle=0.1cm,
bottomtitle=0.1cm,
lefttitle=0.1cm,
righttitle=0.1cm,
top=0.23cm,
bottom=0.23cm,
leftupper=0.3cm,
rightupper=0.2cm + \tcboxedtitlewidth,
leftlower=0.3cm,
rightlower=0.2cm + \tcboxedtitlewidth,
colback=colorbackground,
colframe=colorprimary,
colbacktitle=colorprimary,
coltitle=white,
fonttitle=\ttfamily,
fontupper=\linespread{1.1}\fontsize{10}{12},
fontlower=\linespread{1.1}\fontsize{10}{12},
attach boxed title to top right={yshift=-\tcboxedtitleheight},
boxed title style={
arc=0mm,
outer arc=0mm,
boxrule=0mm,
boxsep=0cm,
top=0.125cm,
bottom=0.125cm,
left=0.075cm,
right=0.075cm,
opacityframe=0,
},
segmentation style={
draw=colorprimary,
solid,
line width=0.4mm,
postaction={decoration={text effects along path, text align={left indent=0em}, text color=colorprimary, text format delimiters={[}{]}, text={OUTPUT},
text effects/.cd, path from text, group letters, every word separator/.style={fill=none}, characters={fill=colorprimary, xshift=0em, yshift=-0.2cm, font=\ttfamily\bfseries\color{colorbackground}},
}, decorate},
},
title={\thetcbcounter},
#1
}

\newenvironment{codetitled}[2]{
\begin{code}[
title={\textsf{\textbf{\textit{\small #1}}} \hfill \thetcbcounter},
attach boxed title to top right=false,
attach boxed title to top,
boxed title style={
top=0.1cm,
bottom=0.1cm,
},
rightupper=0.2cm,
rightlower=0.2cm,
lefttitle=0.075cm,
righttitle=0.075cm,
label={#2},
]}{\end{code}
}

\makeatletter
\appto\tcb@use@after@lastbox{\@endparenv\@doendpe}
\makeatother

\newcommand{\tcblowerspaced}{\vspace{0.00\tcbtextheight}\tcblower\vspace{10.00\tcbtextheight}}

\usepackage{sphinxhighlight}

\usepackage[%
    bibstyle=phys,
    citestyle=numeric-comp,
    sorting=none,
    backend=biber,
    url=true
]{biblatex}
\defbibheading{bibliography}[\bibname]{\subsection*{\Large References}\vspace{0.35em}}
\DeclareFieldFormat{labelnumberwidth}{[#1]\quad}
\DeclareFieldFormat{pubstate}{}

\DeclareFieldFormat[article]{url}{\url{#1}\addperiod}

\usepackage{titling}
\newcommand{\subtitle}[1]{%
    \posttitle{%
        \par\end{center}
    \begin{center}\large#1\end{center}
    \vskip0.5em}%
}

\usepackage{placeins}
\usepackage{flafter}
\usepackage{multicol} 

\let\multicolmulticols\multicols
\let\endmulticolmulticols\endmulticols
\RenewDocumentEnvironment{multicols}{mO{}}
 {%
  \ifnum#1=1
    #2%
  \else 
    \multicolmulticols{#1}[#2]
  \fi
 }
 {%
  \ifnum#1=1
  \else 
    \endmulticolmulticols
  \fi
 }

\usepackage{enumitem}
\setlist{topsep=0.35em,itemsep=0.25em,partopsep=0.25em,parsep=0.25em}
\setlist[1]{leftmargin=2em}
\setlist[2]{leftmargin=1em}
\setlist[3]{leftmargin=1em}
\setlist[4]{leftmargin=1em}
\setlist[itemize,1]{label={\scriptsize{$\blacktriangleright$}}}
\setlist[itemize,2]{label={\footnotesize{$\bullet$}}}
\setlist[itemize,3]{label={\fontsize{4}{6}{$\blacksquare$}}}
\setlist[itemize,4]{label={\scriptsize{$\blacklozenge$}}}

\definecolor{accentcolor}{HTML}{2266C0}

\hypersetup{
    unicode = true,
    colorlinks = true,
    linkcolor = accentcolor,
    anchorcolor = accentcolor,
    citecolor = accentcolor,
    filecolor = accentcolor,
    menucolor = accentcolor,
    runcolor = accentcolor,
    urlcolor = accentcolor,
}

\makeatletter
\patchcmd{\@maketitle}{\LARGE}{\Large}{}{}
\makeatother

\makeatletter
\renewcommand{\section}{
    \@startsection%
    {section}{1}{0mm}%
    {0.75\baselineskip}%
    {0.5\baselineskip}%
    {\sffamily\Large\bfseries\scshape}%
}%
\makeatother

\makeatletter
\renewcommand{\subsection}{
    \@startsection%
    {subsection}{1}{0mm}%
    {0.75\baselineskip}%
    {0.35\baselineskip}%
    {\sffamily\large\bfseries}%
}%
\makeatother

\makeatletter
\renewcommand{\subsubsection}{
    \@startsection%
    {subsubsection}{1}{0mm}%
    {0.75\baselineskip}%
    {0.35\baselineskip}%
    {\sffamily\normalsize\bfseries}%
}%
\makeatother

\makeatletter
\renewcommand{\paragraph}{
    \@startsection%
    {paragraph}{1}{0mm}%
    {0.75\baselineskip}%
    {0.35\baselineskip}%
    {\sffamily\normalsize\bfseries}%
}%
\makeatother

\makeatletter
\renewcommand{\subparagraph}{
    \@startsection%
    {subparagraph}{1}{0mm}%
    {0.75\baselineskip}%
    {0.35\baselineskip}%
    {\sffamily\normalsize}%
}%
\makeatother

\usepackage{abstract}

\makeatletter
\renewcommand{\@pnumwidth}{1em} 
\renewcommand{\@tocrmarg}{1em}
\renewcommand{\@dotsep}{1}
\makeatother

\addto\captionsenglish{
}
\makeatletter
\patchcmd{\ps@plain}{\thepage}{\fontsize{10}{12}\sffamily\bfseries\thepage}{}{}
\makeatother
\usepackage{xstring}

\def\stopreplace#1\stopreplace{}

\newcommand{\ChangeGreek}[1]{\begingroup
\let\theta\Theta
#1
\endgroup}

\renewcommand{\footnotesize}{\fontsize{9pt}{11pt}\selectfont}
\let\originalleft\left
\let\originalright\right
\renewcommand{\left}{\mathopen{}\mathclose\bgroup\originalleft}
\renewcommand{\right}{\aftergroup\egroup\originalright}

\newcommand{\trace}{\text{tr}}

\newcommand{\eye}{{\mathrm{i}\mkern1mu}}
\newcommand{\op}[1]{\hat{#1}}
\renewcommand{\indices}[1]{\left[#1\right]}
\newcommand{\conj}[2]{#1^*#2}

\newcommand{\e}{{\mkern1mu\mathrm{e}\mkern1mu}}
\renewcommand{\ket}[1]{\left| #1 \right\rangle} 
\renewcommand{\bra}[1]{\left\langle #1 \right|} 
\renewcommand{\braket}[2]{\left\langle #1 \vphantom{#2} \vphantom{#2} \vert #2 \vphantom{#1} \right\rangle}
\newcommand{\abs}[1]{\left| #1 \right|} 

\let\baraccent=\= 
\renewcommand{\=}[1]{\stackrel{#1}{=}} 

\newcommand{\Dimension}{d}
\newcommand{\Base}{b}

\newcommand{\Complexes}{\mathbb{C}}
\newcommand{\Integers}{\mathbb{Z}}

\newcommand{\Purity}{\gamma}
\newcommand{\Entropy}{S}
\newcommand{\MutualInformation}{I}
\newcommand{\TraceDistance}{D}
\newcommand{\Fidelity}{F}

\newcommand{\Unitary}{\op{U}}

\newcommand{\Kraus}{\op{K}}
\newcommand{\Observable}{\op{O}}
\newcommand{\Identity}{\op{I}}

\newcommand{\Rotation}{\op{R}}
\newcommand{\Phase}{\op{P}}
\newcommand{\Diagonal}{\op{D}}
\newcommand{\Swap}{\op{S}}

\newcommand{\SUM}{\op{\Sigma}}
\newcommand{\Pauli}{\op{\sigma}}
\newcommand{\GellMann}{\op{\lambda}}

\newcommand{\NOT}{\op{N}}

\newcommand{\QFT}{\op{F}}

\newcommand{\Hadamard}{\op{H}}

\newcommand{\via}{via} 
\newcommand{\ie}{i.e.} 
\newcommand{\eg}{e.g.} 
\newcommand{\etc}{etc.} 

\newcommand{\secn}{Section}

\newcommand{\fig}{Figure}
\newcommand{\subfig}{Subfigure}
\newcommand{\tbl}{Table}
\newcommand{\cdb}{Code Block}
\newcommand{\cbs}{Code Blocks}

\DeclareUnicodeCharacter{0391}{$\Alpha$}
\DeclareUnicodeCharacter{0392}{$\Beta$}
\DeclareUnicodeCharacter{0393}{$\Gamma$}
\DeclareUnicodeCharacter{0394}{$\Delta$}
\DeclareUnicodeCharacter{0395}{$\Epsilon$}
\DeclareUnicodeCharacter{0396}{$\Zeta$}
\DeclareUnicodeCharacter{0397}{$\Eta$}
\DeclareUnicodeCharacter{0398}{$\Theta$}
\DeclareUnicodeCharacter{0399}{$\Iota$}
\DeclareUnicodeCharacter{039A}{$\Kappa$}
\DeclareUnicodeCharacter{039B}{$\Lambda$}
\DeclareUnicodeCharacter{039C}{$\Mu$}
\DeclareUnicodeCharacter{039D}{$\Nu$}
\DeclareUnicodeCharacter{039E}{$\Xi$}
\DeclareUnicodeCharacter{039F}{$\Omicron$}
\DeclareUnicodeCharacter{03A0}{$\Pi$}
\DeclareUnicodeCharacter{03A1}{$\Rho$}
\DeclareUnicodeCharacter{03A3}{$\Sigma$}
\DeclareUnicodeCharacter{03A4}{$\Tau$}
\DeclareUnicodeCharacter{03A5}{$\Upsilon$}
\DeclareUnicodeCharacter{03A6}{$\Phi$}
\DeclareUnicodeCharacter{03A7}{$\Chi$}
\DeclareUnicodeCharacter{03A8}{$\Psi$}
\DeclareUnicodeCharacter{03A9}{$\Omega$}

\DeclareUnicodeCharacter{03B1}{$\alpha$}
\DeclareUnicodeCharacter{03B2}{$\beta$}
\DeclareUnicodeCharacter{03B3}{$\gamma$}
\DeclareUnicodeCharacter{03B4}{$\delta$}
\DeclareUnicodeCharacter{03B5}{$\epsilon$}
\DeclareUnicodeCharacter{03B6}{$\zeta$}
\DeclareUnicodeCharacter{03B7}{$\eta$}
\DeclareUnicodeCharacter{03B8}{$\theta$}
\DeclareUnicodeCharacter{03B9}{$\iota$}
\DeclareUnicodeCharacter{03BA}{$\kappa$}
\DeclareUnicodeCharacter{03BB}{$\lambda$}
\DeclareUnicodeCharacter{03BC}{$\mu$}
\DeclareUnicodeCharacter{03BD}{$\nu$}
\DeclareUnicodeCharacter{03BE}{$\xi$}
\DeclareUnicodeCharacter{03BF}{$\omicron$}
\DeclareUnicodeCharacter{03C0}{$\pi$}
\DeclareUnicodeCharacter{03C1}{$\rho$}
\DeclareUnicodeCharacter{03C2}{$\varsigma$}
\DeclareUnicodeCharacter{03C3}{$\sigma$}
\DeclareUnicodeCharacter{03C4}{$\tau$}
\DeclareUnicodeCharacter{03C5}{$\upsilon$}
\DeclareUnicodeCharacter{03C6}{$\phi$}
\DeclareUnicodeCharacter{03C7}{$\chi$}
\DeclareUnicodeCharacter{03C8}{$\psi$}
\DeclareUnicodeCharacter{03C9}{$\omega$}

\DeclareUnicodeCharacter{03B8}{$\theta$}
\DeclareUnicodeCharacter{03D1}{$\vartheta$}
\DeclareUnicodeCharacter{03D5}{$\phi$}
\DeclareUnicodeCharacter{03D6}{$\varpi$}
\DeclareUnicodeCharacter{03F0}{$\varkappa$}
\DeclareUnicodeCharacter{03F1}{$\varrho$}
\DeclareUnicodeCharacter{03F4}{$\varphi$}
\DeclareUnicodeCharacter{03F5}{$\varpi$}
\DeclareUnicodeCharacter{03F6}{$\varsigma$}
\DeclareUnicodeCharacter{03F7}{$\vartheta$}
\DeclareUnicodeCharacter{03F9}{$\varsigma$}
\DeclareUnicodeCharacter{03FA}{$\vartheta$}
\DeclareUnicodeCharacter{03FB}{$\vartheta$}
\DeclareUnicodeCharacter{03FC}{$\vartheta$}
\DeclareUnicodeCharacter{03FD}{$\vartheta$}
\DeclareUnicodeCharacter{03FE}{$\vartheta$}
\DeclareUnicodeCharacter{03FF}{$\vartheta$}

\DeclareUnicodeCharacter{1F10}{$\alpha$}
\DeclareUnicodeCharacter{1F11}{$\beta$}
\DeclareUnicodeCharacter{1F12}{$\gamma$}
\DeclareUnicodeCharacter{1F13}{$\delta$}
\DeclareUnicodeCharacter{1F14}{$\epsilon$}
\DeclareUnicodeCharacter{1F15}{$\zeta$}
\DeclareUnicodeCharacter{1F16}{$\eta$}
\DeclareUnicodeCharacter{1F17}{$\theta$}
\DeclareUnicodeCharacter{1F18}{$\iota$}
\DeclareUnicodeCharacter{1F19}{$\kappa$}
\DeclareUnicodeCharacter{1F1A}{$\lambda$}
\DeclareUnicodeCharacter{1F1B}{$\mu$}
\DeclareUnicodeCharacter{1F1C}{$\nu$}
\DeclareUnicodeCharacter{1F1D}{$\xi$}
\DeclareUnicodeCharacter{1F1E}{$\omicron$}
\DeclareUnicodeCharacter{1F1F}{$\pi$}
\DeclareUnicodeCharacter{1F20}{$\rho$}
\DeclareUnicodeCharacter{1F21}{$\varsigma$}
\DeclareUnicodeCharacter{1F22}{$\sigma$}
\DeclareUnicodeCharacter{1F23}{$\tau$}
\DeclareUnicodeCharacter{1F24}{$\upsilon$}
\DeclareUnicodeCharacter{1F25}{$\phi$}
\DeclareUnicodeCharacter{1F26}{$\chi$}
\DeclareUnicodeCharacter{1F27}{$\psi$}
\DeclareUnicodeCharacter{1F28}{$\omega$}
\DeclareUnicodeCharacter{1F29}{$\vartheta$}
\DeclareUnicodeCharacter{1F2A}{$\varkappa$}
\DeclareUnicodeCharacter{1F2B}{$\varrho$}
\DeclareUnicodeCharacter{1F2C}{$\varsigma$}
\DeclareUnicodeCharacter{1F2D}{$\varphi$}
\DeclareUnicodeCharacter{1F2E}{$\varpi$}
\DeclareUnicodeCharacter{1F2F}{$\vartheta$}

\DeclareUnicodeCharacter{2295}{$\oplus$}
\DeclareUnicodeCharacter{2296}{$\ominus$}
\DeclareUnicodeCharacter{2297}{$\otimes$}
\DeclareUnicodeCharacter{221A}{$\surd$}
\DeclareUnicodeCharacter{27E8}{$\langle$}
\DeclareUnicodeCharacter{27E9}{$\rangle$}

\DeclareUnicodeCharacter{2032}{${}^{\prime}$}

\DeclareUnicodeCharacter{251C}{$d$}
\DeclareUnicodeCharacter{2500}{$c$}
\DeclareUnicodeCharacter{2502}{$b$}
\DeclareUnicodeCharacter{2514}{$a$}

\begin{document}

\title{\vspace{-0.5cm}\includegraphics[width=0.85\textwidth]{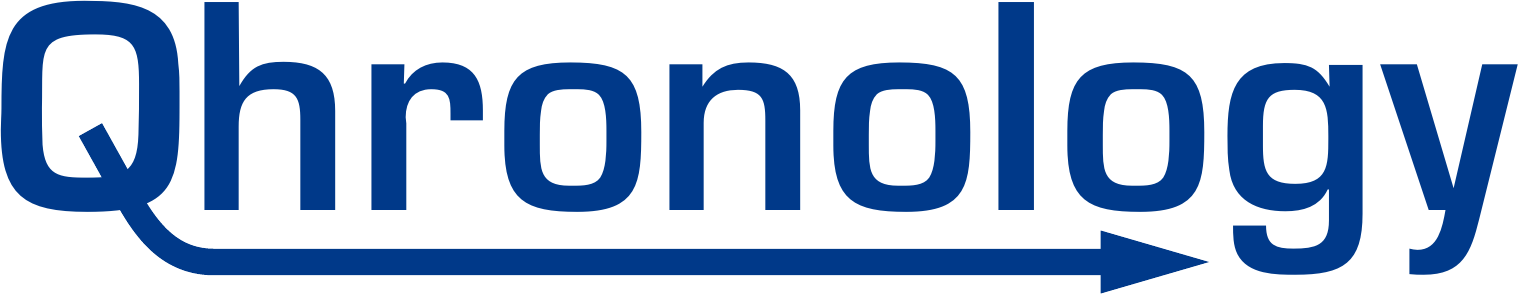}\\\vspace{0.15cm} \sffamily\bfseries\fontsize{12}{16}\selectfont\textcolor{colorlogo}{A Python package for studying quantum models of closed timelike curves}}

\author{\normalsize\sffamily\bfseries Lachlan G. Bishop\thanks{\texttt{lgbishop@protonmail.com}}\vspace*{-0.5em}}
\affil{\href{https://orcid.org/0000-0001-7852-5489}{\includegraphics[scale=0.065]{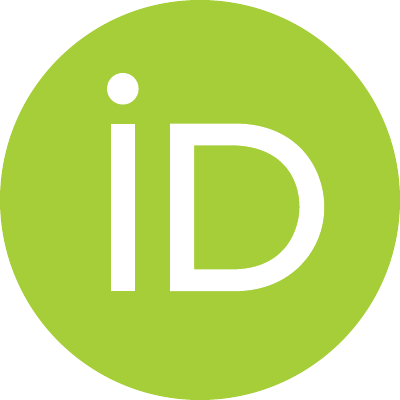} \raisebox{0.35em}{\textnormal{\texttt{0000-0001-7852-5489}}}}}

\date{\vspace{0.15cm}\small\sffamily\bfseries\today\vspace{-0.55cm}}

\maketitle

\begin{abstract}\vspace{0.15cm}
\noindent\emph{Qhronology} is a novel scientific-computing package for studying quantum models of closed timelike curves (CTCs) and simulating general quantum information processing and computation. Written in Python, the program provides a comprehensive framework for analyzing quantum theories of antichronological time travel, including functionality to calculate quantum resolutions to temporal paradoxes. It also operates as a complete quantum circuit simulator, enabling the examination of quantum algorithms and protocols in both numerical and symbolic capacities. In this paper, we formally introduce Qhronology, beginning with discussion on aspects of its design philosophy and architecture. An overview of its basic usage is then presented, along with a collection of examples demonstrating its various capabilities within a variety of distinct contexts. Lastly, the performance of the package's circuit simulation component is characterized by way of some simple empirical benchmarking.
\end{abstract}

\vspace*{-0.25cm}



\vspace{0.15cm}
\begin{center}

\renewcommand{\arraystretch}{1.35}
\begin{NiceTabular}{*{2}{l}}[corners, hlines, first-row, respect-arraystretch]
\Block[c, fill=lightblue]{1-2}{\textbf{\textsf{Project Information}}} \\
\textbf{\textsf{Package name}} & \href{https://github.com/lgbishop/qhronology}{\texttt{qhronology}} \\
\textbf{\textsf{Repository}} & \url{https://github.com/lgbishop/qhronology} \\
\textbf{\textsf{Author}} & Lachlan G. Bishop (\href{https://github.com/lgbishop}{\texttt{lgbishop}})  \\
\textbf{\textsf{Latest version}} & \texttt{1.1.1} (28th July 2026) \\
\textbf{\textsf{Programming language}} & \href{https://www.python.org}{Python} (\texttt{>=3.11}) \\
\textbf{\textsf{Dependencies}} & \Block[l]{}{\href{https://sympy.org}{\texttt{sympy}} (\texttt{>=1.12}) \\ \href{https://numpy.org}{\texttt{numpy}} (\texttt{>=1.26})} \\
\textbf{\textsf{Download}} & \Block[l]{}{PyPI: \url{https://pypi.org/project/qhronology}\\ GitHub: \url{https://github.com/lgbishop/qhronology/releases}} \\
\textbf{\textsf{Documentation}} & \Block[l]{}{Website: \url{https://qhronology.org}\\ PDF: \href{https://raw.githubusercontent.com/lgbishop/qhronology/latest/docs/_build/latex/Qhronology.pdf}{\texttt{Qhronology.pdf}}} \\
\textbf{\textsf{License}} & \Block[l]{}{Non-commercial use: \href{https://github.com/lgbishop/qhronology/blob/latest/licenses/AGPL-3.0-or-later.txt}{\texttt{AGPL-3.0-or-later}} \\ Commercial use: contact the author to arrange a license agreement} \\
\end{NiceTabular}

\end{center}

\newpage


\tableofcontents
\newpage


\begin{center}
\raisebox{1.5\baselineskip}{\includegraphics[width=\textwidth]{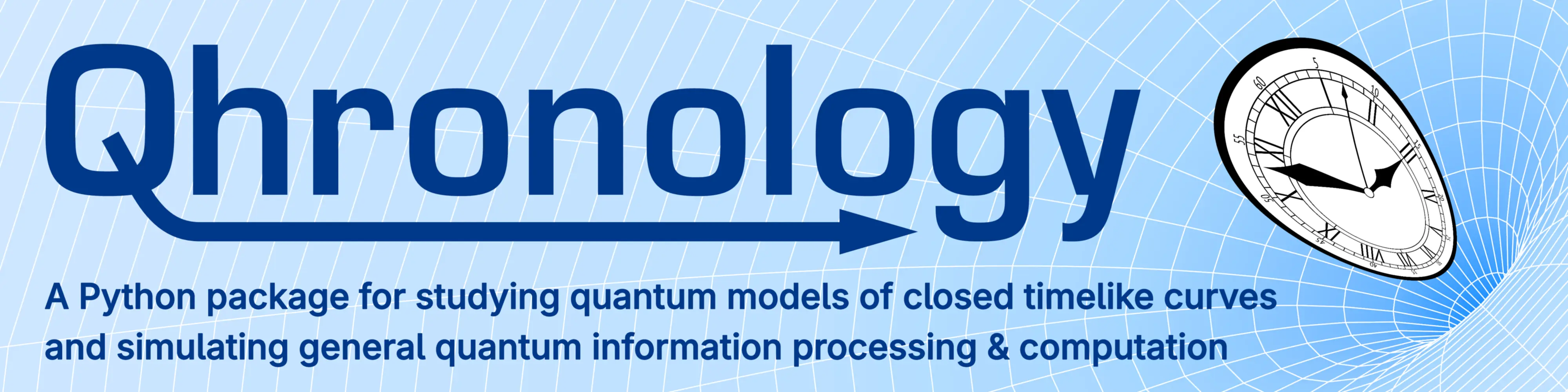}}
\end{center}
\vspace*{-3.00\baselineskip}
\hfill\footnote{The project's banner logo. Visit the official website to see the animated version!}\hspace{-0.5em}
\vspace*{-1.35\baselineskip}

\enlargethispage{1.15\baselineskip}

\section{Introduction}\label{sec:introduction}
\vspace*{-0.15\baselineskip}

Perhaps the most fascinating aspect of the current research on closed timelike curves (CTCs) is that there is more than one way to treat them with quantum mechanics to a sufficient degree of plausibility. While none of the established quantum models (sometimes called \emph{prescriptions}) has been accepted by scientific consensus as the authoritative description of how the Universe may potentially conduct antichronological time travel, all of them can be used to predict evolutions of the time-travelling quantum systems associated with CTCs. Though these predictions often disagree in stark ways, their differences can also be subtle, which may likewise be true for any number of the prescriptions' other aspects, including their mathematical characteristics (such as [non-]linearity and [non-]unitarity in state evolution) and physical consequences (such as the abilities to distinguish non-orthogonal states, clone arbitrary states, signal superluminally, and increase the computational speed and efficiency of both classical and quantum computers). A major part of the research into the various prescriptions therefore is how, despite being formulated with radically distinct (and usually fundamentally incompatible) postulates, they all provide self-consistent, paradox-free evolutions of quantum systems near CTCs (including resolutions to temporal paradoxes), with the associated quantum states themselves always being physically valid, at least in the sense of standard quantum mechanics.

In the absence of physical CTCs on which to perform experiments, one of the more effective ways to investigate the various quantum models of time travel is to compare their theoretical predictions for the states of the quantum systems both internal (\emph{chronology-violating}, or CV) and external (\emph{chronology-respecting}, or CR) to a CTC in the context of specific inter-system interactions. Doing so however can prove to be both tedious and error-prone, especially in cases where the interactions are complex. For example, \emph{Deutsch's model} (giving \emph{Deutschian} CTCs, or D-CTCs) \cite{deutsch_quantum_1991} necessitates the solving of a fixed-point condition for which there is often a parametrized spectrum of (non-unique) CV solutions (and, accordingly, a corresponding spectrum of CR solutions). Similarly, the theory of \emph{postselected teleportation} (giving \emph{postselected} CTCs, or P-CTCs) \cite{lloyd_quantum_2011, lloyd_closed_2011, bennett_simulated_2005, svetlichny_effective_2009, svetlichny_time_2011} requires one to non-unitarily evolve input states and subsequently renormalize the results. Needless to say, the procedures for computing the predictions of these prescriptions are highly involved, and so must be performed with great care.

Born out of the desire for a way to programmatically compute the states of the CR and CV systems according to the foremost quantum prescriptions of antichronological time travel, \emph{Qhronology}\footnote{\url{https://qhronology.org}} \cite{bishop_qhronology-software_2025} was created as a unified computational environment for defining, simulating, and analyzing quantum information processes that incorporate CTCs. Notably, the package can be used to calculate quantum resolutions to any given temporal paradox, thereby enabling users to explore foundational questions regarding the quantum mechanics of time travel. By providing a unique approach to describing general quantum objects (such as states and gates), Qhronology can also operate as a complete quantum circuit simulator, of which a prominent component is its engine for the visualization of quantum circuit diagrams. Its main features include:

\begin{itemize}
    \item Calculation of the states of the CR and CV quantum systems according to quantum-mechanical prescriptions of closed timelike curves
    \begin{itemize}
        \item Deutsch's model (D-CTCs)
        \item Postselected teleportation (P-CTCs)
    \end{itemize}
    \item Simulation of general quantum information processing and computation
    \begin{itemize}
        \item Symbolic calculations involving any number of variables and parameters
        \item Numerical (classical) replication of quantum experiments
    \end{itemize}
    \item Visualization of quantum circuit diagrams
    \begin{itemize}
        \item Text-based semigraphical diagrams constructed from fixed-width typefaces
    \end{itemize}
\end{itemize}

\enlargethispage{\baselineskip}
The primary purpose of Qhronology is to facilitate the study of quantum theories of antichronological time travel and quantum algorithms of quantum computing in both educational and research capacities. As part of this, the project aims to make the expression of quantum states, gates, circuits, and models of CTCs near-limitlessly possible within a framework that is syntactically simple, informationally dense, mathematically powerful, extremely flexible, and easily extensible. Qhronology therefore provides a sufficiently complete and self-contained set of tools with the intention that needing to use external packages and libraries to perform transformations on its quantum constructs should not (usually) be necessary. Its underlying mathematical system accomplishes this using the standard $\Dimension$-dimensional matrix mechanics of discrete-variable quantum theory in a general $\Complexes^\Dimension$-representation.

Qhronology is written entirely in the Python\footnote{\url{https://www.python.org}} programming language. Being high-level, dynamically type-checked, and interpreted (at least within the context of its CPython reference implementation), Python is well-suited for building an accessible framework that emphasizes interactivity and scriptability. Additionally, like any popular language, it has both an extensive standard library and a plethora of powerful community packages available to it. Qhronology is built around features from two such packages: the eminent SymPy\footnote{\url{https://sympy.org}} \cite{meurer_sympy_2017} and NumPy\footnote{\url{https://numpy.org}} \cite{harris_array_2020} projects. In particular, the package greatly leverages the linear algebra capabilities of both of these, and so aims to have a deep compatibility with their respective matrix and array objects. It is therefore hoped that users who possess experience with these projects find Qhronology's interface to be both familiar and intuitive.

Although not designed to produce publication-quality circuit diagrams, Qhronology's visualization engine can create simple diagrams that are reasonably faithful to the canonical quantum circuitry picturalism. A notable benefit of this feature is that it allows the user to verify (visually) that the circuits which they construct within the framework match their intentions. Extensive modification of any diagram's appearance is also possible, with spacing, padding, and style ({\ie}, character set) all completely customizable. Note that all of the circuit diagrams contained within this paper are LaTeX-rendered (but otherwise unaltered) raw text output from the program.

While Qhronology can accomplish all of the aforementioned tasks (albeit to varying degrees of success), it is important to be aware of its limitations. Perhaps the most significant of these is that, while the package possesses a high-level interface to its powerful internal framework, it cannot be used to write programs which execute on physical quantum computing hardware{\textemdash}a limitation mainly due to such functionality ({\ie}, the ability for logical quantum circuits to be compiled [or transpiled] to low-level representations or instruction sets [{\eg}, OpenQASM] for any target quantum architecture) being out of the project's scope. Despite this, Qhronology is well-suited for intensive numerical calculations (such as circuit simulation), as the accelerated floating-point arithmetic of its NumPy backend is computationally fast and efficient, especially for hardware with multithreaded processor architecture.

\enlargethispage{\baselineskip}
Another aspect of Qhronology it is important to acknowledge is that the program itself, at least in its current form, is considered to be highly experimental. Its output may be not always be correct, and some features may not work as intended. Additionally, note that all components of the package, including its functions, methods, classes, modules, and subpackages, may be subject to change in future versions. With this in mind, as its source code is distributed under version 3 of the GNU Affero General Public License (AGPL-3.0-or-later) for non-commercial use, contributions to the project are always welcome, in accordance with the spirit of open-source software development. These can include anything from small fixes to new features, provided they conform to the project's vision and style. Potential contributors are welcome to contact the author to discuss any changes they may wish to propose.

\enlargethispage{-\baselineskip}
We begin this paper with a discussion on Qhronology's internal architecture, including some of the significant decisions that went into its design, in {\secn} \ref{sec:architecture}. Next, in {\secn} \ref{sec:usage}, a brief tutorial on the usage of the program is given, along with numerous code examples of each part of its core functionality. \capitalisewords{\secn} \ref{sec:examples} follows this with a small collection of complete examples showcasing how Qhronology can be used to great success for a variety of different tasks. An examination of the package's performance, as observed in a small set of empirical benchmarks, is then presented in {\secn} \ref{sec:performance}. Finally, concluding remarks are given in {\secn} \ref{sec:conclusion}.

\section{Architecture}\label{sec:architecture}

\subsection{Package structure}\label{sec:structure}

\begin{figure}[h!]
\begin{center}
\begin{forest}
  for tree={
    folder,
    grow'=0,
    fit=band,
    l sep=1.5em,
    s sep=0.0em,
    edge=thick,
    font=\normalsize\ttfamily,
  }
  [qhronology
    [quantum
      [circuits.py]
      [gates.py]
      [prescriptions.py]
      [states.py]
    ]
    [mechanics
      [matrices.py]
      [operations.py]
      [quantities.py]
    ]
    [utilities
      [classification.py]
      [diagrams.py]
      [helpers.py]
      [objects.py]
      [symbolics.py]
    ]
  ]
\end{forest}
\end{center}
\vspace*{-0.25cm}
\caption[Package filesystem structure diagram]{\label{fig:structure_diagram}The directory structure of Qhronology, depicting the hierarchy of the various subpackages and modules within the package.}
\end{figure}
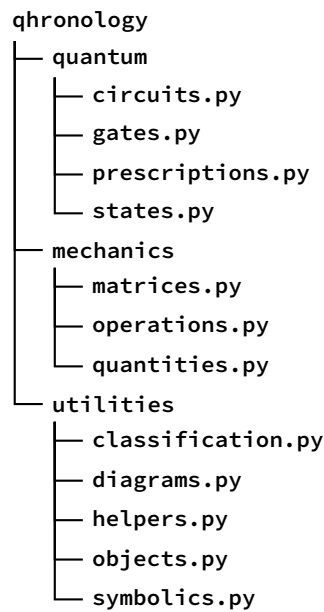

The directory structure of the package appears in {\fig} \ref{fig:structure_diagram}. It consists of three subpackages:
\begin{itemize}
    \item \mintinline{text}{quantum}: most of the package's underlying mathematical framework and its user-facing classes.
    \item \mintinline{text}{mechanics}: core logic for creating fundamental quantum objects, performing operations on them, and computing various quantum-mechanical scalar quantities.
    \item \mintinline{text}{utilities}: a collection of modules containing various functionality intended for internal-use only, including the visualization engine, the core \py{QuantumObject} class, and assorted helper functions.
\end{itemize}
In this paper, we focus solely on the \mintinline{text}{quantum} subpackage, which contains all of the classes intended to be user-facing. Occasionally however, functionality from the \mintinline{text}{mechanics} subpackage may be required in various code snippets and examples, and in such cases, it is simply imported and used. For more detail, please consult the project's official documentation \cite{bishop_qhronology-documentation_2025}.

\vspace*{-1\baselineskip}

\subsection{Classes and their relationships}\label{sec:classes}

\enlargethispage{\baselineskip}
Qhronology presents an innovative approach to describing both simulations of quantum mechanics and the various associated mathematical constructs and processes which collectively form the foundation of contemporary quantum physics. While this is not novel in the world of quantum libraries and toolkits, Qhronology is unique in how it places great emphasis on modularity in its descriptions of quantum states, gates, and circuits. Perhaps the best way by which this aspect can be appreciated is to gain a deeper understanding of the package's interface, specifically that manifested by its user-facing classes.

\begin{figure}[t!]
    \begin{center}
    \includegraphics[scale=1.15]{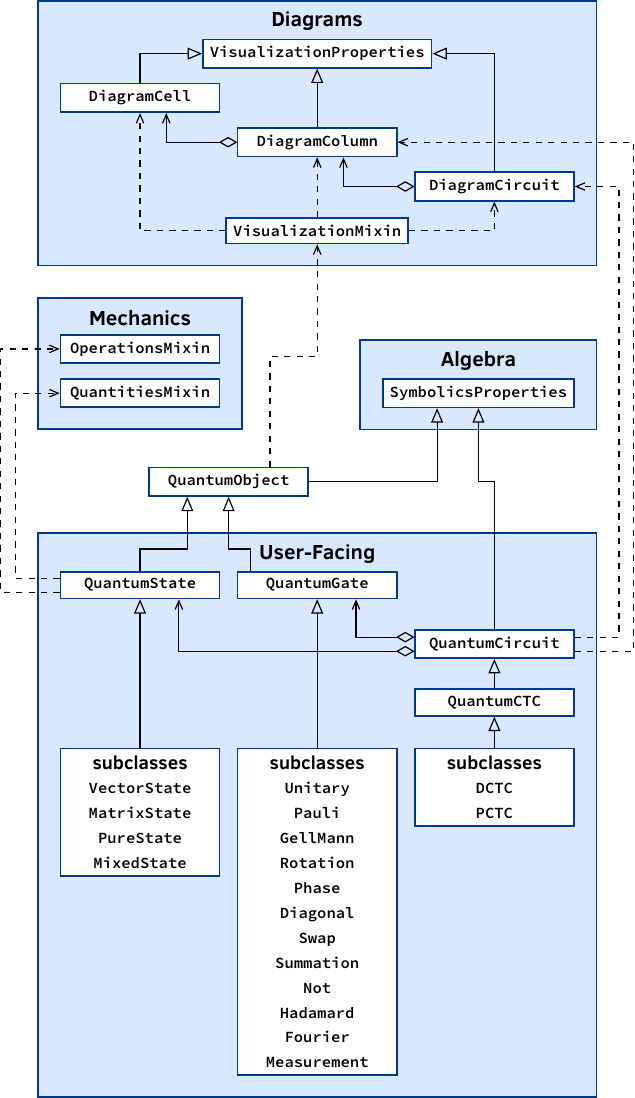}
    \end{center}
    \caption[UML class diagram]{\label{fig:UML_class_diagram}A simplified UML (Unified Modeling Language) class diagram visualizing the relationships between Qhronology's core classes.}
    \vspace*{-0.50cm}
\end{figure}

\capitalisewords{\fig} \ref{fig:UML_class_diagram} contains a simplified UML class diagram, depicting the majority of the package's class objects and the relationships (including inheritance and composition) between them. As shown in this diagram, the \py{QuantumObject} class is central to Qhronology's functionality. Being an \emph{abstract} base class (and so is not meant to be instantiated itself), \py{QuantumObject} constitutes the common substructure upon which all primitive quantum-mechanical objects are built. Specifically, this consists of the two main classes:
\begin{itemize}
    \item \py{QuantumState}: for creating quantum states. Instances have \emph{mutable} internal states.
    \item \py{QuantumGate}: for creating quantum gates. Instances have \emph{immutable} internal states.
\end{itemize}
These are implemented as \emph{extending} subclasses, where each adds to and modifies the functionality of the \py{QuantumObject} base class primarily through class properties and methods. Instances of these derived classes provide exhaustive descriptions of their corresponding quantum constructs: in addition to containing a precise mathematical specification (including metadata regarding symbols and their associated constraints), they can be inspected, visualized, and, in the case of quantum states, transformed (mutated) {\via} quantum operations. Here, \py{QuantumObject} provides the core matrix, symbolic, and visualization machinery (in addition to all other internal implementation details) required by Qhronology's programmatic description of fundamental quantum objects. Thus, as both states and gates are simply just specific types of such objects, then using the \py{QuantumObject} class as a shared foundation is a natural arrangement{\textemdash}one which greatly simplifies the project's source code by directly reducing redundancy.

Having gained an appreciation for Qhronology's implementation of quantum states and gates, we now turn our attention to that of quantum circuits. In standard quantum theory, a circuit can be thought of as being essentially an assembly of states and gates that corresponds to a mathematical description of some specific quantum-mechanical process. This ontology is reflected in Qhronology's circuit construction{\textemdash}facilitated by the \py{QuantumCircuit} class{\textemdash}whereby circuit instances are created by passing \py{QuantumState} and \py{QuantumGate} objects to the appropriate arguments in the class constructor (or, alternatively, to the appropriate properties post-instantiation). In a pragmatic (albeit reductive) sense, \py{QuantumCircuit} objects are merely containers for their various elementary components (while also possessing other necessary functionality). Of all the class's abilities, the most defining is the computation of the output state (of the circuit assembled from its state and gate components), which may be returned as a SymPy matrix, NumPy array, or \py{QuantumState} instance.

Building upon \py{QuantumCircuit}, Qhronology's \py{QuantumCTC} class provides the core scaffolding by which quantum circuits that possess CTCs are described. As far as the programmatic specification of such circuits is concerned, the main practical difference between them and their CTC-free counterparts is that the quantum systems of the former are not exclusively chronology-respecting{\textemdash}those which comprise the CTC subsystem are, by definition, chronology-violating. Qhronology recognizes this sentiment, and accordingly \py{QuantumCTC} extends the \py{QuantumCircuit} class only by the addition of the \py{systems_respecting} and \py{systems_violating} constructor arguments (along with homonymous class properties).

Instances of the \py{QuantumCTC} class, constructed in much the same way as \py{QuantumCircuit} instances ({\eg}, the passing of states and gates to the class constructor's relevant arguments), provide descriptions of specific interactions (defined by the given gates) between the CR and CV subsystems (with input state given on the former). By itself however, the \py{QuantumCTC} class is unable to compute the output states on these subsystems, as this can only be accomplished within the context of a particular quantum prescription. Such models can be implemented simply as subclasses of \py{QuantumCTC}, wherein the appropriate machinery is granted to the class methods which require it. Indeed, Qhronology provides the two foremost prescriptions, D-CTCs and P-CTCs, as the \py{QuantumDCTC} and \py{QuantumPCTC} classes, respectively. It is these subclasses which provide the capabilities for calculating the specific predictions of their associated quantum prescription of antichronological time travel.

\clearpage

\subsection{Modularity and design patterns}\label{sec:modularity}

Qhronology provides a flexible domain-specific framework for theoretical quantum mechanics, with the usage pattern of its circuit creation in particular being a highly modular procedure. As part of this, the program places great emphasis on the ability to define quantum primitives outside of the context of any circuit description. In other words, Qhronology treats individual states and gates as complete, fundamental, and independent objects.

The standard pattern of Qhronology's circuit instantiation (described in {\secn} \ref{sec:circuits}) consists primarily of the passing of pre-existing state (\py{QuantumState}) and gate (\py{QuantumGate}) instances to the circuit (\py{QuantumCircuit}) constructor's arguments. With an assembled circuit, one can then extract its total (composite) input and output states as \py{QuantumState} instances, as well as the entire gate sequence as a single amalgamated \py{QuantumGate} instance. These extracted objects are no different from any other user-created states or gates in Qhronology, and so can be inspected, modified, and incorporated into other circuits. Qhronology's claim to extensive modularity stems primarily from its ability to both assemble and extract elementary objects, which themselves can be subsequently used in a variety of contexts.

Although the typical procedure of creating circuits within the framework does not precisely resemble any single design pattern, it does possess some traits from what is canonically known as the \emph{composite pattern}{\textemdash}a software-engineering design pattern \cite{gamma_design_1994} characterized by the aggregation of elementary objects into a single, more complex object. Qhronology's particular structure however diverges from this pattern in one important way: while the \py{QuantumCircuit} class facilitates the composition of fundamental objects (states and gates) into a more complex container (from which both all kinds of fundamental objects can be obtained), it itself cannot be further composed into other objects, nor does it share the same interface as its constituent objects. Nonetheless, the program's process of creating circuits (almost) completely from pre-defined primitives at instantiation is a distinctly novel approach, one that is in stark contrast with the \emph{builder pattern}{\textemdash}quantum circuits constructed {\via} a succession of methods called on an initially empty quantum circuit class instance{\textemdash}used in other Python-based quantum projects, including IBM's \emph{Qiskit} \cite{ibm_qiskit_2017, javadi-abhari_quantum_2024}, Google's \emph{Cirq} \cite{google_cirq_2018}, Xanadu's \emph{PennyLane} \cite{xanadu_pennylane_2018, bergholm_pennylane_2022}, and the community-run \emph{QuTiP} \cite{the_qutip_community_qutip_2012, lambert_qutip_2026}.

\enlargethispage{1.5\baselineskip}
\vspace*{-0.15\baselineskip}

\section{Usage}\label{sec:usage}

Qhronology's functionality revolves around a few central concepts{\textemdash}states, gates, and circuits{\textemdash}all of which should be familiar to those with experience in quantum information science. In this section, the implementation and usage of each are briefly detailed. Note that in the following examples, while not always explicitly shown, the SymPy and NumPy packages are imported into the namespace in the conventional manner:
\begin{code}
\begin{minted}{python}
>>> import sympy as sp
>>> import numpy as np
\end{minted}
\end{code}

Being a Python project, Qhronology is most accessible to users who possess at least intermediate proficiency in the programming language. However, even in the complete absence of knowledge of Python, much can still be achieved simply by copying the provided examples and adapting them to desire. Note that the features and functionality presented here are not exhaustive{\textemdash}please see Qhronology's official documentation\footnote{\url{https://qhronology.org/text/documentation/part-documentation.html}} \cite{bishop_qhronology-documentation_2025} for comprehensive information on the package, including descriptions of all properties and methods available to every user-facing class.

As with any quantum program, Qhronology follows a few conventions of which it is important to be aware. The most significant of these is that the indexing of systems begins at 0, as customary for the labelling of elements in a sequence within computer science. Furthermore, the standard correspondence (as per canonical quantum theory) in the ordering of systems between the mathematical representation's tensor products (left-to-right) and circuit's wires (top-to-bottom) is followed. Additionally, in all quantum circuit diagrams, time advances horizontally from left to right, and so the input states (forming a collection often known as the circuit's \emph{register}) are on the left, while the output states are on the right.

\subsection{States}\label{sec:states}

In Qhronology, quantum states are described in the \emph{computational} basis (also known as the \emph{standard} basis or the $z$-basis) and represented by instances of the \py{QuantumState} class:
\begin{code}
\begin{minted}{python}
>>> from qhronology.quantum.states import QuantumState
\end{minted}
\end{code}
In the construction of an instance of this class, the characterization of the quantum state is facilitated primarily by four arguments:
\begin{itemize}
    \item \py{spec}: quantifies the values ({\ie}, amplitudes or probabilities) corresponding to specific components of the state's mathematical representation.
    \item \py{form}: describes the state as being either a \py{"vector"} or a \py{"matrix"}. Defaults to \py{"vector"}.
    \item \py{kind}: describes the state as being either \py{"pure"} or \py{"mixed"}. Defaults to \py{"pure"}.
    \item \py{dim}: quantifies the state dimensionality as an integer greater than or equal to \py{2}. Defaults to \py{2}.
\end{itemize}
Note that, of the four combinations (pairs) of the values which may be passed to \py{form} and \py{kind}, all are valid except for the pairing of \py{"vector"} and \py{"mixed"}.

The object passed to \py{spec} can be any of the following four types:
\begin{itemize}
    \item a SymPy matrix
    \item a NumPy array
    \item a list of lists (describing a matrix)
    \item a list of 2-tuples
\end{itemize}
The data type of the elements contained within the first three of these options can be any of the following: numerical (including all scalars from SymPy, NumPy, and the standard library), SymPy symbolic (including expressions), or string representations of such scalar types. However, the fourth option{\textemdash}the bespoke list-of-tuples format{\textemdash}is intended to be the primary way of characterizing quantum states in Qhronology. Its structure is reasonably straightforward: each 2-tuple contains an amplitude or probability (a scalar expression as a numerical, symbolic, or string value) followed by a list of non-negative integers corresponding to the levels of the number states of the desired basis vector. The resulting quantum state is simply the sum of the components corresponding to each individual tuple. For example, passing the list \py{[("α", [0, 0]), ("β", [1, 1])]} to \py{spec} in a \py{QuantumState} construction yields a state which corresponds to one of the following forms (depending on the values passed to the other core arguments):
\begin{itemize}
    \item \py{form} is \py{"vector"}: $\alpha\ket{0,0} + \beta\ket{1,1}$
    \item \py{form} is \py{"matrix"}:
    \begin{itemize}
        \item \py{kind} is \py{"pure"}: $\abs{\alpha}^2\ket{0,0}\bra{0,0} + \alpha\conj{\beta}\ket{0,0}\bra{1,1} + \conj{\alpha}\beta\ket{1,1}\bra{0,0} + \abs{\beta}^2\ket{1,1}\bra{1,1}$
        \item \py{kind} is \py{"mixed"}: $\alpha\ket{0,0}\bra{0,0} + \beta\ket{1,1}\bra{1,1}$
    \end{itemize}
\end{itemize}
Thus, despite being an unfamiliar way to describe the values and structure of quantum states, the list-of-tuples format is concise and powerful, making it highly useful.

\enlargethispage{2\baselineskip}

The usage of both the \py{QantumState} class and much of the functionality available to it can be elucidated with some examples. We begin by creating perhaps the simplest non-trivial state: a qubit describing an equal superposition of the primitive basis vectors $\ket{0}$ and $\ket{1}$. In this case, \py{form} must be \py{"vector"}, and so we do not need to specify the \py{kind} argument (as only a value of \py{"pure"} is valid here, which is assumed by default). Thus, the state can be instantiated as:
\begin{code}[label=code:state_simplest]
\begin{minted}{python}
>>> vector_state = QuantumState(
...     spec=[(1, [0]), (1, [1])],
...     form="vector",
... )
\end{minted}
\end{code}

\subsubsection{Inspection}\label{sec:states_inspection}

In order to verify that a quantum state such as that in {\cdb} \ref{code:state_simplest} is as intended, we can inspect the object instance \py{vector_state} using a few class methods. For its matrix representation, there is the \py{output()} method:
\begin{code}
\begin{minted}{python}
>>> vector_state.output()
Matrix([
[1],
[1]])
\end{minted}
\end{code}
For the mathematical expression ({\ie}, bra-ket notation), we can use the \py{print()} method:
\begin{code}
\begin{minted}{python}
>>> vector_state.print()
|ψ⟩ = |0⟩ + |1⟩
\end{minted}
\end{code}
The state can also be visualized {\via} the \py{diagram()} method:
\begin{code}
\begin{minted}{python}
>>> vector_state.diagram()
$\includegraphics[scale=1.25, trim=-0.02cm 0 0 -0.15cm]{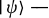}$
\end{minted}
\end{code}
All three of these methods take modifying arguments{\textemdash}see the documentation \cite{bishop_qhronology-documentation_2025} for more detail.

\subsubsection{Normalization}\label{sec:states_normalization}

One significant aspect of the state in {\cdb} \ref{code:state_simplest} is that it is not normalized, which is of course contrary to the requirements of any valid quantum state. This can be rectified in two distinct ways: either manually by specifying appropriate values in \py{spec}, or automatically by using the \py{norm} argument (or property). For the manual approach, one simply passes a \py{spec} whose amplitudes (or probabilities) inherently describe a normalized state, {\eg},
\begin{code}
\begin{minted}{python}
>>> vector_state = QuantumState(
...     spec=[(1/sp.sqrt(2), [0]), (1/sp.sqrt(2), [1])],
...     form="vector",
... )
>>> vector_state.print()
|ψ⟩ = sqrt(2)/2|0⟩ + sqrt(2)/2|1⟩
\end{minted}
\end{code}
A better practice is to instead use the functionality provided by the \py{norm} argument, to which a normalization value is specified. This can be either a non-negative scalar or \py{bool} (\py{True} to normalize to unity):
\begin{code}
\begin{minted}{python}
>>> vector_state = QuantumState(
...     spec=[(1, [0]), (1, [1])],
...     form="vector",
...     norm=1,
... )
>>> vector_state.print()
|ψ⟩ = sqrt(2)/2|0⟩ + sqrt(2)/2|1⟩
\end{minted}
\end{code}
In any case, please be aware that normalization does not occur unless explicitly specified{\textemdash}Qhronology considers the creation of valid (physical) quantum states (including any normalization) to be the responsibility of the user.

\subsubsection{Labels and notation}\label{sec:states_labelling}

In the inspection of states (such as printing or diagramming), the default symbol used to represent vector states is the (lowercase) Greek letter \emph{psi} ($\psi$), while for matrix (density) states it is (lowercase) \emph{rho} ($\rho$). To use a specific symbol as a label for a state, simply pass the string representation of the desired character(s) to the \py{label} argument at creation. For example:
\begin{code}
\begin{minted}{python}
>>> vector_state = QuantumState(
...     spec=[(1, [0]), (1, [1])],
...     form="vector",
...     norm=1,
...     label="φ",
... )
>>> vector_state.print()
|φ⟩ = sqrt(2)/2|0⟩ + sqrt(2)/2|1⟩
>>> vector_state.diagram()
$\includegraphics[scale=1.25, trim=-0.02cm 0 0 -0.15cm]{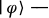}$
\end{minted}
\end{code}
Alternatively, the state's \py{label} property can be set after instantiation. Note that the \py{notation} argument (and property) can be used to completely override the (formatted) label in both diagrams and prints.

\subsubsection{Symbols and substitutions}\label{sec:states_symbols}

If we desire not an equiprobabilistic state but rather one with some parametrized (symbolic) amplitudes, we can create such a state simply by including the appropriate strings (which are subsequently converted into SymPy variables by the class) as the amplitudes within the \py{spec} argument, {\eg},
\begin{code}
\begin{minted}{python}
>>> vector_state = QuantumState(
...     spec=[("a", [0]), ("b", [1])],
...     form="vector",
... )
>>> vector_state.print()
|ψ⟩ = a|0⟩ + b|1⟩
\end{minted}
\end{code}
Note that these strings can contain any scalar expression (recognized by SymPy), meaning that many mathematical functions can also be used:\enlargethispage{\baselineskip}
\begin{code}
\begin{minted}{python}
>>> vector_state = QuantumState(
...     spec=[("sin(θ)", [0]), ("exp(I*φ)*cos(θ)", [1])],
...     form="vector",
... )
>>> vector_state.print()
|ψ⟩ = sin(θ)|0⟩ + exp(I*φ)*cos(θ)|1⟩
\end{minted}
\end{code}
For any non-numerical or non-functional characters in the state's specification, a corresponding SymPy \py{Symbol} object (of the same notation) is created and stored in the working memory. The mathematical properties of these symbols can be specified as a dictionary to the \py{symbols} argument, in which the keys are individual symbols (contained within the object's matrix representation), while the values are dictionaries that contain SymPy keyword-argument \emph{assumptions} (known also as \emph{predicates}). For example:\enlargethispage{\baselineskip}
\begin{code}[label=code:state_parametrized]
\begin{minted}{python}
>>> vector_state = QuantumState(
...     spec=[("a", [0]), ("b", [1])],
...     form="vector",
...     symbols={"a": {"complex": True}, "b": {"complex": True}},
... )
>>> vector_state.print()
|ψ⟩ = a|0⟩ + b|1⟩
\end{minted}
\end{code}
By default, all symbols are assumed to be both complex and commutative. The package's documentation \cite{bishop_qhronology-documentation_2025} contains a full list of currently supported predicates.

The information bestowed to instances of the \py{QuantumState} class by the \py{symbols} property can be very useful in defining quantum states with certain behaviours. Consider the following example, in which the symbols (\py{x} and \py{y}) in the expressions passed to \py{spec} are undefined (and so both are assumed to be complex), resulting in the state being unable to simplify itself when inspected:
\begin{code}
\begin{minted}{python}
>>> vector_state = QuantumState(
...     spec=[("conjugate(x)", [0]), ("sqrt(y**2)", [1])],
...     form="vector",
... )
>>> vector_state.print()
|ψ⟩ = conjugate(x)|0⟩ + sqrt(y**2)|1⟩
\end{minted}
\end{code}
If we instead pass the appropriate predicate(s) to each symbol in the \py{symbols} argument, we obtain the alternative form:
\begin{code}
\begin{minted}{python}
>>> vector_state = QuantumState(
...     spec=[("conjugate(x)", [0]), ("sqrt(y**2)", [1])],
...     form="vector",
...     symbols={"x": {"real": True}, "y": {"positive": True}},
... )
>>> vector_state.print()
|ψ⟩ = x|0⟩ + y|1⟩
\end{minted}
\end{code}
Naturally, the particular predicates specified for any given symbol should be guided by what the user wishes to achieve.

For complicated states, especially those possessing multiple symbols, adjunct (external) conditions are often required in order to fully encapsulate them. This includes algebraic constraints, identities, and other similar relations typically associated with quantum states. Such conditions can be imposed simply by assembling their scalar representations into equation 2-tuples (the equation's left-hand side and right-hand side expressions as an ordered pair) within a list that is passed to the constructor's \py{substitutions} argument (or the class property). For example, while normalization of our parametrized vector state ({\cdb} \ref{code:state_parametrized}) {\via} the \py{norm} argument does result in a normalized state,
\begin{code}
\begin{minted}{python}
>>> vector_state = QuantumState(
...     spec=[("a", [0]), ("b", [1])],
...     form="vector",
...     symbols={"a": {"complex": True}, "b": {"complex": True}},
...     norm=1,
... )
>>> vector_state.print()
|ψ⟩ = a/sqrt(a*conjugate(a) + b*conjugate(b))|0⟩ + b/sqrt(a*conjugate(a) + b*conjugate(b))|1⟩
\end{minted}
\end{code}\enlargethispage{\baselineskip}
it has no knowledge of any constraints or relations (inherent or otherwise) between its parameters, and so cannot present in a form which is any simpler. This can be amended by endowing it with an appropriate trace condition ($\abs{a}^2 + \abs{b}^2 = 1$) and using the \py{simplify} argument in the \py{print()} method:
\begin{code}
\begin{minted}{python}
>>> vector_state = QuantumState(
...     spec=[("a", [0]), ("b", [1])],
...     form="vector",
...     symbols={"a": {"complex": True}, "b": {"complex": True}},
...     substitutions=[("a*conjugate(a) + b*conjugate(b)", 1)],
...     norm=1,
... )
>>> vector_state.print(simplify=True)
|ψ⟩ = a|0⟩ + b|1⟩
\end{minted}
\end{code}
Thus, supplying a normalization constraint to \py{substitutions} can be used in conjuction with the \py{norm} argument to great effect.

\subsubsection{Conjugation}\label{sec:states_conjugation}

The last important argument (and property) of the \py{QuantumState} class is \py{conjugate}, which is used simply to perform Hermitian conjugation on the state whenever it is called, {\eg},
\begin{code}
\begin{minted}{python}
>>> vector_state = QuantumState(
...     spec=[("a", [0]), ("b", [1])],
...     form="vector",
...     conjugate=True,
... )
>>> vector_state.print()
⟨ψ| = conjugate(a)⟨0| + conjugate(b)⟨1|
\end{minted}
\end{code}
Note that when a vector state is used as input to a quantum circuit, it is necessarily converted into its ket (column) vector form ({\ie}, conjugation is removed if required) automatically, if it is not already in such a form.

\subsubsection{Density matrices}\label{sec:states_matrices}

\enlargethispage{2\baselineskip}

Having focused solely on vector states thus far, we turn our attention now to matrix states. There are of course two main archetypes: pure states and mixed states. To obtain either, one simply has to pass the appropriate string to the \py{kind} argument (while making sure that \py{"matrix"} is given to \py{form}):
\begin{code}
\begin{minted}{python}
>>> pure_state = QuantumState(
...     spec=[("a", [0]), ("b", [1])],
...     form="matrix",
...     kind="pure",
... )
>>> pure_state.print()
|ψ⟩⟨ψ| = a*conjugate(a)|0⟩⟨0| + a*conjugate(b)|0⟩⟨1| + b*conjugate(a)|1⟩⟨0| + b*conjugate(b)|1⟩⟨1|
\end{minted}
\end{code}
\begin{code}
\begin{minted}{python}
>>> mixed_state = QuantumState(
...     spec=[("a", [0]), ("b", [1])],
...     form="matrix",
...     kind="mixed",
... )
>>> mixed_state.print()
ρ = a|0⟩⟨0| + b|1⟩⟨1|
\end{minted}
\end{code}
Such states have identical usage and functionality to their vector counterparts.

\subsubsection{Multipartite (composite) states}\label{sec:states_multipartite}

So far, we have discussed only single-system (\emph{unipartite} or \emph{monopartite}) quantum states. As we shall see, composite vector states, {\ie}, those spanning more than one system (\emph{multipartite}), are just as easy to create, especially when using the list-of-tuples format for \py{QuantumState}'s \py{spec} argument. For example, a bipartite vector state may be constructed as follows:
\begin{code}
\begin{minted}{python}
>>> bipartite_vector = QuantumState(
...     spec=[("a", [0, 1]), ("b", [1, 0])],
...     form="vector",
...     symbols={"a": {"complex": True}, "b": {"complex": True}},
...     substitutions=[("a*conjugate(a) + b*conjugate(b)", 1)],
...     norm=1,
...     label="Ψ",
... )
>>> bipartite_vector.print(simplify=True)
|Ψ⟩ = a|0,1⟩ + b|1,0⟩
>>> bipartite_vector.print(simplify=True, delimiter="")
|Ψ⟩ = a|01⟩ + b|10⟩
>>> bipartite_vector.print(simplify=True, product=True)
|Ψ⟩ = a|0⟩⊗|1⟩ + b|1⟩⊗|0⟩
\end{minted}
\end{code}
Likewise, a general bipartite (mixed) density matrix state can be created as:
\begin{code}
\begin{minted}{python}
>>> bipartite_density = QuantumState(
...     spec=[("a", [0, 1]), ("b", [1, 0])],
...     kind="mixed",
...     symbols={"a": {"real": True}, "b": {"real": True}},
...     substitutions=[("a + b", 1)],
...     norm=1,
...     label="ρ",
... )
>>> bipartite_density.print(simplify=True)
ρ = a|0,1⟩⟨0,1| + b|1,0⟩⟨1,0|
>>> bipartite_density.print(simplify=True, delimiter="")
ρ = a|01⟩⟨01| + b|10⟩⟨10|
>>> bipartite_density.print(simplify=True, product=True)
ρ = a|0⟩⟨0|⊗|1⟩⟨1| + b|1⟩⟨1|⊗|0⟩⟨0|
\end{minted}
\end{code}

\subsubsection{Numerical representations}\label{sec:states_numerical}

For completely non-symbolic states ({\ie}, those possessing no SymPy symbols), their representations can be cast to inexact (floating-point) form (if not already) {\via} the \py{numerical} (\py{bool}) argument in all inspection methods (including both \py{print()} and \py{output()}). Additionally, using the \py{array} (\py{bool}) argument, the output of the \py{output()} method can be formatted as a NumPy array (with NumPy floating-point types), as opposed to the default of a SymPy matrix (with numerical SymPy types).

\newpage
For example:
\begin{code}
\begin{minted}{python}
>>> numerical_state = VectorState(spec=[(1/sp.sqrt(2),[0]),(1/sp.sqrt(2),[1])])
>>> numerical_state.print()
|ψ⟩ = sqrt(2)/2|0⟩ + sqrt(2)/2|1⟩
>>> numerical_state.print(numerical=True)
|ψ⟩ = 0.707106781186548|0⟩ + 0.707106781186548|1⟩
>>> numerical_state.output()
Matrix([
[sqrt(2)/2],
[sqrt(2)/2]])
>>> numerical_state.output(numerical=True)
Matrix([
[0.707106781186548],
[0.707106781186548]])
>>> numerical_state.output(array=True)
array([[sqrt(2)/2],
       [sqrt(2)/2]], dtype=object)
>>> numerical_state.output(numerical=True, array=True)
array([[0.70710678+0.j],
       [0.70710678+0.j]])
\end{minted}
\end{code}

Note also that both the \py{numerical} and \py{array} arguements have associated homonymous class properties (defaulting to \py{False}), the values of which serve as argument defaults in all class methods.

\subsubsection{Examples}\label{sec:states_examples}

Presented here is a collection of examples showcasing how different kinds of states can be created.

\begin{code}
\begin{minted}{python}
>>> qubit_vector = QuantumState(
...     spec=[("a", [0]), ("b", [1])],
...     form="vector",
...     symbols={"a": {"complex": True}, "b": {"complex": True}},
...     substitutions=[("a*conjugate(a) + b*conjugate(b)", 1)],
...     norm=1,
...     label="ψ",
... )
>>> qubit_vector.print(simplify=True)
|ψ⟩ = a|0⟩ + b|1⟩
>>> qubit_vector.diagram()
$\includegraphics[scale=1.25, trim=-0.02cm 0 0 -0.15cm]{./text_examples_docstrings_state_unipartite_qubit_vector.pdf}$
\end{minted}
\end{code}

\begin{code}
\begin{minted}{python}
>>> qutrit_vector = QuantumState(
...     spec=[("a", [0]), ("b", [1]), ("c", [2])],
...     form="vector",
...     dim=3,
...     symbols={
...         "a": {"complex": True},
...         "b": {"complex": True},
...         "c": {"complex": True},
...     },
...     substitutions=[("a*conjugate(a) + b*conjugate(b) + c*conjugate(c)", 1)],
...     norm=1,
...     label="φ",
... )
>>> qutrit_vector.print(simplify=True)
|φ⟩ = a|0⟩ + b|1⟩ + c|2⟩
>>> qutrit_vector.diagram()
$\includegraphics[scale=1.25, trim=-0.02cm 0 0 -0.15cm]{./text_examples_docstrings_state_unipartite_qutrit_vector.pdf}$
\end{minted}
\end{code}

\begin{code}
\begin{minted}{python}
>>> qubit_pure = QuantumState(
...     spec=[("α", [0]), ("β", [1])],
...     form="matrix",
...     kind="pure",
...     symbols={"a": {"complex": True}, "b": {"complex": True}},
...     substitutions=[("α*conjugate(α) + β*conjugate(β)", 1)],
...     norm=1,
...     label="ξ",
... )
>>> qubit_pure.print(simplify=True)
|ξ⟩⟨ξ| = α*conjugate(α)|0⟩⟨0| + α*conjugate(β)|0⟩⟨1| + β*conjugate(α)|1⟩⟨0| + β*conjugate(β)|1⟩⟨1|
>>> qubit_pure.diagram()
$\includegraphics[scale=1.25, trim=-0.02cm 0 0 -0.15cm]{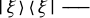}$
\end{minted}
\end{code}

\begin{code}
\begin{minted}{python}
>>> qubit_mixed = QuantumState(
...     spec=[("p", [0]), ("1 - p", [1])],
...     form="matrix",
...     kind="mixed",
...     symbols={"p": {"real": True, "nonnegative": True}},
...     norm=1,
...     label="τ",
... )
>>> qubit_mixed.print()
τ = p|0⟩⟨0| + (1 - p)|1⟩⟨1|
>>> qubit_mixed.diagram()
$\includegraphics[scale=1.25, trim=-0.02cm 0 0 -0.15cm]{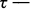}$
\end{minted}
\end{code}

\begin{code}
\begin{minted}{python}
>>> custom_vector = QuantumState(
...     spec=[["μ"], ["ν"]],
...     kind="mixed",
...     label="η",
... )
>>> custom_vector.print()
η = μ*conjugate(μ)|0⟩⟨0| + μ*conjugate(ν)|0⟩⟨1| + ν*conjugate(μ)|1⟩⟨0| + ν*conjugate(ν)|1⟩⟨1|
>>> custom_vector.diagram()
$\includegraphics[scale=1.25, trim=-0.02cm 0 0 -0.15cm]{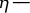}$
\end{minted}
\end{code}

\begin{code}
\begin{minted}{python}
>>> custom_matrix = QuantumState(
...     spec=[["w", "x"], ["y", "z"]],
...     kind="mixed",
...     label="ω",
... )
>>> custom_matrix.print()
ω = w|0⟩⟨0| + x|0⟩⟨1| + y|1⟩⟨0| + z|1⟩⟨1|
>>> custom_matrix.diagram()
$\includegraphics[scale=1.25, trim=-0.02cm 0 0 -0.15cm]{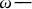}$
\end{minted}
\end{code}

\begin{code}
\begin{minted}{python}
>>> bell_state = QuantumState(
...     spec=[(1, [0, 0]), (1, [1, 1])],
...     form="vector",
...     norm=1,
...     label="Φ",
... )
>>> bell_state.print()
|Φ⟩ = sqrt(2)/2|0,0⟩ + sqrt(2)/2|1,1⟩
>>> bell_state.diagram()
$\includegraphics[scale=1.25, trim=-0.02cm 0 0 -0.15cm]{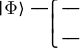}$
\end{minted}
\end{code}

\begin{code}
\begin{minted}{python}
>>> ghz_state = QuantumState(
...     spec=[(1, [0, 0, 0]), (1, [1, 1, 1])],
...     form="vector",
...     norm=1,
...     label="GHZ",
... )
>>> ghz_state.print()
|GHZ⟩ = sqrt(2)/2|0,0,0⟩ + sqrt(2)/2|1,1,1⟩
>>> ghz_state.diagram()
$\includegraphics[scale=1.25, trim=-0.02cm 0 0 -0.15cm]{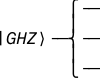}$
\end{minted}
\end{code}

\enlargethispage{2\baselineskip}

\begin{code}
\begin{minted}{python}
>>> w_state = QuantumState(
...     spec=[(1, [0, 0, 1]), (1, [0, 1, 0]), (1, [1, 0, 0])],
...     form="vector",
...     norm=1,
...     label="W",
... )
>>> w_state.print()
|W⟩ = sqrt(3)/3|0,0,1⟩ + sqrt(3)/3|0,1,0⟩ + sqrt(3)/3|1,0,0⟩
>>> w_state.diagram()
$\includegraphics[scale=1.25, trim=-0.02cm 0 0 -0.15cm]{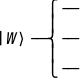}$
\end{minted}
\end{code}

\subsubsection{Subclasses}\label{sec:states_subclasses}

To simplify state creation, the following four subclasses of the base class \py{QuantumState} are provided:
\begin{code}
\begin{minted}{python}
>>> from qhronology.quantum.states import VectorState, MatrixState, PureState, MixedState
\end{minted}
\end{code}
These classes are \emph{specialized} (or \emph{restricting}) subclasses, meaning that they do not extend the base class in any way, and instead merely constrain its functionality in order to enforce the desired behaviour. They represent all possible valid combinations \py{form} and \py{kind} arguments, and therefore allow for quantum state objects to be initialized more concisely than using the general \py{QuantumState} class.

\subsubsection{Quantities}\label{sec:states_quantities}

The \py{QuantumState} class possesses a collection of methods with which various quantum quantities can be calculated from state instances.

\paragraph{\colorbox{colorbackground}{\fontsize{11}{9}\texttt{trace()}}}

Calculate the trace $\trace[\op{\rho}]$ of the internal state ($\op{\rho}$).

\enlargethispage{\baselineskip}

\begin{code}
\begin{minted}{python}
>>> state = QuantumState(
...     spec=[("a", [0]), ("b", [1])],
...     form="vector",
...     symbols={"a": {"complex": True}, "b": {"complex": True}},
...     substitutions=[("a*conjugate(a) + b*conjugate(b)", 1)],
...     norm=1,
... )
>>> state.trace()
1
\end{minted}
\end{code}

\begin{code}
\begin{minted}{python}
>>> state = QuantumState(
...     spec=[(1, [0]), (1, [1])],
...     kind="mixed",
...     symbols={"d": {"real": True}},
...     norm="1/d",
... )
>>> state.trace()
1/d
\end{minted}
\end{code}

\paragraph{\colorbox{colorbackground}{\fontsize{11}{9}\texttt{purity()}}}

Calculate the purity
\begin{equation}
\Purity(\op{\rho}) = \trace[\op{\rho}^2]
\end{equation}
of the internal state ($\op{\rho}$).

\begin{code}
\begin{minted}{python}
>>> state = QuantumState(
...     spec=[("a", [0]), ("b", [1])],
...     form="vector",
...     symbols={"a": {"complex": True}, "b": {"complex": True}},
...     substitutions=[("a*conjugate(a) + b*conjugate(b)", 1)],
...     norm=1,
... )
>>> state.purity()
1
\end{minted}
\end{code}

\begin{code}
\begin{minted}{python}
>>> state = QuantumState(
...     spec=[("p", [0]), ("1 - p", [1])],
...     kind="mixed",
...     norm=1,
... )
>>> state.purity()
p**2 + (1 - p)**2
\end{minted}
\end{code}

\paragraph{\colorbox{colorbackground}{\fontsize{11}{9}\texttt{distance(\emph{state})}}}

Calculate the trace distance
\begin{equation}
\TraceDistance(\op{\rho}, \op{\tau}) = \frac{1}{2}\trace{\abs{\op{\rho} - \op{\tau}}}
\end{equation}
between the internal state ($\op{\rho}$) and the given \py{state} ($\op{\tau}$).

\enlargethispage{\baselineskip}
\begin{code}
\begin{minted}{python}
>>> state_A = QuantumState(
...     spec=[("a", [0]), ("b", [1])],
...     form="vector",
...     symbols={"a": {"complex": True}, "b": {"complex": True}},
...     substitutions=[("a*conjugate(a) + b*conjugate(b)", 1)],
...     norm=1,
... )
>>> state_B = QuantumState(
...     spec=[("c", [0]), ("d", [1])],
...     form="vector",
...     symbols={"c": {"complex": True}, "d": {"complex": True}},
...     substitutions=[("c*conjugate(c) + d*conjugate(d)", 1)],
...     norm=1,
... )
>>> state_A.distance(state_A)
0
>>> state_B.distance(state_B)
0
>>> state_A.distance(state_B)
sqrt((a*conjugate(b) - c*conjugate(d))*(b*conjugate(a) - d*conjugate(c)) + (b*conjugate(b) - d*conjugate(d))**2)/2 + sqrt((a*conjugate(a) - c*conjugate(c))**2 + (a*conjugate(b) - c*conjugate(d))*(b*conjugate(a) - d*conjugate(c)))/2
\end{minted}
\end{code}

\begin{code}
\begin{minted}{python}
>>> state_A = QuantumState(
...     spec=[("p", [0]), ("1 - p", [1])],
...     kind="mixed",
...     symbols={"p": {"positive": True}},
...     norm=1,
... )
>>> state_B = QuantumState(
...     spec=[("q", [0]), ("1 - q", [1])],
...     kind="mixed",
...     symbols={"q": {"positive": True}},
...     norm=1,
... )
>>> state_A.distance(state_B)
Abs(p - q)
\end{minted}
\end{code}

\begin{code}
\begin{minted}{python}
>>> plus_state = QuantumState(
...     spec=[(1, [0]), (1, [1])],
...     form="vector",
...     norm=1,
... )
>>> minus_state = QuantumState(
...     spec=[(1, [0]), (-1, [1])],
...     form="vector",
...     norm=1,
... )
>>> plus_state.distance(minus_state)
1
\end{minted}
\end{code}

\enlargethispage{\baselineskip}
\paragraph{\colorbox{colorbackground}{\fontsize{11}{9}\texttt{fidelity(\emph{state})}}}

Calculate the fidelity
\begin{equation}
\Fidelity(\op{\rho}, \op{\tau}) = \left(\trace{\sqrt{\sqrt{\op{\rho}}\,\op{\tau}\sqrt{\op{\rho}}}}\right)^2
\end{equation}
between the internal state ($\op{\rho}$) and the given \py{state} ($\op{\tau}$).

\begin{code}
\begin{minted}{python}
>>> state_A = QuantumState(
...     spec=[("a", [0]), ("b", [1])],
...     form="vector",
...     symbols={"a": {"complex": True}, "b": {"complex": True}},
...     substitutions=[("a*conjugate(a) + b*conjugate(b)", 1)],
...     norm=1,
... )
>>> state_B = QuantumState(
...     spec=[("c", [0]), ("d", [1])],
...     form="vector",
...     symbols={"c": {"complex": True}, "d": {"complex": True}},
...     substitutions=[("c*conjugate(c) + d*conjugate(d)", 1)],
...     norm=1,
... )
>>> state_A.fidelity(state_A)
1
>>> state_B.fidelity(state_B)
1
>>> state_A.fidelity(state_B)
(a*conjugate(c) + b*conjugate(d))*(c*conjugate(a) + d*conjugate(b))
\end{minted}
\end{code}

\begin{code}
\begin{minted}{python}
>>> state_A = QuantumState(
...     spec=[("p", [0]), ("1 - p", [1])],
...     kind="mixed",
...     symbols={"p": {"positive": True}},
...     norm=1,
... )
>>> state_B = QuantumState(
...     spec=[("q", [0]), ("1 - q", [1])],
...     kind="mixed",
...     symbols={"q": {"positive": True}},
...     norm=1,
... )
>>> state_A.fidelity(state_B)
(sqrt(p)*sqrt(q) + sqrt((1 - p)*(1 - q)))**2
\end{minted}
\end{code}

\begin{code}
\begin{minted}{python}
>>> plus_state = QuantumState(
...     spec=[(1, [0]), (1, [1])],
...     form="vector",
...     norm=1,
... )
>>> minus_state = QuantumState(
...     spec=[(1, [0]), (-1, [1])],
...     form="vector",
...     norm=1,
... )
>>> plus_state.fidelity(minus_state)
0
\end{minted}
\end{code}

\paragraph{\colorbox{colorbackground}{\fontsize{11}{9}\texttt{entropy(\emph{state}, \emph{base}=\textcolor{colorcomment}{2})}}}

Calculate the relative von Neumann entropy
\begin{equation}
\Entropy(\op{\rho} \mathbin{\Vert} \op{\tau}) = \trace\bigl[\op{\rho} (\log_\Base\op{\rho} - \log_\Base\op{\tau})\bigr]
\end{equation}
between the internal state ($\op{\rho}$) and the given \py{state} ($\op{\tau}$). If \py{state} is not specified ({\ie}, \py{None}), calculate instead the ordinary von Neumann entropy
\begin{equation}
\Entropy(\op{\rho}) = \trace[\op{\rho}\log_\Base\op{\rho}]
\end{equation}
of the internal state ($\op{\rho}$). Here, $\Base$ represents the argument \py{base} (defaults to \py{2}), which is the dimensionality of the unit of information in which the entropy is measured.

\enlargethispage{\baselineskip}
\begin{code}
\begin{minted}{python}
>>> state_A = QuantumState(
...     spec=[("a", [0]), ("b", [1])],
...     form="vector",
...     symbols={"a": {"complex": True}, "b": {"complex": True}},
...     substitutions=[("a*conjugate(a) + b*conjugate(b)", 1)],
...     norm=1,
... )
>>> state_B = QuantumState(
...     spec=[("c", [0]), ("d", [1])],
...     form="vector",
...     symbols={"c": {"complex": True}, "d": {"complex": True}},
...     substitutions=[("c*conjugate(c) + d*conjugate(d)", 1)],
...     norm=1,
... )
>>> state_A.entropy()
0
>>> state_B.entropy()
0
>>> state_A.entropy(state_B)
0
\end{minted}
\end{code}

\begin{code}
\begin{minted}{python}
>>> state_A = QuantumState(
...     spec=[("p", [0]), ("1 - p", [1])],
...     kind="mixed",
...     symbols={"p": {"positive": True}},
...     norm=1,
... )
>>> state_B = QuantumState(
...     spec=[("q", [0]), ("1 - q", [1])],
...     kind="mixed",
...     symbols={"q": {"positive": True}},
...     norm=1,
... )
>>> state_A.entropy()
(-p*log(p) + (p - 1)*log(1 - p))/log(2)
>>> state_B.entropy()
(-q*log(q) + (q - 1)*log(1 - q))/log(2)
>>> state_A.entropy(state_B, base="d")
(-(p - 1)*(log(1 - p) - log(1 - q)) + log((p/q)**p))/log(d)
\end{minted}
\end{code}

\paragraph{\colorbox{colorbackground}{\fontsize{11}{9}\texttt{mutual(\emph{systems\_A}, \emph{systems\_B}, \emph{base}=\textcolor{colorcomment}{self.dim})}}}

Calculate the mutual information
\begin{equation}
\MutualInformation(A : B) = \Entropy(\op{\rho}^{\indices{A}}) + \Entropy(\op{\rho}^{\indices{B}}) - \Entropy(\op{\rho}^{\indices{A,B}})
\end{equation}
between two subsystems \py{systems_A} ($A$) and \py{systems_B} ($B$) of the internal state ($\rho^{\indices{A,B}}$). Here, $\Entropy(\op{\rho})$ is the von Neumann entropy of a state $\op{\rho}$. The dimensionality of the unit of information (in which the mutual information is measured) can be changed by supplying a value to the \py{base} argument (which defaults to the dimensionality of the state itself).

\begin{code}
\begin{minted}{python}
>>> state_AB = QuantumState(
...     spec=[("a", [0, 0]), ("b", [1, 1])],
...     form="vector",
...     symbols={"a": {"complex": True}, "b": {"complex": True}},
...     substitutions=[("a*conjugate(a) + b*conjugate(b)", 1)],
...     norm=1,
... )
>>> state_AB.mutual([0], [1])
2*(-a*log(a*conjugate(a))*conjugate(a) - b*log(b*conjugate(b))*conjugate(b))/log(2)
\end{minted}
\end{code}

\begin{code}
\begin{minted}{python}
>>> state_AB = QuantumState(
...     spec=[("a", [0, 0]), ("b", [1, 1])],
...     kind="mixed",
...     symbols={"a": {"positive": True}, "b": {"positive": True}},
...     substitutions=[("a + b", 1)],
...     norm=1,
... )
>>> state_AB.mutual([0], [1], base="d")
-log(a**a*b**b)/log(d)
\end{minted}
\end{code}

\begin{code}
\begin{minted}{python}
>>> state_ABC = QuantumState(
...     spec=[("a", [1, 0, 0]), ("b", [0, 1, 0]), ("c", [0, 0, 1])],
...     kind="mixed",
...     symbols={
...         "a": {"positive": True},
...         "b": {"positive": True},
...         "c": {"positive": True},
...     },
...     substitutions=[("a + b + c", 1)],
...     norm=1,
... )
>>> state_ABC.mutual([0], [1])
-log((a**a*b**b*c**c)**(1/log(2)))
\end{minted}
\end{code}

\enlargethispage{-5\baselineskip}
\subsubsection{Operations}\label{sec:states_operations}

The \py{QuantumState} class possesses a collection of methods with which the internal states of instances can be transformed (mutated) in various ways.

\paragraph{\colorbox{colorbackground}{\fontsize{11}{9}\texttt{reset()}}}

Reset the quantum state to its initial state. Undoes all method operations performed on the state.

\paragraph{\colorbox{colorbackground}{\fontsize{11}{9}\texttt{densify()}}}

Convert the state to its equivalent (density) matrix representation. Note that states that are already in density matrix form are unmodified.

\begin{code}
\begin{minted}{python}
>>> psi = QuantumState(
...     spec=[("a", [0]), ("b", [1])],
...     form="vector",
...     label="ψ",
... )
>>> psi.print()
|ψ⟩ = a|0⟩ + b|1⟩
>>> psi.densify()
>>> psi.print()
|ψ⟩⟨ψ| = a*conjugate(a)|0⟩⟨0| + a*conjugate(b)|0⟩⟨1| + b*conjugate(a)|1⟩⟨0| + b*conjugate(b)|1⟩⟨1|
\end{minted}
\end{code}

\paragraph{\colorbox{colorbackground}{\fontsize{11}{9}\texttt{dagger()}}}

Perform conjugate transposition on the state.

\begin{code}
\begin{minted}{python}
>>> psi = QuantumState(
...     spec=[("a", [0]), ("b", [1])],
...     form="vector",
...     label="ψ",
... )
>>> psi.print()
|ψ⟩ = a|0⟩ + b|1⟩
>>> psi.dagger()
>>> psi.print()
⟨ψ| = conjugate(a)⟨0| + conjugate(b)⟨1|
\end{minted}
\end{code}

\paragraph{\colorbox{colorbackground}{\fontsize{11}{9}\texttt{simplify(\emph{comprehensive}=\textcolor{colorcomment}{False})}}}

Apply a forced simplification to the state using the values of its \py{symbols} and \py{substitutions} properties. This is useful if intermediate simplification is required during a sequence of mutating operations in order to process the state into a more desirable form. The \py{comprehensive} argument (which defaults to \py{False}) can be used to specify whether the simplifying algorithm should use a relatively efficient subset of simplifying operations (\py{False}), or alternatively use a larger, more powerful (but slower) set (\py{True}).
\enlargethispage{-3\baselineskip}

\begin{code}
\begin{minted}{python}
>>> matrix = sp.Matrix(
...     [
...         ["(a**2 - 1)/(a - 1) - 1",
...          "log(cos(b) + I*sin(b))/I"],
...         ["acos((exp(I*c) + exp(-I*c))/2)",
...          "d**log(E*(sin(d)**2 + cos(d)**2))"],
...     ]
... )
>>> rho = QuantumState(spec=matrix, form="matrix", label="ρ")
>>> rho.print()
ρ = (-1 + (a**2 - 1)/(a - 1))|0⟩⟨0| + -I*log(I*sin(b) + cos(b))|0⟩⟨1| + acos(exp(I*c)/2 + exp(-I*c)/2)|1⟩⟨0| + d**log(E*(sin(d)**2 + cos(d)**2))|1⟩⟨1|
>>> rho.simplify()
>>> rho.print()
ρ = a|0⟩⟨0| + b|0⟩⟨1| + c|1⟩⟨0| + d|1⟩⟨1|
\end{minted}
\end{code}

\paragraph{\colorbox{colorbackground}{\fontsize{11}{9}\texttt{apply(\emph{function}, \emph{arguments}=\textcolor{colorcomment}{dict()})}}}

Apply a Python function (\py{function}) to the state.

Useful when used with SymPy's symbolic-manipulation functions, such as:\enlargethispage{\baselineskip}
\begin{itemize}
    \item \py{simplify()}
    \item \py{expand()}
    \item \py{factor()}
    \item \py{collect()}
    \item \py{cancel()}
    \item \py{apart()}
    \item \py{separatevars()}
    \item \py{rewrite()} (though the \py{QuantumState.rewrite()} method should be used instead)
\end{itemize}
More can be found in SymPy's official documentation, in both the Simplification\footnote{\url{https://docs.sympy.org/latest/tutorials/intro-tutorial/simplification.html}} and Simplify\footnote{\url{https://docs.sympy.org/latest/modules/simplify/simplify.html}} sections. Arguments to these functions can be supplied as a dictionary (with keywords as strings) {\via} the \py{arguments} argument.

\enlargethispage{0.15\baselineskip}
\begin{code}
\begin{minted}{python}
>>> psi = QuantumState(
...     spec=[("a*b + b*c + c*a", [0]), ("x*y + y*z + z*x", [1])],
...     form="vector",
...     label="ψ",
... )
>>> psi.print()
|ψ⟩ = (a*b + a*c + b*c)|0⟩ + (x*y + x*z + y*z)|1⟩
>>> psi.apply(sp.collect, {"syms": ["a", "x"]})
>>> psi.print()
|ψ⟩ = (a*(b + c) + b*c)|0⟩ + (x*(y + z) + y*z)|1⟩
>>> psi.apply(sp.expand)
>>> psi.print()
|ψ⟩ = (a*b + a*c + b*c)|0⟩ + (x*y + x*z + y*z)|1⟩
\end{minted}
\end{code}

\paragraph{\colorbox{colorbackground}{\fontsize{11}{9}\texttt{rewrite(\emph{function})}}}

Rewrite the elements of the state using the given mathematical function (\py{function}). Useful when used with SymPy's mathematical functions, such as:
\begin{itemize}
    \item \py{exp()}
    \item \py{log()}
    \item \py{sin()}
    \item \py{cos()}
\end{itemize}

\begin{code}
\begin{minted}{python}
>>> psi = QuantumState(
...     spec=[("cos(θ)", [0]), ("sin(θ)", [1])],
...     form="vector",
...     label="ψ",
... )
>>> psi.print()
|ψ⟩ = cos(θ)|0⟩ + sin(θ)|1⟩
>>> psi.rewrite(sp.exp)
>>> psi.print()
|ψ⟩ = (exp(I*θ)/2 + exp(-I*θ)/2)|0⟩ + -I*(exp(I*θ) - exp(-I*θ))/2|1⟩
\end{minted}
\end{code}

\paragraph{\colorbox{colorbackground}{\fontsize{11}{9}\texttt{normalize(\emph{norm}=\textcolor{colorcomment}{1})}}}

Perform a forced (re)normalization on the state to the value specified (\py{norm}, defaults to \py{1}). Useful when applied to a quantum state prior to any simplification performed on its processed output (obtained {\via} the \py{state()} method).

\begin{code}
\begin{minted}{python}
>>> psi = QuantumState(
...     spec=[("a", [0]), ("b", [1])],
...     form="vector",
...     label="ψ",
... )
>>> psi.print()
|ψ⟩ = a|0⟩ + b|1⟩
>>> psi.normalize()
>>> psi.print()
|ψ⟩ = a/sqrt(a*conjugate(a) + b*conjugate(b))|0⟩ + b/sqrt(a*conjugate(a) + b*conjugate(b))|1⟩
\end{minted}
\end{code}

\begin{code}
\begin{minted}{python}
>>> identity = QuantumState(
...     spec=[(1, [0]), (1, [1])],
...     symbols={"d": {"real": True}},
...     label="I",
... )
>>> identity.print()
I = |0⟩⟨0| + |1⟩⟨1|
>>> identity.normalize("2/d")
>>> identity.print()
I = 1/d|0⟩⟨0| + 1/d|1⟩⟨1|
\end{minted}
\end{code}

\paragraph{\colorbox{colorbackground}{\fontsize{11}{9}\texttt{coefficient(\emph{scalar}=\textcolor{colorcomment}{1})}}}

Multiply the state by a scalar value (\py{scalar}, defaults to \py{1}). Can be useful to manually (re)normalize states, or introduce a phase factor.

\begin{code}
\begin{minted}{python}
>>> psi = QuantumState(
...     spec=[(1, [0]), (1, [1])],
...     form="vector",
...     label="ψ",
... )
>>> psi.print()
|ψ⟩ = |0⟩ + |1⟩
>>> psi.coefficient(1 / sp.sqrt(2))
>>> psi.print()
|ψ⟩ = sqrt(2)/2|0⟩ + sqrt(2)/2|1⟩
\end{minted}
\end{code}

\begin{code}
\begin{minted}{python}
>>> phi = QuantumState(
...     spec=[("cos(θ)", [0]), ("sin(θ)", [1])],
...     form="vector",
...     label="φ",
... )
>>> phi.print()
|φ⟩ = cos(θ)|0⟩ + sin(θ)|1⟩
>>> phi.coefficient("exp(I*ξ)")
>>> phi.print()
|φ⟩ = exp(I*ξ)*cos(θ)|0⟩ + exp(I*ξ)*sin(θ)|1⟩
\end{minted}
\end{code}

\paragraph{\colorbox{colorbackground}{\fontsize{11}{9}\texttt{partial\_trace(\emph{targets}=\textcolor{colorcomment}{list()}, \emph{discard}=\textcolor{colorcomment}{True}, \emph{optimize}=\textcolor{colorcomment}{True})}}}

Perform a partial trace operation on the state. Use the \py{discard} (\py{bool}) argument to specify whether to keep or discard the system indices passed to \py{targets} (as a list of integers).

\begin{code}
\begin{minted}{python}
>>> psi = QuantumState(
...     spec=[
...         ("a*u", [0, 0]),
...         ("b*u", [1, 0]),
...         ("a*v", [0, 1]),
...         ("b*v", [1, 1]),
...     ],
...     form="vector",
...     substitutions=[
...         ("a*conjugate(a) + b*conjugate(b)", 1),
...         ("u*conjugate(u) + v*conjugate(v)", 1),
...     ],
...     label="Ψ",
... )
>>> psi.print()
|Ψ⟩ = a*u|0,0⟩ + a*v|0,1⟩ + b*u|1,0⟩ + b*v|1,1⟩
>>> psi.partial_trace([1])
>>> psi.simplify()
>>> psi.notation = "ρ"
>>> psi.print()
ρ = a*conjugate(a)|0⟩⟨0| + a*conjugate(b)|0⟩⟨1| + b*conjugate(a)|1⟩⟨0| + b*conjugate(b)|1⟩⟨1|
\end{minted}
\end{code}

\begin{code}
\begin{minted}{python}
>>> bell = QuantumState(
...     spec=[(1, [0, 0]), (1, [1, 1])],
...     form="vector",
...     norm=1,
...     label="Φ",
... )
>>> bell.print()
|Φ⟩ = sqrt(2)/2|0,0⟩ + sqrt(2)/2|1,1⟩
>>> bell.partial_trace([0])
>>> bell.notation = "ρ"
>>> bell.print()
ρ = 1/2|0⟩⟨0| + 1/2|1⟩⟨1|
\end{minted}
\end{code}

\paragraph{\colorbox{colorbackground}{\fontsize{11}{9}\texttt{measure(\emph{operators}, \emph{targets}=\textcolor{colorcomment}{self.systems}, \emph{observable}=\textcolor{colorcomment}{False}, \emph{statistics}=\textcolor{colorcomment}{False})}}}\label{sec:states_operations_measure}

Perform a quantum measurement on one or more systems (indicated in \py{targets}) of the state. This method has multiple modes of operation, which are selected according to the values passed to \py{statistics} and \py{observable} (both \py{bool}):
\begin{itemize}
    \item When \py{statistics} is \py{True}, the (reduced) state ($\op{\rho}$) (residing on the systems indicated in \py{targets}) is measured and the set of resulting statistics is returned. This takes the form of an ordered list of values $\{p_i\}_i$ associated with each given operator, where:
    \begin{itemize}
        \item $p_i = \trace[\Kraus_i^\dagger \Kraus_i \op{\rho}]$ (measurement probabilities) when \py{observable} is \py{False} \newline (\py{operators} is a list of Kraus operators or projectors $\Kraus_i$)
        \item $p_i = \trace[\Observable_i \op{\rho}]$ (expectation values) when \py{observable} is \py{True} \newline (\py{operators} is a list of observables $\Observable_i$)
    \end{itemize}
    \item When \py{statistics} is \py{False}, the (reduced) state ($\op{\rho}$) (residing on the systems indicated in \py{targets}) is measured and mutated it according to its predicted post-measurement form ({\ie}, the sum of all possible measurement outcomes). This yields the transformed states:
    \begin{itemize}
        \item When \py{observable} is \py{False}:
        \begin{equation}
            \op{\rho}^\prime = \sum_i \Kraus_i \op{\rho} \Kraus_i^\dagger.
        \end{equation}
        \item When \py{observable} is \py{True}:
        \begin{equation}
            \op{\rho}^\prime = \sum_i \trace[\Observable_i \op{\rho}]\Observable_i.
        \end{equation}
    \end{itemize}
\end{itemize}
The systems on which to perform the measurement, specified in \py{targets}, defaults to the state's entire system space. All other (unspecified) systems are discarded (traced over) in the course of performing the measurement.

As hinted above, the measurement operator(s) are supplied to the \py{operators} argument in the form of a list of \py{QuantumState} instances, SymPy matrices, or NumPy arrays. The elements in this list would typically be a (complete) set of Kraus operators forming a positive operator-valued measure (POVM), a (complete) set of (orthogonal) projectors forming a projection-valued measure (PVM), or a set of observables constituting a complete basis for the relevant state space.\enlargethispage{-3\baselineskip}

In the case where \py{operators} contains only a single item ($\Kraus$) and the current state ($\ket{\psi}$) is a vector form, the transformation of the state is in accordance with the rule
\begin{equation}
    \ket{\psi^\prime} = \frac{\Kraus \ket{\psi}}{\sqrt{\bra{\psi} \Kraus^\dagger \Kraus \ket{\psi}}}
\end{equation}
when \py{observable} is \py{False}. In all other mutation cases, the post-measurement state is a matrix, even if the pre-measurement state was a vector.

The items in the list \py{operators} can also be vectors ({\eg}, $\ket{\xi_i}$), in which case each is converted into its corresponding operator matrix representation ({\eg}, $\ket{\xi_i}\bra{\xi_i}$) prior to any measurements.

Please note that this method does not check for validity of supplied POVMs or the completeness of sets of observables, nor does it renormalize the post-measurement state in general.

\begin{code}
\begin{minted}{python}
>>> psi = QuantumState(
...     spec=[("a", [0]), ("b", [1])],
...     form="vector",
...     label="ψ",
... )
>>> psi.print()
|ψ⟩ = a|0⟩ + b|1⟩
>>> I = Pauli(index=0)
>>> X = Pauli(index=1)
>>> Y = Pauli(index=2)
>>> Z = Pauli(index=3)
>>> psi.measure(
...     operators=[I, X, Y, Z],
...     observable=True,
...     statistics=True,
... )
[a*conjugate(a) + b*conjugate(b),
 a*conjugate(b) + b*conjugate(a),
 I*(a*conjugate(b) - b*conjugate(a)),
 a*conjugate(a) - b*conjugate(b)]
>>> psi.measure(
...     operators=[I, X, Y, Z],
...     observable=True,
...     statistics=False,
... )
>>> psi.simplify()
>>> psi.coefficient(sp.Rational(1, 2))
>>> psi.label += "′"
>>> psi.print()
|ψ′⟩⟨ψ′| = a*conjugate(a)|0⟩⟨0| + a*conjugate(b)|0⟩⟨1| + b*conjugate(a)|1⟩⟨0| + b*conjugate(b)|1⟩⟨1|
\end{minted}
\end{code}

\enlargethispage{-\baselineskip}
\begin{code}
\begin{minted}{python}
>>> from qhronology.mechanics.matrices import ket
>>> psi = QuantumState(
...     spec=[("a", [0]), ("b", [1])],
...     form="vector",
...     label="ψ",
... )
>>> psi.print()
|ψ⟩ = a|0⟩ + b|1⟩
>>> psi.measure(
...     operators=[ket(0), ket(1)],
...     observable=False,
...     statistics=True,
... )
[a*conjugate(a), b*conjugate(b)]
>>> psi.measure(
...     operators=[ket(0), ket(1)],
...     observable=False,
...     statistics=False,
... )
>>> psi.notation = "ρ′"
>>> psi.print()
ρ′ = a*conjugate(a)|0⟩⟨0| + b*conjugate(b)|1⟩⟨1|
\end{minted}
\end{code}
Note that these examples use some of Qhronology's quantum gates{\textemdash}see {\secn} \ref{sec:gates} for more detail.

\paragraph{\colorbox{colorbackground}{\fontsize{11}{9}\texttt{postselect(\emph{postselections})}}}

Perform postselection on the state against the operators(s) specified in \py{postselections}. The precise format of the object passed to this argument consists of a list of 2-tuples, with each tuple containing an operator (as a \py{QuantumState} object, SymPy matrix, or NumPy array), followed by the first (smallest) index (as an integer) of the set of postselection target systems for its corresponding operator. While unusual, this bespoke arrangement means that multiple systems can be postselected (effectively simultaneously) using a single method call.

The postselections can be given in either vector or matrix form. For the former, the transformation of the composite vector state $\ket{\Psi}$ follows the standard rule
\begin{equation}
    \ket{\Psi^\prime} = \braket{\phi}{\Psi}
\end{equation}
where $\ket{\phi}$ is the postselection vector (spanning a proper subset of the systems spanned by $\ket{\Psi}$). In the case of a matrix form $\op{\omega}$, the notion of postselection of a density matrix state $\op{\rho}$ naturally generalizes to
\begin{equation}
    \op{\rho}^\prime = \trace_{\{i\}}[\op{\omega} \op{\rho}]
\end{equation}
where $\{i\}$ is the set of indices corresponding to the subsystem(s) upon which the postselection is performed. Note that all indices must be relative to the state's current composite space.

If multiple postselections are supplied, the state will be successively postselected in the order in which they are specified. If a vector state is postselected against a matrix form, it will automatically be transformed into its matrix form, as necessary.

\enlargethispage{-\baselineskip}
\begin{code}
\begin{minted}{python}
>>> psi = QuantumState(
...     spec=[("a", [0, 0]), ("b", [1, 1])],
...     form="vector",
...     label="Ψ",
... )
>>> phi = QuantumState(
...     spec=[("c", [0]), ("d", [1])],
...     form="vector",
...     label="φ",
... )
>>> psi.print()
|Ψ⟩ = a|0,0⟩ + b|1,1⟩
>>> phi.print()
|φ⟩ = c|0⟩ + d|1⟩
>>> psi.postselect([(phi, [0])])
>>> psi.label += "′"
>>> psi.print()
|Ψ′⟩ = a*conjugate(c)|0⟩ + b*conjugate(d)|1⟩
\end{minted}
\end{code}

\begin{code}
\begin{minted}{python}
>>> from qhronology.mechanics.matrices import ket
>>> psi = QuantumState(
...     spec=[("a", [0, 0]), ("b", [1, 1])],
...     form="vector",
...     label="Ψ",
... )
>>> psi.print()
|Ψ⟩ = a|0,0⟩ + b|1,1⟩
>>> psi.label += "′"
>>> psi.postselect([(ket(0), [0])])
>>> psi.print()
|Ψ′⟩ = a|0⟩
>>> psi.reset()
>>> psi.postselect([(ket(1), [0])])
>>> psi.print()
|Ψ′⟩ = b|1⟩
\end{minted}
\end{code}

\subsection{Gates}\label{sec:gates}

Quantum logic gates provide the building blocks for describing quantum operations, which are usually either (unitary) interactions between two or more (sub)systems, or (linear) transformations of any number of systems. In Qhronology, they are represented by instances of the \py{QuantumGate} base class,
\begin{code}
\begin{minted}{python}
>>> from qhronology.quantum.gates import QuantumGate
\end{minted}
\end{code}
and its derivatives (subclasses) (see {\secn} \ref{sec:gates_subclasses}). These objects describe a distinct vertical ``slice'' in the quantum circuit picturalism, and so include information about the locations of both control and anticontrol nodes, in addition to the presence of any empty wires. They also possess other metadata associated with the gate such as parameter values, symbolic assumptions, and algebraic substitutions.

The base class \py{QuantumGate}, from which all other gates are derived, is the one to be used for creating general gates corresponding to user-specified (unitary) matrices. For example, using SymPy to create an arbitrary $2 \times 2$ matrix,
\begin{code}
\begin{minted}{python}
>>> unitary = sp.MatrixSymbol("U", 2, 2).as_mutable()
\end{minted}
\end{code}
we can pass this to \py{QuantumGate}'s \py{spec} argument at construction:
\begin{code}
\begin{minted}{python}
>>> U = QuantumGate(spec=unitary, label="U")
\end{minted}
\end{code}

\enlargethispage{-2\baselineskip}
Inspecting an instance of this class (and any of its subclasses) can be carried out in the same way as the \py{QuantumState} class, {\eg},
\begin{code}
\begin{minted}{python}
>>> U.output()
Matrix([
[U[0, 0], U[0, 1]],
[U[1, 0], U[1, 1]]])
\end{minted}
\end{code}
\begin{code}
\begin{minted}{python}
>>> U.print()
U = U[0, 0]|0⟩⟨0| + U[0, 1]|0⟩⟨1| + U[1, 0]|1⟩⟨0| + U[1, 1]|1⟩⟨1|
\end{minted}
\end{code}
\begin{code}
\begin{minted}{python}
>>> U.diagram()
$\includegraphics[scale=1.25, trim=-0.02cm 0 0 -0.15cm]{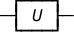}$
\end{minted}
\end{code}
\newpage
Higher-order gates, such as those for qutrits, can be created in much the same way, except with the \py{dim} argument (which defaults to \py{2}, corresponding to qubits) needing to be specified, {\eg},
\begin{code}
\begin{minted}{python}
>>> unitary = sp.MatrixSymbol("U", 3, 3).as_mutable()
>>> U3 = QuantumGate(spec=unitary, dim=3, label="U")
>>> U3.output()
Matrix([
[U[0, 0], U[0, 1], U[0, 2]],
[U[1, 0], U[1, 1], U[1, 2]],
[U[2, 0], U[2, 1], U[2, 2]]])
>>> U3.diagram()
$\includegraphics[scale=1.25, trim=-0.02cm 0 0 -0.15cm]{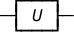}$
\end{minted}
\end{code}
Gates corresponding to transformations which affect more than one system can be created by using the \py{targets} argument (a list of system indices, with indexing beginning at \py{0}) to specify the gate's structure:
\begin{code}
\begin{minted}{python}
>>> unitary = sp.MatrixSymbol("U", 4, 4).as_mutable()
>>> UU = QuantumGate(spec=unitary, targets=[0, 1], dim=2, label="U")
>>> UU.output()
Matrix([
[U[0, 0], U[0, 1], U[0, 2], U[0, 3]],
[U[1, 0], U[1, 1], U[1, 2], U[1, 3]],
[U[2, 0], U[2, 1], U[2, 2], U[2, 3]],
[U[3, 0], U[3, 1], U[3, 2], U[3, 3]]])
>>> UU.diagram()
$\includegraphics[scale=1.25, trim=-0.02cm 0 0 -0.15cm]{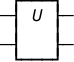}$
\end{minted}
\end{code}

Note that the matrix passed to \py{spec} must be contiguous, {\ie}, the systems upon which it acts must be consecutive.

\subsubsection{Empty wires}\label{sec:gates_empty}

For a gate acting on a (true) subset of subsystems of a composite system, the \py{num_systems} argument can be used to specify the number of total systems which make up the global space. We can construct a few examples:
\begin{code}
\begin{minted}{python}
>>> unitary = sp.MatrixSymbol("U", 2, 2).as_mutable()
>>> UI = QuantumGate(spec=unitary, targets=[0], num_systems=2, label="U")
>>> UI.output()
Matrix([
[U[0, 0],       0, U[0, 1],       0],
[      0, U[0, 0],       0, U[0, 1]],
[U[1, 0],       0, U[1, 1],       0],
[      0, U[1, 0],       0, U[1, 1]]])
>>> UI.diagram()
$\includegraphics[scale=1.25, trim=-0.02cm 0 0 -0.15cm]{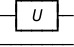}$
\end{minted}
\end{code}
\begin{code}
\begin{minted}{python}
>>> unitary = sp.MatrixSymbol("U", 2, 2).as_mutable()
>>> IU = QuantumGate(spec=unitary, targets=[1], num_systems=2, label="U")
>>> IU.output()
Matrix([
[U[0, 0], U[0, 1],       0,       0],
[U[1, 0], U[1, 1],       0,       0],
[      0,       0, U[0, 0], U[0, 1]],
[      0,       0, U[1, 0], U[1, 1]]])
>>> IU.diagram()
$\includegraphics[scale=1.25, trim=-0.02cm 0 0 -0.15cm]{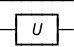}$
\end{minted}
\end{code}
Any combination of \py{targets} and \py{num_systems} is valid provided the indices specified in the former match composition of the given \py{spec} while also not exceeding the bounds set by the latter ({\eg}, for $N$ systems, the largest a system's index can be is $N-1$).

\subsubsection{Symbols and substitutions}\label{sec:gates_symbols}

Like the \py{QuantumState} class, \py{QuantumGate} accepts the \py{symbols} argument as a way to specify the mathematical properties of any parameters or variables contained within the gate's matrix representation. Similarly, the \py{substitutions} argument can be used to impose any adjunct constraints onto the gate when it is called. For example, to grant unitarity to an elementary construction of a single-system (unipartite) gate, we can construct: \enlargethispage{\baselineskip}
\begin{code}
\begin{minted}{python}
>>> U = QuantumGate(
...     spec=[["a", "b"], ["-exp(I*φ)*conjugate(b)", "exp(I*φ)*conjugate(a)"]],
...     symbols={
...         "a": {"complex": True},
...         "b": {"complex": True},
...         "φ": {"real": True},
...     },
...     substitutions=[("a*conjugate(a) + b*conjugate(b)", 1)],
...     label="U",
... )
>>> U.output()
Matrix([
[                     a,                     b],
[-exp(I*φ)*conjugate(b), exp(I*φ)*conjugate(a)]])
\end{minted}
\end{code}
See {\cdb} \ref{code:circuit_unitarity} for another example of unitarity enforced on a general symbolic gate.

\subsubsection{Mathematical modifiers}\label{sec:gates_modifiers}

Other arguments (and properties) that can be used to modify gates include \py{conjugate}, \py{coefficient} and \py{exponent}. The first can be used in much the same way as in the \py{QuantumState} class{\textemdash}to perform Hermitian conjugation on the gate when it is called, {\eg},\enlargethispage{\baselineskip}
\begin{code}
\begin{minted}{python}
>>> unitary = sp.MatrixSymbol("U", 2, 2).as_mutable()
>>> U = QuantumGate(spec=unitary, conjugate=True, label="U")
>>> U.output()
Matrix([
[conjugate(U[0, 0]), conjugate(U[1, 0])],
[conjugate(U[0, 1]), conjugate(U[1, 1])]])
\end{minted}
\end{code}

The \py{coefficient} argument simply multiplies the gate's core matrix $\op{U}$ (occupying the systems specified in \py{targets}) by a scalar value $\lambda$ (given as a numerical, symbolic, or string representation), {\eg}, $\lambda\op{U}$. For example:
\begin{code}
\begin{minted}{python}
>>> unitary = sp.MatrixSymbol("U", 2, 2).as_mutable()
>>> U = QuantumGate(spec=unitary, coefficient="λ", label="U")
>>> U.output()
Matrix([
[λ*U[0, 0], λ*U[0, 1]],
[λ*U[1, 0], λ*U[1, 1]]])
\end{minted}
\end{code}

Similarly, the scalar value passed to the \py{exponent} argument is used as the power $p$ by which the gate's matrix $\op{U}$ is raised, {\eg}, $\op{U}^p$. For example:
\begin{code}
\begin{minted}{python}
>>> unitary = sp.MatrixSymbol("U", 2, 2).as_mutable()
>>> U = QuantumGate(spec=unitary, exponent="p", label="U")
>>> U.output()
Matrix([
[(1/2 - exp(I*pi*p)/2)*U[0, 0] + exp(I*pi*p)/2 + 1/2,                       (1/2 - exp(I*pi*p)/2)*U[0, 1]],
[                      (1/2 - exp(I*pi*p)/2)*U[1, 0], (1/2 - exp(I*pi*p)/2)*U[1, 1] + exp(I*pi*p)/2 + 1/2]])
\end{minted}
\end{code}
Note that Qhronology's exponentiation functionality is guaranteed to be valid only for gates whose matrix representations ({\eg}, $\op{U}$) are involutory ({\ie}, $\op{U}^2 = \op{I}$, where $\op{I}$ is the identity matrix of the same shape as $\op{U}$).

\subsubsection{Control and anticontrol nodes}\label{sec:gates_control}

An important mechanism in quantum logic theory is the ability to modify the operation of a gate according to the state of a system different from those upon which it acts. This manifests in the form of both \emph{control nodes} and \emph{anticontrol nodes}, which are implemented in Qhronology {\via} arguments (and properties) to the \py{QuantumGate} class. To endow such nodes to any gate instance, one simply passes the indices of the systems (as a list of integers) to the \py{controls} and \py{anticontrols} arguments at instantiation (or, alternatively, to the homonymous properties after instantiation). Naturally, the \py{controls}, \py{anticontrols}, and \py{targets} lists cannot contain any conflicting system indices ({\ie}, they cannot have any elements in common). Some examples:
\enlargethispage{3\baselineskip}
\begin{code}
\begin{minted}{python}
>>> unitary = sp.MatrixSymbol("U", 2, 2).as_mutable()
>>> CU = QuantumGate(
...     spec=unitary,
...     targets=[1],
...     controls=[0],
...     label="U",
... )
>>> CU.output()
Matrix([
[1, 0,       0,       0],
[0, 1,       0,       0],
[0, 0, U[0, 0], U[0, 1]],
[0, 0, U[1, 0], U[1, 1]]])
>>> CU.diagram()
$\includegraphics[scale=1.25, trim=-0.02cm 0 0 -0.15cm]{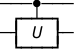}$
\end{minted}
\end{code}

\begin{code}
\begin{minted}{python}
>>> unitary = sp.MatrixSymbol("U", 2, 2).as_mutable()
>>> UC = QuantumGate(
...     spec=unitary,
...     targets=[0],
...     controls=[1],
...     label="U",
... )
>>> UC.output()
Matrix([
[1,       0, 0,       0],
[0, U[0, 0], 0, U[0, 1]],
[0,       0, 1,       0],
[0, U[1, 0], 0, U[1, 1]]])
>>> UC.diagram()
$\includegraphics[scale=1.25, trim=-0.02cm 0 0 -0.15cm]{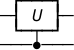}$
\end{minted}
\end{code}

\enlargethispage{-2\baselineskip}
\begin{code}
\begin{minted}{python}
>>> unitary = sp.MatrixSymbol("U", 2, 2).as_mutable()
>>> CCU = QuantumGate(
...     spec=unitary,
...     targets=[2],
...     controls=[0, 1],
...     label="U",
... )
>>> CCU.output()
Matrix([
[1, 0, 0, 0, 0, 0,       0,       0],
[0, 1, 0, 0, 0, 0,       0,       0],
[0, 0, 1, 0, 0, 0,       0,       0],
[0, 0, 0, 1, 0, 0,       0,       0],
[0, 0, 0, 0, 1, 0,       0,       0],
[0, 0, 0, 0, 0, 1,       0,       0],
[0, 0, 0, 0, 0, 0, U[0, 0], U[0, 1]],
[0, 0, 0, 0, 0, 0, U[1, 0], U[1, 1]]])
>>> CCU.diagram()
$\includegraphics[scale=1.25, trim=-0.02cm 0 0 -0.15cm]{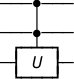}$
\end{minted}
\end{code}
\vspace*{-0.25\baselineskip}

\begin{code}
\begin{minted}{python}
>>> unitary = sp.MatrixSymbol("U", 2, 2).as_mutable()
>>> AU = QuantumGate(
...     spec=unitary,
...     targets=[1],
...     anticontrols=[0],
...     label="U",
... )
>>> AU.output()
Matrix([
[U[0, 0], U[0, 1], 0, 0],
[U[1, 0], U[1, 1], 0, 0],
[      0,       0, 1, 0],
[      0,       0, 0, 1]])
>>> AU.diagram()
$\includegraphics[scale=1.25, trim=-0.02cm 0 0 -0.15cm]{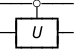}$
\end{minted}
\end{code}
\vspace*{-0.25\baselineskip}

\enlargethispage{2.5\baselineskip}
\begin{code}
\begin{minted}{python}
>>> unitary = sp.MatrixSymbol("U", 2, 2).as_mutable()
>>> AUC = QuantumGate(
...     spec=unitary,
...     targets=[1],
...     controls=[2],
...     anticontrols=[0],
...     label="U",
... )
>>> AUC.output()
Matrix([
[1,       0, 0,       0, 0, 0, 0, 0],
[0, U[0, 0], 0, U[0, 1], 0, 0, 0, 0],
[0,       0, 1,       0, 0, 0, 0, 0],
[0, U[1, 0], 0, U[1, 1], 0, 0, 0, 0],
[0,       0, 0,       0, 1, 0, 0, 0],
[0,       0, 0,       0, 0, 1, 0, 0],
[0,       0, 0,       0, 0, 0, 1, 0],
[0,       0, 0,       0, 0, 0, 0, 1]])
>>> AC.diagram()
$\includegraphics[scale=1.25, trim=-0.02cm 0 0 -0.15cm]{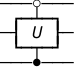}$
\end{minted}
\end{code}

\subsubsection{Subclasses}\label{sec:gates_subclasses}

Most of the canonical gates in standard quantum computing theory are implemented in Qhronology as an assortment of subclasses of the \py{QuantumGate} class:
\begin{code}
\begin{minted}{python}
>>> from qhronology.quantum.gates import Unitary, Pauli, GellMann, Rotation, Phase, Diagonal, Swap, Summation, Not, Hadamard, Fourier, Measurement
\end{minted}
\end{code}
It is important to be aware of how, even though they share many of the same arguments and properties, the usage of these classes can differ greatly. This is especially true among a few distinct categorizations of gates, namely \emph{dimensionality} and \emph{compositionality}, which are summarized in \linebreak {\tbl} \ref{tbl:gate_classes}.

For the classes of fixed dimensionality, their constructors do not take \py{dim} as an argument, nor can the associated property be set. For the classes describing unipartite gates, more than one system can still be targeted, in which case the gate's elementary matrix will simply be duplicated onto each system.

\renewcommand{\arraystretch}{1.25}
\renewcommand\cellgape{\Gape[2pt]}
\renewcommand\cellset{\renewcommand\arraystretch{1}
\setlength\extrarowheight{2pt}}
\begin{center}
\begin{NiceTabular}{*{4}{c}}[corners,hvlines]
 & & \Block[c, fill=lightblue]{1-2}{\textbf{\textsf{Compositionality}}} \\
 & & \textit{\textsf{Unipartite}} & \textit{\textsf{Multipartite}} \\
\Block[c, fill=lightblue]{2-1}{\textbf{\textsf{Dimensionality}}} & \textit{\textsf{Fixed}} &  \Block[l, respect-arraystretch]{}{\py{Unitary} \\ \py{Pauli} \\ \py{GellMann} \\ \py{Rotation} \\ \py{Not}} & {\small \textsf{(none)}} \\
 & \textit{\textsf{Variable}} & \Block[l, respect-arraystretch]{}{\py{Phase} \\ \py{Diagonal} \\ \py{Hadamard} \\ \py{Summation}} & \Block[l, respect-arraystretch]{}{\py{Swap} \\ \py{Fourier} \\ \py{Measurement}} \\
\end{NiceTabular}
\captionof{table}{Classification of Qhronology's \py{QuantumGate} subclasses. Note that the \py{Swap} class can only describe bipartite gates, and so is not multipartite for any general number of systems. Also note that gates of the \py{Measurement} class can act on systems of any dimension but do not themselves possess a dimensionality.}\label{tbl:gate_classes}
\end{center}

Here, we provide a small description for each, along with some examples. Please note that, as indicated by the presence of the \py{*args} and \py{**kwargs} arguments in their constructors' signatures, these subclasses take the same arguments as their \py{QuantumGate} parent class (with the exception of fixed-dimensionality subclasses having their \py{dim} argument unable to be set). Additionally, within said signatures, equality signs are used to specify the arguments' default values, wherein \py{self} references the instance itself (as conventional in Python).

\paragraph{\colorbox{colorbackground}{\fontsize{11}{9}\texttt{Not(\emph{*args}, \emph{**kwargs})}}}

We begin with perhaps the simplest of all gates: the \emph{NOT gate}. Corresponding to the logical \emph{negation} operation (also known as a ``bit-flip''), this gate has the elementary matrix representation
\begin{equation}
\begin{aligned}
\NOT &= \ket{0}\bra{1} + \ket{1}\bra{0} \\
&= \begin{bmatrix} 0 & 1 \\ 1 & 0 \end{bmatrix},
\end{aligned}
\end{equation}
and is implemented as the \py{Not} class:
\begin{code}
\begin{minted}{python}
>>> NOT = Not()
>>> NOT.output()
Matrix([
[0, 1],
[1, 0]])
>>> NOT.diagram()
$\includegraphics[scale=1.25, trim=-0.02cm 0 0 -0.15cm]{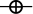}$
\end{minted}
\end{code}
Endowing a NOT gate with a control node produces a \emph{controlled-NOT} (\emph{CNOT}) \emph{gate}:
\begin{code}
\begin{minted}{python}
>>> CNOT = Not(targets=[1], controls=[0])
>>> CNOT.output()
Matrix([
[1, 0, 0, 0],
[0, 1, 0, 0],
[0, 0, 0, 1],
[0, 0, 1, 0]])
>>> CNOT.diagram()
$\includegraphics[scale=1.25, trim=-0.02cm 0 0 -0.15cm]{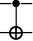}$
\end{minted}
\end{code}
Using an anticontrol node (instead of a control node) yields an \emph{anticontrolled-NOT} (\emph{ANOT}) \emph{gate}:
\begin{code}
\begin{minted}{python}
>>> ANOT = Not(targets=[1], anticontrols=[0])
>>> ANOT.output()
Matrix([
[0, 1, 0, 0],
[1, 0, 0, 0],
[0, 0, 1, 0],
[0, 0, 0, 1]])
>>> ANOT.diagram()
$\includegraphics[scale=1.25, trim=-0.02cm 0 0 -0.15cm]{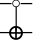}$
\end{minted}
\end{code}
Bestowing a NOT gate with a pair of control nodes results in the \emph{Toffoli} (\emph{controlled-controlled-NOT}, or \emph{CCNOT}) \emph{gate}:
\begin{code}
\begin{minted}{python}
>>> CCNOT = Not(targets=[2], controls=[0, 1])
>>> CCNOT.output()
Matrix([
[1, 0, 0, 0, 0, 0, 0, 0],
[0, 1, 0, 0, 0, 0, 0, 0],
[0, 0, 1, 0, 0, 0, 0, 0],
[0, 0, 0, 1, 0, 0, 0, 0],
[0, 0, 0, 0, 1, 0, 0, 0],
[0, 0, 0, 0, 0, 1, 0, 0],
[0, 0, 0, 0, 0, 0, 0, 1],
[0, 0, 0, 0, 0, 0, 1, 0]])
>>> CCNOT.diagram()
$\includegraphics[scale=1.25, trim=-0.02cm 0 0 -0.15cm]{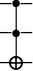}$
\end{minted}
\end{code}
Passing a value of $\tfrac{1}{2}$ to the \py{exponent} argument yields a \emph{root-NOT} ($\sqrt{\text{NOT}}$) \emph{gate}:\enlargethispage{2\baselineskip}
\begin{code}
\begin{minted}{python}
>>> RNOT = Not(
...     exponent=sp.Rational(1, 2),
...     label="√NOT",
...     family="GATE",
... )
>>> RNOT.output()
Matrix([
[1/2 + I/2, 1/2 - I/2],
[1/2 - I/2, 1/2 + I/2]])
>>> RNOT.diagram()
$\includegraphics[scale=1.25, trim=-0.02cm 0 0 -0.15cm]{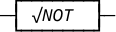}$
\end{minted}
\end{code}

\paragraph{\colorbox{colorbackground}{\fontsize{11}{9}\texttt{Pauli(\emph{*args}, \emph{index}, \emph{**kwargs})}}}

The \emph{Pauli gates}, corresponding to the \emph{Pauli matrices} $\Pauli_i$,
\begin{equation}
\begin{aligned}
    \Pauli_1 &= \Pauli_x \equiv \ket{0}\bra{1} + \ket{1}\bra{0} = \begin{bmatrix} 0 & 1 \\ 1 & 0 \end{bmatrix}, \\
    \Pauli_2 &= \Pauli_y \equiv -\eye \ket{0}\bra{1} + \eye \ket{1}\bra{0} = \begin{bmatrix} 0 & -\eye \\ \eye & 0 \end{bmatrix}, \\
    \Pauli_3 &= \Pauli_z \equiv \ket{0}\bra{0} - \ket{1}\bra{1} = \begin{bmatrix} 1 & 0 \\ 0 & -1 \end{bmatrix},
\end{aligned}
\end{equation}
(indexed by $i$, which includes the 2-dimensional identity matrix for $i=0$), can be easily created using the \py{Pauli} class. To obtain the desired gate, simply pass any of the four valid integers (\py{0}, \py{1}, \py{2}, \py{3}) to its \py{index} argument:
\begin{code}
\begin{minted}{python}
>>> X = Pauli(index=1)
>>> X.output()
Matrix([
[0, 1],
[1, 0]])
>>> X.diagram()
$\includegraphics[scale=1.25, trim=-0.02cm 0 0 -0.15cm]{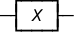}$
\end{minted}
\end{code}
\begin{code}
\begin{minted}{python}
>>> Y = Pauli(index=2)
>>> Y.output()
Matrix([
[0, -I],
[I,  0]])
>>> Y.diagram()
$\includegraphics[scale=1.25, trim=-0.02cm 0 0 -0.15cm]{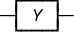}$
\end{minted}
\end{code}
\begin{code}
\begin{minted}{python}
>>> Z = Pauli(index=3)
>>> Z.output()
Matrix([
[1,  0],
[0, -1]])
>>> Z.diagram()
$\includegraphics[scale=1.25, trim=-0.02cm 0 0 -0.15cm]{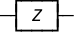}$
\end{minted}
\end{code}\enlargethispage{2.5\baselineskip}
\begin{code}
\begin{minted}{python}
>>> I = Pauli(index=0)
>>> I.output()
Matrix([
[1, 0],
[0, 1]])
>>> I.diagram()
$\includegraphics[scale=1.25, trim=-0.02cm 0 0 -0.15cm]{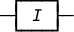}$
\end{minted}
\end{code}
\begin{code}
\begin{minted}{python}
>>> ZZ = Pauli(index=3, targets=[0, 1], label="Z⊗Z")
>>> ZZ.output()
Matrix([
[1,  0,  0, 0],
[0, -1,  0, 0],
[0,  0, -1, 0],
[0,  0,  0, 1]])
>>> ZZ.diagram()
$\includegraphics[scale=1.25, trim=-0.02cm 0 0 -0.15cm]{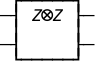}$
\end{minted}
\end{code}
\begin{code}
\begin{minted}{python}
>>> CZ = Pauli(index=3, targets=[1], controls=[0])
>>> CZ.output()
Matrix([
[1, 0, 0,  0],
[0, 1, 0,  0],
[0, 0, 1,  0],
[0, 0, 0, -1]])
>>> CZ.diagram()
$\includegraphics[scale=1.25, trim=-0.02cm 0 0 -0.15cm]{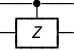}$
\end{minted}
\end{code}

Clever use of the \py{exponent} and \py{coefficient} arguments allows for the creation of more complex gates, such as the (parametric) \emph{Ising XX-interaction} (or \emph{XX-coupling}) \emph{gate}:
\begin{code}
\begin{minted}{python}
>>> R_xx = Pauli(
...     index=1,
...     targets=[0, 1],
...     exponent="θ/pi",
...     coefficient="exp(-I*θ/2)",
...     label="R_xx(θ)",
... )
>>> R_xx.output(simplify=True)
Matrix([
[   cos(θ/2),           0,           0, -I*sin(θ/2)],
[          0,    cos(θ/2), -I*sin(θ/2),           0],
[          0, -I*sin(θ/2),    cos(θ/2),           0],
[-I*sin(θ/2),           0,           0,    cos(θ/2)]])
>>> R_xx.diagram()
$\includegraphics[scale=1.25, trim=-0.02cm 0 0 -0.15cm]{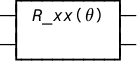}$
\end{minted}
\end{code}

\paragraph{\colorbox{colorbackground}{\fontsize{11}{9}\texttt{GellMann(\emph{*args}, \emph{index}, \emph{**kwargs})}}}

The \emph{Gell-Mann gates}, corresponding to the \emph{Gell-Mann matrices} $\GellMann_i$,
\begin{equation}
\begin{aligned}
    &\GellMann_1 \equiv \ket{0}\bra{1} + \ket{1}\bra{0} = \begin{bmatrix} 0 & 1 & 0 \\ 1 & 0 & 0 \\ 0 & 0 & 0 \end{bmatrix}, \\
    &\GellMann_3 \equiv \ket{0}\bra{0} - \ket{1}\bra{1} = \begin{bmatrix} 1 & 0 & 0 \\ 0 & -1 & 0 \\ 0 & 0 & 0 \end{bmatrix}, \\
    &\GellMann_5 \equiv -\eye\ket{0}\bra{2} + \eye\ket{2}\bra{0} = \begin{bmatrix} 0 & 0 & -\eye \\ 0 & 0 & 0 \\ \eye & 0 & 0 \end{bmatrix}, \\
    &\GellMann_7 \equiv -\eye\ket{2}\bra{3} + \eye\ket{3}\bra{2} = \begin{bmatrix} 0 & 0 & 0 \\ 0 & 0 & -\eye \\ 0 & \eye & 0 \end{bmatrix},
\end{aligned}
\qquad
\begin{aligned}
    &\GellMann_2 \equiv -\eye\ket{0}\bra{1} + \eye \ket{1}\bra{0} = \begin{bmatrix} 0 & -\eye & 0 \\ \eye & 0 & 0 \\ 0 & 0 & 0 \end{bmatrix}, \\
    &\GellMann_4 \equiv \ket{0}\bra{2} + \ket{2}\bra{0} = \begin{bmatrix} 0 & 0 & 1 \\ 0 & 0 & 0 \\ 1 & 0 & 0 \end{bmatrix}, \\
    &\GellMann_6 \equiv \ket{2}\bra{3} + \ket{3}\bra{2} = \begin{bmatrix} 0 & 0 & 0 \\ 0 & 0 & 1 \\ 0 & 1 & 0 \end{bmatrix}, \\
    &\GellMann_8 \equiv \frac{1}{\sqrt{3}}\bigl(\ket{0}\bra{0} + \ket{1}\bra{1} - 2\ket{2}\bra{2}\bigr) = \frac{1}{\sqrt{3}}\begin{bmatrix} 1 & 0 & 0 \\ 0 & 1 & 0 \\ 0 & 0 & -2 \end{bmatrix},
\end{aligned}
\end{equation}
(indexed by $i$, which includes the 3-dimensional identity matrix for $i=0$), are accessible {\via} the \py{GellMann} class. Each can be selected using the \py{index} argument (with an integer from \py{0} to \py{8}), {\eg},
\begin{code}
\begin{minted}{python}
>>> L = GellMann(index=8)
>>> L.output()
Matrix([
[sqrt(3)/3,         0,            0],
[        0, sqrt(3)/3,            0],
[        0,         0, -2*sqrt(3)/3]])
>>> L.diagram()
$\includegraphics[scale=1.25, trim=-0.02cm 0 0 -0.15cm]{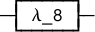}$
\end{minted}
\end{code}
\begin{code}
\begin{minted}{python}
>>> I = GellMann(index=0)
>>> I.output()
Matrix([
[1, 0, 0],
[0, 1, 0],
[0, 0, 1]])
>>> I.diagram()
$\includegraphics[scale=1.25, trim=-0.02cm 0 0 -0.15cm]{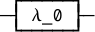}$
\end{minted}
\end{code}

\newpage

\paragraph{\colorbox{colorbackground}{\fontsize{11}{9}\texttt{Hadamard(\emph{*args}, \emph{**kwargs})}}}

An important gate in the quantum computing canonical toolbox is the \emph{Hadamard gate}. It is an example of a single-system gate that acts on qudits (states of any dimensionality greater than 2), and corresponds to the (generalized) \emph{Hadamard matrix}
\begin{equation}
\begin{aligned}
\Hadamard_\Dimension &= \frac{1}{\sqrt{\Dimension}}\sum\limits_{j,k=0}^{\Dimension - 1} \omega_\Dimension^{k(\Dimension - j)} \ket{j}\bra{k} \\
&= \begin{bmatrix} 1 & 1 & 1 & \ldots & 1 \\ 1 & \omega^{\Dimension - 1} & \omega^{2(\Dimension - 1)} & \ldots & \omega^{(\Dimension - 1)^2} \\ 1 & \omega^{\Dimension - 2} & \omega^{2(\Dimension - 2)} & \ldots & \omega^{(\Dimension - 1)(\Dimension - 2)} \\ \vdots & \vdots & \vdots & \ddots & \vdots \\ 1 & \omega & \omega^{2} & \ldots & \omega^{\Dimension - 1} \end{bmatrix},
\end{aligned}
\end{equation}
where $\omega_\Dimension \equiv \e^{2\pi\eye/\Dimension}$. Implemented as the \py{Hadamard} class, the simplest possible example of this gate is the familiar 2-dimensional form:
\begin{code}
\begin{minted}{python}
>>> H = Hadamard()
>>> H.output()
Matrix([
[sqrt(2)/2,  sqrt(2)/2],
[sqrt(2)/2, -sqrt(2)/2]])
>>> H.diagram()
$\includegraphics[scale=1.25, trim=-0.02cm 0 0 -0.15cm]{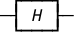}$
\end{minted}
\end{code}\enlargethispage{\baselineskip}
\begin{code}
\begin{minted}{python}
>>> HH = Hadamard(targets=[0, 1], label="H⊗H")
>>> HH.output()
Matrix([
[1/2,  1/2,  1/2,  1/2],
[1/2, -1/2,  1/2, -1/2],
[1/2,  1/2, -1/2, -1/2],
[1/2, -1/2, -1/2,  1/2]])
>>> HH.diagram()
$\includegraphics[scale=1.25, trim=-0.02cm 0 0 -0.15cm]{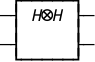}$
\end{minted}
\end{code}
Higher-dimensional variants can be created simply by passing any integer (greater than \py{2}) to the \py{dim} argument, {\eg},
\begin{code}
\begin{minted}{python}
>>> H3 = Hadamard(dim=3)
>>> H3.output()
Matrix([
[sqrt(3)/3,                sqrt(3)/3,                sqrt(3)/3],
[sqrt(3)/3, sqrt(3)*exp(-2*I*pi/3)/3,  sqrt(3)*exp(2*I*pi/3)/3],
[sqrt(3)/3,  sqrt(3)*exp(2*I*pi/3)/3, sqrt(3)*exp(-2*I*pi/3)/3]])
>>> H3.diagram()
$\includegraphics[scale=1.25, trim=-0.02cm 0 0 -0.15cm]{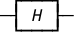}$
\end{minted}
\end{code}

\paragraph{\colorbox{colorbackground}{\fontsize{11}{9}\texttt{Rotation(\emph{*args}, \emph{axis}, \emph{angle}=\textcolor{colorcomment}{0}, \emph{**kwargs})}}}

The elementary \emph{rotation matrices} $\Rotation_i$ (axes indexed by $i$) are a set of three $2 \times 2$ matrices,
\begin{equation}
\begin{aligned}
\Rotation_x = \Rotation_1 = \e^{-\eye\Pauli_{x}\theta/2} &= \begin{bmatrix} \cos(\theta/2) & -\eye\sin(\theta/2) \\ -\eye\sin(\theta/2) & \cos(\theta/2)  \end{bmatrix}, \\
\Rotation_y = \Rotation_2 = \e^{-\eye\Pauli_{y}\theta/2} &= \begin{bmatrix} \cos(\theta/2) & -\sin(\theta/2) \\ \sin(\theta/2) & \cos(\theta/2) \end{bmatrix}, \\
\Rotation_z = \Rotation_3 = \e^{-\eye\Pauli_{z}\theta/2} &= \begin{bmatrix} \e^{-\eye\theta/2} & 0 \\ 0 & \e^{\eye\theta/2} \end{bmatrix},
\end{aligned}
\end{equation}
where $\theta$ is the \emph{rotation angle}. The corresponding 2-dimensional \emph{rotation gates} are implemented {\via} the \mintinline{python}{Rotation} class, for which the axis index ($i$) and rotation angle ($\theta$) can both be specified by the eponymic \mintinline{python}{axis} and \mintinline{python}{angle} arguments:
\begin{code}
\begin{minted}{python}
>>> R_x = Rotation(axis=1, angle="θ", label="R_x")
>>> R_x.output()
Matrix([
[   cos(θ/2), -I*sin(θ/2)],
[-I*sin(θ/2),    cos(θ/2)]])
>>> R_x.diagram()
$\includegraphics[scale=1.25, trim=-0.02cm 0 0 -0.15cm]{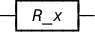}$
\end{minted}
\end{code}
\begin{code}
\begin{minted}{python}
>>> R_y = Rotation(axis=2, angle="φ", label="R_y")
>>> R_y.output()
Matrix([
[cos(φ/2), -sin(φ/2)],
[sin(φ/2),  cos(φ/2)]])
>>> R_y.diagram()
$\includegraphics[scale=1.25, trim=-0.02cm 0 0 -0.15cm]{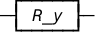}$
\end{minted}
\end{code}
\begin{code}
\begin{minted}{python}
>>> R_z = Rotation(axis=3, angle="t", label="R_z")
>>> R_z.output()
Matrix([
[exp(-I*t/2),          0],
[          0, exp(I*t/2)]])
>>> R_z.diagram()
$\includegraphics[scale=1.25, trim=-0.02cm 0 0 -0.15cm]{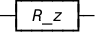}$
\end{minted}
\end{code}

\newpage

\paragraph{\colorbox{colorbackground}{\fontsize{11}{9}\texttt{Phase(\emph{*args}, \emph{phase}=\textcolor{colorcomment}{sp.exp(2 * sp.pi * sp.I / self.dim)}, \emph{**kwargs})}}}

The \py{Phase} class can be used to create \emph{phase gates}, which describe various phase transformations, both relative and global. These correspond to the $\Dimension$-dimensional operator $\Phase$, which itself may be represented as a $\Dimension \times \Dimension$ diagonal matrix
\begin{equation}
\begin{aligned}
\Phase(\omega) &= \sum\limits_{k=0}^{\Dimension - 1} \omega^k \ket{k}\bra{k} \\
&= \begin{bmatrix} 1 & 0 & 0 & \ldots & 0 \\ 0 & \omega & 0 & \ldots & 0 \\ 0 & 0 & \omega^2 & \ldots & 0 \\ \vdots & \vdots & \vdots & \ddots & \vdots \\ 0 & 0 & 0 & \ldots & \omega^{\Dimension - 1} \end{bmatrix}.
\end{aligned}
\end{equation}
Here, the \emph{phase factor} $\omega \in \Complexes$ is implemented {\via} the \py{phase} argument, which takes any valid scalar (numerical, symbolic, or string expression) value and defaults to $\e^{2\pi\eye/\Dimension}$ (as a SymPy expression).

An example:
\begin{code}
\begin{minted}{python}
>>> P = Phase()
>>> P.output()
Matrix([
[1,  0],
[0, -1]])
>>> P.diagram()
$\includegraphics[scale=1.25, trim=-0.02cm 0 0 -0.15cm]{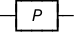}$
\end{minted}
\end{code}

The canonical $\op{S}$ (square root of $\op{Z}$) and $\op{T}$ (fourth root of $\op{Z}$) gates can each be easily defined by passing the appropriate values to the \py{exponent} argument:
\begin{code}
\begin{minted}{python}
>>> S = Phase(exponent=sp.Rational(1, 2), label="S")
>>> S.output()
Matrix([
[1, 0],
[0, I]])
>>> S.diagram()
$\includegraphics[scale=1.25, trim=-0.02cm 0 0 -0.15cm]{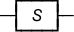}$
\end{minted}
\end{code}
\begin{code}
\begin{minted}{python}
>>> T = Phase(exponent=sp.Rational(1, 4), label="T")
>>> T.output()
Matrix([
[1,           0],
[0, exp(I*pi/4)]])
>>> T.diagram()
$\includegraphics[scale=1.25, trim=-0.02cm 0 0 -0.15cm]{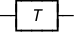}$
\end{minted}
\end{code}

\newpage
Higher-dimensional phase gates are also supported, {\eg},
\begin{code}
\begin{minted}{python}
>>> P3 = Phase(dim=3)
>>> P3.output()
Matrix([
[1,             0,              0],
[0, exp(2*I*pi/3),              0],
[0,             0, exp(-2*I*pi/3)]])
>>> P3.diagram()
$\includegraphics[scale=1.25, trim=-0.02cm 0 0 -0.15cm]{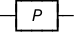}$
\end{minted}
\end{code}
\begin{code}
\begin{minted}{python}
>>> W = Phase(
...     phase="w",
...     dim=3,
...     label="W",
... )
>>> W.output()
Matrix([
[1, 0,    0],
[0, w,    0],
[0, 0, w**2]])
>>> W.diagram()
$\includegraphics[scale=1.25, trim=-0.02cm 0 0 -0.15cm]{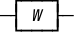}$
\end{minted}
\end{code}

\paragraph{\colorbox{colorbackground}{\fontsize{11}{9}\texttt{Diagonal(\emph{*args}, \emph{entries}, \emph{exponentiation}=\textcolor{colorcomment}{False}, \emph{**kwargs})}}}

The \py{Diagonal} class facilitates the construction of so-called \emph{diagonal gates}. These correspond to the $\Dimension$-dimensional operator $\Diagonal$, which may be represented as a $\Dimension \times \Dimension$ diagonal matrix
\begin{equation}
\begin{aligned}
\Diagonal(\lambda_0, \lambda_1, \ldots, \lambda_{\Dimension - 1}) &= \sum\limits_{k=0}^{\Dimension - 1} \lambda_k \ket{k}\bra{k} \\
&= \begin{bmatrix} \lambda_0 & 0 & 0 & \ldots & 0 \\ 0 & \lambda_1 & 0 & \ldots & 0 \\ 0 & 0 & \lambda_2 & \ldots & 0 \\ \vdots & \vdots & \vdots & \ddots & \vdots \\ 0 & 0 & 0 & \ldots & \lambda_{\Dimension - 1} \end{bmatrix}.
\end{aligned}
\end{equation}
Here, the main diagonal's \emph{entries} $\bigl\{\lambda_k \in \Complexes : \abs{\lambda_k} = 1\bigr\}_{k=0}^{\Dimension - 1}$ are implemented {\via} the \py{entries} argument, to which a dictionary containing level specifications (non-negative integer or a list of such types) (as the keys) and appropriate scalars (as the values) should be passed. Additionally, exponentiation (with imaginary unit) of the supplied values can be controlled by passing a Boolean value to the of the eponymic \py{exponentiation} argument. Note that levels which are unspecified in the \py{entries} argument all have a corresponding matrix element of \py{1}, regardless of the value of exponentiation.

\newpage
Some assorted examples:
\begin{code}
\begin{minted}{python}
>>> D = Diagonal(entries={0: "u", 1: "v"})
>>> D.output()
Matrix([
[u, 0],
[0, v]])
>>> D.diagram()
$\includegraphics[scale=1.25, trim=-0.02cm 0 0 -0.15cm]{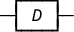}$
\end{minted}
\end{code}
\begin{code}
\begin{minted}{python}
>>> D3 = Diagonal(
...     entries={0: "a", 1: "b", 2: "c"},
...     dim=3,
... )
>>> D3.output()
Matrix([
[a, 0, 0],
[0, b, 0],
[0, 0, c]])
>>> D3.diagram()
$\includegraphics[scale=1.25, trim=-0.02cm 0 0 -0.15cm]{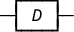}$
\end{minted}
\end{code}
\begin{code}
\begin{minted}{python}
>>> P = Diagonal(
...     entries={1: "p"},
...     exponentiation=True,
...     symbols={"p": {"real": True}},
...     label="P",
... )
>>> P.output()
Matrix([
[1,        0],
[0, exp(I*p)]])
>>> P.diagram()
$\includegraphics[scale=1.25, trim=-0.02cm 0 0 -0.15cm]{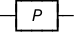}$
\end{minted}
\end{code}
In essence, as it grants the user full control over the elements along the resulting gate's main diagonal, the \py{Diagonal} gate can simply be thought of as a more general version of the \py{Phase} gate.

\paragraph{\colorbox{colorbackground}{\fontsize{11}{9}\texttt{Summation(\emph{*args}, \emph{shift}=\textcolor{colorcomment}{1}, \emph{**kwargs})}}}

The \emph{SUM gate} is essentially a parametric generalization of the NOT gate to higher dimensions. In $\Dimension$ dimensions, it is defined as the operator
\begin{equation}
\SUM(n) = \sum\limits_{k=0}^{\Dimension - 1} \: \ket{k \oplus n}\bra{k},
\end{equation}
where $n \in \Integers_{\geq 0}$ is the \emph{shift} parameter, and $k \oplus n \equiv k + n\ (\mathrel{\mathrm{mod}} \Dimension)$ is the modular addition of $k$ and $n$. Qhronology's implementation of this gate is provided by the \py{Sum} class, to which non-negative integer values can be passed to its \py{shift} argument.

A couple of examples:
\begin{code}
\begin{minted}{python}
>>> SUM = Summation(shift=1)
>>> SUM.output()
Matrix([
[0, 1],
[1, 0]])
>>> SUM.diagram()
$\includegraphics[scale=1.25, trim=-0.02cm 0 0 -0.15cm]{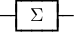}$
\end{minted}
\end{code}
\begin{code}
\begin{minted}{python}
>>> SUM3 = Summation(shift=1, dim=3)
>>> SUM3.output()
Matrix([
[0, 0, 1],
[1, 0, 0],
[0, 1, 0]])
>>> SUM3.diagram()
$\includegraphics[scale=1.25, trim=-0.02cm 0 0 -0.15cm]{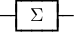}$
\end{minted}
\end{code}

As it is defined here, the matrix representation of the SUM operator corresponds to Sylvester's \emph{shift matrix} when \py{shift=1} (coinciding with $n=1$), and so represents one possible choice for a (non-Hermitian) generalization of the Pauli-$X$ $\Pauli_x$ matrix to $\Dimension$ dimensions. In the same vein, Qhronology's \py{Phase} gate (for specific, non-symbolic values of the phase factor, {\eg}, $\omega = -1$) can be likened to Sylvester's \emph{clock matrix}, which similarly generalizes the Pauli-$Z$ $\Pauli_z$ matrix.

\paragraph{\colorbox{colorbackground}{\fontsize{11}{9}\texttt{Swap(\emph{*args}, \emph{**kwargs})}}}

The \py{Swap} class can be used to define $\Dimension$-dimensional \emph{SWAP gates}, which represent the exchange of values (amplitudes) between any two quantum (sub)systems. Labelling such systems as $A$ and $B$, the associated \emph{SWAP operator} has the form
\begin{equation}
\Swap^{\indices{A,B}} = \sum\limits_{j,k=0}^{\Dimension - 1} {\ket{j}\bra{k}}^{\indices{A}} \otimes {\ket{k}\bra{j}}^{\indices{B}},
\end{equation}
which corresponds to a $\Dimension^2 \times \Dimension^2$ matrix in the computational basis. For SWAP gates in composite spaces of more than two systems, the identity operator acts on all non-targeted systems. Accordingly, the SWAP gate is an example of a bi-qudit gate, and so exactly two unique indices must be specified in the list passed to its \py{targets} argument, {\eg},
\begin{code}
\begin{minted}{python}
>>> SWAP = Swap(targets=[0, 1])
>>> SWAP.output()
Matrix([
[1, 0, 0, 0],
[0, 0, 1, 0],
[0, 1, 0, 0],
[0, 0, 0, 1]])
>>> SWAP.diagram()
$\includegraphics[scale=1.25, trim=-0.02cm 0 0 -0.15cm]{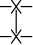}$
\end{minted}
\end{code}
\begin{code}
\begin{minted}{python}
>>> SWAP3 = Swap(targets=[0, 1], dim=3)
>>> SWAP3.output()
Matrix([
[1, 0, 0, 0, 0, 0, 0, 0, 0],
[0, 0, 0, 1, 0, 0, 0, 0, 0],
[0, 0, 0, 0, 0, 0, 1, 0, 0],
[0, 1, 0, 0, 0, 0, 0, 0, 0],
[0, 0, 0, 0, 1, 0, 0, 0, 0],
[0, 0, 0, 0, 0, 0, 0, 1, 0],
[0, 0, 1, 0, 0, 0, 0, 0, 0],
[0, 0, 0, 0, 0, 1, 0, 0, 0],
[0, 0, 0, 0, 0, 0, 0, 0, 1]])
>>> SWAP3.diagram()
$\includegraphics[scale=1.25, trim=-0.02cm 0 0 -0.15cm]{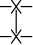}$
\end{minted}
\end{code}
A \emph{controlled-SWAP} (\emph{CSWAP}) \emph{gate}, also known as the \emph{Fredkin gate}, can be easily created with the inclusion of a control node:
\begin{code}
\begin{minted}{python}
>>> CSWAP = Swap(targets=[1, 2], controls=[0])
>>> CSWAP.output()
Matrix([
[1, 0, 0, 0, 0, 0, 0, 0],
[0, 1, 0, 0, 0, 0, 0, 0],
[0, 0, 1, 0, 0, 0, 0, 0],
[0, 0, 0, 1, 0, 0, 0, 0],
[0, 0, 0, 0, 1, 0, 0, 0],
[0, 0, 0, 0, 0, 0, 1, 0],
[0, 0, 0, 0, 0, 1, 0, 0],
[0, 0, 0, 0, 0, 0, 0, 1]])
>>> CSWAP.diagram()
$\includegraphics[scale=1.25, trim=-0.02cm 0 0 -0.15cm]{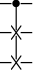}$
\end{minted}
\end{code}
\enlargethispage{\baselineskip}
A \emph{root-SWAP} ($\sqrt{\text{SWAP}}$) \emph{gate} can be constructed simply by specifying the appropriate value to the \py{exponent} argument:
\begin{code}
\begin{minted}{python}
>>> RSWAP = Swap(
...     targets=[0, 1],
...     exponent=sp.Rational(1, 2),
...     label="√SWAP",
...     family="GATE",
... )
>>> RSWAP.output()
Matrix([
[1,         0,         0, 0],
[0, 1/2 + I/2, 1/2 - I/2, 0],
[0, 1/2 - I/2, 1/2 + I/2, 0],
[0,         0,         0, 1]])
>>> RSWAP.diagram()
$\includegraphics[scale=1.25, trim=-0.02cm 0 0 -0.15cm]{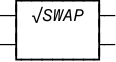}$
\end{minted}
\end{code}
A similar technique can be used to obtain a \emph{PSWAP} ($\text{SWAP}^p$) \emph{gate}, {\eg},
\begin{code}
\begin{minted}{python}
>>> PSWAP = Swap(
...     targets=[0, 1],
...     exponent="p",
...     symbols={"p": {"real": True}},
...     label="SWAP^p",
...     family="GATE",
... )
>>> PSWAP.output()
Matrix([
[1,                   0,                   0, 0],
[0, exp(I*pi*p)/2 + 1/2, 1/2 - exp(I*pi*p)/2, 0],
[0, 1/2 - exp(I*pi*p)/2, exp(I*pi*p)/2 + 1/2, 0],
[0,                   0,                   0, 1]])
>>> PSWAP.diagram()
$\includegraphics[scale=1.25, trim=-0.02cm 0 0 -0.15cm]{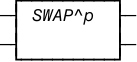}$
\end{minted}
\end{code}

\paragraph{\colorbox{colorbackground}{\fontsize{11}{9}\texttt{Fourier(\emph{*args}, \emph{composite}=\textcolor{colorcomment}{True}, \emph{reverse}=\textcolor{colorcomment}{False}, \emph{**kwargs})}}}

\enlargethispage{\baselineskip}

The \emph{Fourier gate} is a genuinely multipartite gate in the sense that it acts on multiple systems, with the order in which it does so being important. Corresponding to the \emph{Fourier operator} $\QFT$, the matrix representation of this gate has different forms depending on the number of systems it targets. For a single $\Dimension$-dimensional qudit, the Fourier operator may be represented as the $\Dimension \times \Dimension$ matrix
\begin{equation}
\begin{aligned}
\QFT &= \frac{1}{\sqrt{\Dimension}} \sum\limits_{j,k=0}^{\Dimension - 1} \omega_{\Dimension}^{jk} \ket{j}\bra{k} \\
&= \frac{1}{\sqrt{\Dimension}} \begin{bmatrix} 1 & 1 & 1 & 1 & \ldots & 1 \\
1 & \omega & \omega^2 & \omega^3 & \ldots & \omega^{\Dimension - 1} \\
1 & \omega^2 & \omega^4 & \omega^6 & \ldots & \omega^{2(\Dimension - 1)} \\
1 & \omega^3 & \omega^6 & \omega^9 & \ldots & \omega^{3(\Dimension - 1)} \\
\vdots & \vdots & \vdots & \vdots & \ddots & \vdots \\
1 & \omega^{\Dimension - 1} & \omega^{2(\Dimension - 1)} & \omega^{3(\Dimension - 1)} & \ldots & \omega^{(\Dimension - 1)(\Dimension - 1)}
\end{bmatrix},
\end{aligned}
\end{equation}
where $\omega_{\Dimension} \equiv \e^{2\pi\eye/\Dimension} = \omega$. In the case of $N$ qudits, it is easier to characterize the \emph{multipartite Fourier operator} $\QFT_N$ not by its matrix form but by the transformation it imposes, to which its action on the basis state $\bigotimes\limits_{\ell=1}^{N} \ket{j_\ell} \equiv \ket{j_1, \ldots, j_N}$ (where $j_\ell \in \Integers_{0}^{\Dimension - 1}$) is
\begin{equation}        
\ket{j_1, \ldots, j_N} \stackrel{\QFT_N}{\longrightarrow} \frac{1}{\sqrt{\Dimension^N}} \bigotimes\limits_{\ell=1}^{N} \sum\limits_{k_\ell=0}^{\Dimension - 1} \e^{2\pi\eye j k_\ell \Dimension^{-\ell}} \ket{k_\ell},
\end{equation}
where $j \equiv \sum\limits_{\ell=1}^{N} j_\ell \Dimension^{N - \ell}$.

Both of these operators are implemented by the \py{Fourier} class, with the desired behaviour obtained {\via} the \py{composite} argument. If \py{True} is given, the composite form $\QFT_N$ is applied to the \py{targets} subsystems in the order specified by the class argument \py{reverse}:
\begin{itemize}
    \item \emph{ascending} order if \py{reverse} is \py{False}
    \item \emph{descending} order if \py{reverse} is \py{True}
\end{itemize}
Alternatively, if \py{False} is given, a copy of the elementary form $\QFT$ is placed on each of the subsystems corresponding to the indices in the \py{targets} property.

Some examples:
\begin{code}
\begin{minted}{python}
>>> F = Fourier()
>>> F.output()
Matrix([
[sqrt(2)/2,  sqrt(2)/2],
[sqrt(2)/2, -sqrt(2)/2]])
>>> F.diagram()
$\includegraphics[scale=1.25, trim=-0.02cm 0 0 -0.15cm]{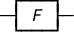}$
\end{minted}
\end{code}\enlargethispage{\baselineskip}
\begin{code}
\begin{minted}{python}
>>> F3 = Fourier(dim=3)
>>> F3.output()
Matrix([
[sqrt(3)/3,                sqrt(3)/3,                sqrt(3)/3],
[sqrt(3)/3, sqrt(3)*exp(-2*I*pi/3)/3,  sqrt(3)*exp(2*I*pi/3)/3],
[sqrt(3)/3,  sqrt(3)*exp(2*I*pi/3)/3, sqrt(3)*exp(-2*I*pi/3)/3]])
>>> F3.diagram()
$\includegraphics[scale=1.25, trim=-0.02cm 0 0 -0.15cm]{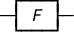}$
\end{minted}
\end{code}
\begin{code}
\begin{minted}{python}
>>> FF = Fourier(targets=[0, 1])
>>> FF.output()
Matrix([
[1/2,  1/2,  1/2,  1/2],
[1/2, -1/2,  1/2, -1/2],
[1/2,  I/2, -1/2, -I/2],
[1/2, -I/2, -1/2,  I/2]])
>>> FF.diagram()
$\includegraphics[scale=1.25, trim=-0.02cm 0 0 -0.15cm]{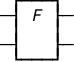}$
\end{minted}
\end{code}

\paragraph{\colorbox{colorbackground}{\fontsize{11}{9}\texttt{Unitary(\emph{*args}, \emph{parameters}=\textcolor{colorcomment}{(0, 0, 0)}, \emph{**kwargs})}}}

Any single-system \emph{unitary gate} acting on qubits can be written as a 2-dimensional unitary operator $\Unitary$, with one parametrization of its matrix representation being the $2 \times 2$ matrix
\begin{equation}
\Unitary(\theta,\phi,\lambda) =
\begin{bmatrix}
\cos(\theta/2) & -\e^{\eye\lambda}\sin(\theta/2) \\
\e^{\eye\phi}\sin(\theta/2) & \e^{\eye(\phi + \lambda)}\cos(\theta/2)
\end{bmatrix}.
\end{equation}
Here, the parameters $\theta$, $\phi$, and $\lambda$, interpretable as angles, collectively characterize the corresponding gate, which is implemented as the \py{Unitary} class:

\begin{code}
\begin{minted}{python}
>>> U = Unitary(parameters=("θ", "φ", "λ"))
>>> U.output()
Matrix([
[         cos(θ/2),      -exp(I*λ)*sin(θ/2)],
[exp(I*φ)*sin(θ/2), exp(I*(λ + φ))*cos(θ/2)]])
>>> U.diagram()
$\includegraphics[scale=1.25, trim=-0.02cm 0 0 -0.15cm]{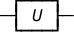}$
\end{minted}
\end{code}
\begin{code}
\begin{minted}{python}
>>> I = Unitary(parameters=(0, 0, 0), label="I")
>>> I.output()
Matrix([
[1, 0],
[0, 1]])
>>> I.diagram()
$\includegraphics[scale=1.25, trim=-0.02cm 0 0 -0.15cm]{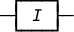}$
\end{minted}
\end{code}
\begin{code}
\begin{minted}{python}
>>> H = Unitary(parameters=(sp.pi/2, 0, sp.pi), label="H")
>>> H.output()
Matrix([
[sqrt(2)/2,  sqrt(2)/2],
[sqrt(2)/2, -sqrt(2)/2]])
>>> H.diagram()
$\includegraphics[scale=1.25, trim=-0.02cm 0 0 -0.15cm]{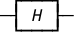}$
\end{minted}
\end{code}

\paragraph{\colorbox{colorbackground}{\fontsize{11}{9}\texttt{Measurement(\emph{*args}, \emph{operators}, \emph{observable}=\textcolor{colorcomment}{False}, \emph{**kwargs})}}}

Last of Qhronology's gates is the \emph{measurement gate}, used to perform quantum measurement operations on its target states. While technically not a gate (as it is generally not a linear transformation on its input states, and so does not constitute a reversible process), its ability to function as a circuit element is similar to that of true gates, and so it can be implemented as such.

A measurement gate can be created with the \py{Measurement} class, which possesses the primary arguments of \py{operators} and \py{observable}. Note that the operation of these is identical to that of the homonymous arguments of \py{QuantumState}'s \py{measure()} method (in {\secn} \ref{sec:states_operations}), so please consult its description for more detail about their usage.

\enlargethispage{-2\baselineskip}
A few useful examples:
\begin{code}
\begin{minted}{python}
>>> from qhronology.mechanics.matrices import ket
>>> basis_vectors = [ket(i) for i in [0, 1]]
>>> M_basis = Measurement(operators=basis_vectors, observable=False)
>>> M_basis.diagram()
$\includegraphics[scale=1.25, trim=-0.02cm 0 0 -0.15cm]{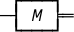}$
\end{minted}
\end{code}
\begin{code}
\begin{minted}{python}
>>> pauli_matrices = [Pauli(index=i) for i in [1, 2, 3]]
>>> M_pauli = Measurement(
...     operators=pauli_matrices,
...     observable=True,
...     targets=[0],
...     num_systems=2,
... )
>>> M_pauli.diagram()
$\includegraphics[scale=1.25, trim=-0.02cm 0 0 -0.15cm]{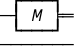}$
\end{minted}
\end{code}
\begin{code}
\begin{minted}{python}
>>> from qhronology.mechanics.matrices import ket
>>> plus = (ket(0) + ket(1)) / sp.sqrt(2)
>>> minus = (ket(0) - ket(1)) / sp.sqrt(2)
>>> M_pm = Measurement(
...     operators=[plus, minus],
...     observable=False,
...     targets=[1],
...     num_systems=2,
... )
>>> M_pm.diagram()
$\includegraphics[scale=1.25, trim=-0.02cm 0 0 -0.15cm]{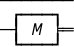}$
\end{minted}
\end{code}

\subsubsection{Gate compositions}\label{sec:gates_compositions}

In addition to implementations of many quantum gates, Qhronology provides the ability to combine them, which can prove particularly useful for the simplification of circuit diagrams. The functionality to achieve this takes two forms: ``interleaved'' compositions {\via} the \py{GateInterleave} class and ``stacked'' compositions {\via} the \py{GateStack} class:
\begin{code}
\begin{minted}{python}
>>> from qhronology.quantum.gates import GateInterleave, GateStack
\end{minted}
\end{code}
Both of these classes possess the (starred) \py{*gates} argument, meaning that the \py{QuantumGate} objects to be composed may be given as an unpacked list directly to the constructor. Given such gates, the \py{GateInterleave} class computes the matrix multiplication of their matrix representations, while \py{GateStack} computes their tensor product. This means that the constituent gates used in the former must all span the same number of systems (with no overlapping target, control, and anticontrol systems), while those in the latter can span any number of systems.

The composition classes also take a few other useful arguments. Principal of these is \py{merge} (\py{bool}, defaults to \py{False}), which determines whether to merge the individual gates together diagrammatically. The \py{conjugate}, \py{exponent}, and \py{coefficient} arguments are used in exactly the same way as in the \py{QuantumGate} class, while the \py{label} argument specifies the character(s) to be used to represent the composed gate in both prints and diagrams.

On the topic of combining gates together, it can be useful to know that temporal (``horizontal'') compositions ({\ie}, gates wired in serial and merged into single object instances) are easily obtained in Qhronology. By simply by combining the individual gates sequentially in a circuit, one may extract the total gate (as a single \py{QuantumGate} instance) {\via} the \py{gate()} method (see {\secn} \ref{sec:circuits_gates}), thereby yielding a gate object describing the product of a sequence of gates.

\newpage
Presented below is an example of each type of composition, both of which yield the same gate:
\begin{code}
\begin{minted}{python}
>>> XII = Pauli(index=1, targets=[0], num_systems=3)
>>> IYI = Pauli(index=2, targets=[1], num_systems=3)
>>> IIZ = Pauli(index=3, targets=[2], num_systems=3)
>>> XYZ = GateInterleave(XII, IYI, IIZ)
>>> XYZ.output()
Matrix([
[0,  0,  0, 0, 0,  0, -I, 0],
[0,  0,  0, 0, 0,  0,  0, I],
[0,  0,  0, 0, I,  0,  0, 0],
[0,  0,  0, 0, 0, -I,  0, 0],
[0,  0, -I, 0, 0,  0,  0, 0],
[0,  0,  0, I, 0,  0,  0, 0],
[I,  0,  0, 0, 0,  0,  0, 0],
[0, -I,  0, 0, 0,  0,  0, 0]])
>>> XYZ.diagram(sep=(1, 2))
$\includegraphics[scale=1.25, trim=-0.02cm 0 0 -0.15cm]{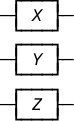}$
\end{minted}
\end{code}
\begin{code}
\begin{minted}{python}
>>> X = Pauli(index=1)
>>> Y = Pauli(index=2)
>>> Z = Pauli(index=3)
>>> XYZ = GateStack(X, Y, Z)
>>> XYZ.output()
Matrix([
[0,  0,  0, 0, 0,  0, -I, 0],
[0,  0,  0, 0, 0,  0,  0, I],
[0,  0,  0, 0, I,  0,  0, 0],
[0,  0,  0, 0, 0, -I,  0, 0],
[0,  0, -I, 0, 0,  0,  0, 0],
[0,  0,  0, I, 0,  0,  0, 0],
[I,  0,  0, 0, 0,  0,  0, 0],
[0, -I,  0, 0, 0,  0,  0, 0]])
>>> XYZ.diagram(sep=(1, 2))
$\includegraphics[scale=1.25, trim=-0.02cm 0 0 -0.15cm]{./text_examples_docstrings_gate_composition_xyz.pdf}$
\end{minted}
\end{code}

\newpage
\subsection{Circuits}\label{sec:circuits}

In Qhronology, quantum circuits are created as instances of the \py{QuantumCircuit} class:
\begin{code}
\begin{minted}{python}
>>> from qhronology.quantum.circuits import QuantumCircuit
\end{minted}
\end{code}
In the circuit diagram picturalism, time increases from left to right. Accordingly, the preparation of input states begins in the past (on the left), while post-processing (such as postselections and partial traces) occurs in the future (on the right). Intermediary operations on these states are represented by quantum gates, created as instances of the \py{QuantumGate} class and its derivatives. All of these events are connected by quantum wires describing the flow of quantum information ({\ie}, quantum probabilities) through time. In the following sections, specific detail on Qhronology's implementation of each of these aspects is given.

\subsubsection{Inputs}\label{sec:circuits_inputs}

The input states to be used in a circuit instance are passed in the form of a list of \py{QuantumState} instances to its constructor's \py{inputs} argument. The order of the list should reflect the order in which the states appear in the composition's tensor product (left-to-right) or, equivalently, the circuit diagram (top-to-bottom). Note that both the \py{symbols} and \py{substitutions} properties from each state are merged into the same properties of the circuit instance.

The entire input composition (as a tensor product of all individual states in the order in which they appear in the \py{inputs} property), in the form of a \py{QuantumState} instance, can be obtained from any \py{QuantumCircuit} {\via} the \py{input()} method. This state is a vector only if all of its component states are vectors, otherwise it is a matrix.

As a simple, consider the following bipartite circuit in which two single-system gates are be passed to the \py{inputs} argument, resulting in a composite input vector state. The total input state, retrieved using the \py{input()} method, can be inspected in the usual ways:\enlargethispage{2\baselineskip}
\begin{code}[label=code:circuit_input_vector]
\begin{minted}{python}
>>> input_upper = QuantumState(
...     spec=[("a", [0]), ("b", [1])],
...     form="vector",
...     substitutions=[("a*conjugate(a) + b*conjugate(b)", 1)],
...     label="ψ",
... )
>>> input_lower = QuantumState(
...     spec=[("c", [0]), ("d", [1])],
...     form="vector",
...     substitutions=[("c*conjugate(c) + d*conjugate(d)", 1)],
...     label="φ",
... )
>>> input_upper.print()
|ψ⟩ = a|0⟩ + b|1⟩
>>> input_lower.print()
|φ⟩ = c|0⟩ + d|1⟩
>>> bipartite_inputs = QuantumCircuit(inputs=[input_upper, input_lower])
>>> bipartite_inputs.diagram()
$\includegraphics[scale=1.25, trim=-0.02cm 0 0 -0.15cm]{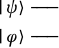}$
>>> input_both = bipartite_inputs.input(merge=False)
>>> input_both.diagram()
$\includegraphics[scale=1.25, trim=-0.02cm 0 0 -0.15cm]{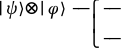}$
>>> input_both.print(product=True)
|ψ⟩⊗|φ⟩ = a*c|0⟩⊗|0⟩ + a*d|0⟩⊗|1⟩ + b*c|1⟩⊗|0⟩ + b*d|1⟩⊗|1⟩
\end{minted}
\end{code}
Combining both vector and matrix states is perfectly valid{\textemdash}all vector states will simply be converted to their density matrix forms prior to transformation by the gates (the calculations of which will be performed in matrix form). For example:\enlargethispage{2\baselineskip}
\begin{code}[label=code:circuit_input_matrix]
\begin{minted}{python}
>>> input_upper = QuantumState(
...     spec=[("a", [0]), ("b", [1])],
...     kind="mixed",
...     symbols={"a": {"real": True}, "b": {"real": True}},
...     substitutions=[("a + b", 1)],
...     label="ρ",
... )
>>> input_lower = QuantumState(
...     spec=[("c", [0]), ("d", [1])],
...     form="vector",
...     substitutions=[("c*conjugate(c) + d*conjugate(d)", 1)],
...     label="φ",
... )
>>> input_upper.print()
ρ = a|0⟩⟨0| + b|1⟩⟨1|
>>> input_lower.print()
|φ⟩ = c|0⟩ + d|1⟩
>>> bipartite_inputs = QuantumCircuit(inputs=[input_upper, input_lower])
>>> bipartite_inputs.diagram()
$\includegraphics[scale=1.25, trim=-0.02cm 0 0 -0.15cm]{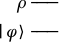}$
>>> input_both = bipartite_inputs.input()
>>> input_both.diagram()
$\includegraphics[scale=1.25, trim=-0.02cm 0 0 -0.15cm]{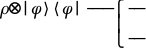}$
>>> input_both.print(product=True)
ρ⊗|φ⟩⟨φ| = a*c*conjugate(c)|0⟩⟨0|⊗|0⟩⟨0| + a*c*conjugate(d)|0⟩⟨0|⊗|0⟩⟨1| + a*d*conjugate(c)|0⟩⟨0|⊗|1⟩⟨0| + a*d*conjugate(d)|0⟩⟨0|⊗|1⟩⟨1| + b*c*conjugate(c)|1⟩⟨1|⊗|0⟩⟨0| + b*c*conjugate(d)|1⟩⟨1|⊗|0⟩⟨1| + b*d*conjugate(c)|1⟩⟨1|⊗|1⟩⟨0| + b*d*conjugate(d)|1⟩⟨1|⊗|1⟩⟨1|
\end{minted}
\end{code}

\subsubsection{Gates}\label{sec:circuits_gates}

The quantum gates to be used in a circuit instance are passed in the form of a list of \py{QuantumGate} instances to its constructor's \py{gates} argument. The order of the list should reflect the order in which the gates are applied to the inputs states (as per the circuit diagram, {\eg}, left-to-right), which is the reverse of the order of operations in the mathematical representation. The \py{symbols} and \py{substitutions} properties of each gate are merged into the same properties of the circuit instance.

The entire gate sequence, in the form of a single \py{QuantumGate} instance (describing the operator matrix-product of all individual gates in the order in which they appear in the \py{gates} property), can be obtained from any \py{QuantumCircuit} object {\via} the \py{gate()} method, whereby it can be inspected, modified, and incorporated into another circuit. This functionality can prove to be especially useful for computing the matrix forms of sequences of gates.

\newpage
Note that input states need not be specified to a circuit: systems without an input state automatically assume a the zero state $\ket{0}$ of the same dimensionality of the given gate(s), which does not interfere with any gate calculations.

As an example, to verify the identity
\begin{equation}
\Pauli_z \Pauli_y \Pauli_x = -\eye\Identity,
\end{equation}
(where $\Identity$ is the $2 \times 2$ identity matrix), we can construct a sequence of Pauli gates in a circuit, and subsequently extract its total gate. This can be accomplished as follows:
\begin{code}[label=code:circuit_gate_xyz]
\begin{minted}{python}
>>> X = Pauli(index=1)
>>> Y = Pauli(index=2)
>>> Z = Pauli(index=3)
>>> pauli_sequence = QuantumCircuit(gates=[X, Y, Z])
>>> pauli_sequence.diagram(visible={"gates"})
$\includegraphics[scale=1.25, trim=-0.02cm 0 0 -0.15cm]{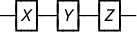}$
>>> XYZ = pauli_sequence.gate(label="XYZ")
>>> XYZ.diagram()
$\includegraphics[scale=1.25, trim=-0.02cm 0 0 -0.15cm]{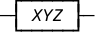}$
>>> XYZ.output()
Matrix([
[-I,  0],
[ 0, -I]])
\end{minted}
\end{code}

Similarly, we can verify that a sequence of exactly three CNOT gates (with alternating target and control nodes) is equivalent to a SWAP gate:
\begin{code}[label=code:circuit_gate_swapcnots]
\begin{minted}{python}
>>> CN = Not(targets=[1], controls=[0])
>>> NC = Not(targets=[0], controls=[1])
>>> CNOTs = QuantumCircuit(gates=[CN, NC, CN])
>>> CNOTs.diagram(sep=(2, 1), force_separation=True, visible={"gates"})
$\includegraphics[scale=1.25, trim=-0.02cm 0 0 -0.15cm]{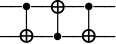}$
>>> SWAP = CNOTs.gate(label="S")
>>> SWAP.diagram()
$\includegraphics[scale=1.25, trim=-0.02cm 0 0 -0.15cm]{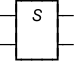}$
>>> SWAP.output()
Matrix([
[1, 0, 0, 0],
[0, 0, 1, 0],
[0, 1, 0, 0],
[0, 0, 0, 1]])
\end{minted}
\end{code}

\newpage
\subsubsection{Outputs}\label{sec:circuits_outputs}

There are a couple of methods by which a circuit's output state may be obtained. First is \py{output()}, which simply returns the total output state as a SymPy matrix or NumPy array (chosen using the \py{array} argument or property). The second (and more important) method is \py{state()}, which returns the output state of the current quantum circuit as a \py{QuantumState} instance. Note that, for both of these, the \py{numerical} argument (defaulting to the value of the homonymous class property) may be used to cast the output as floating-point values instead of exact SymPy data types (in the case of purely non-symbolic output).

The \py{state()} method is intended to be the primary way by which evolved states are extracted from the circuit. Of its many arguments, only a select few will be discussed here. Perhaps the most useful of these is \py{traces}, which performs partial traces on the systems corresponding to the list of indices passed to it. Note that these indices must be relative to the composite system space of the state, not the circuit. This is because if any partial traces ({\secn} \ref{sec:circuits_traces}) or postselections ({\secn} \ref{sec:circuits_postselections}) are specified, the Hilbert space to which the output state belongs is necessarily smaller than the space in which the circuit (in its entirety, prior to any reduction) resides. In this vein, one may use the \py{postprocessing} argument (passing either \py{True} or \py{False}) to choose whether to ignore the circuit's partial traces and postselections when obtaining the output state.

As the \py{state()} method returns an instance of the \py{QuantumState} class, it has a plethora of properties which can be set. For a few of the most significant, including \py{norm} and \py{label}, this can be achieved at instantiation by specifying the appropriate values to the eponymic arguments in the \py{state()} method itself. Note that any values contained within the circuit's \py{symbols} and \py{substitutions} properties are automatically copied to the corresponding properties of the output state upon creation.

To demonstrate the extraction of output states from a quantum circuit, we can combine the previous examples in {\cbs} \ref{code:circuit_input_vector} and \ref{code:circuit_gate_swapcnots} to give a simple bipartite circuit describing a pair of single-system states that are transformed by a sequence of alternating CNOT gates (equivalent to a SWAP gate):\enlargethispage{2\baselineskip}
\begin{codetitled}{CNOTs equivalent to SWAP}{code:circuit_swapcnots}
\begin{minted}{python}
from qhronology.quantum.states import VectorState
from qhronology.quantum.gates import Not
from qhronology.quantum.circuits import QuantumCircuit

# Input
input_upper = VectorState(
    spec=[("a", [0]), ("b", [1])],
    substitutions=[("a*conjugate(a) + b*conjugate(b)", 1)],
    label="ψ",
)
input_lower = VectorState(
    spec=[("c", [0]), ("d", [1])],
    substitutions=[("c*conjugate(c) + d*conjugate(d)", 1)],
    label="φ",
)

# Gates
CN = Not(targets=[1], controls=[0])
NC = Not(targets=[0], controls=[1])

# Circuit
swapcnots = QuantumCircuit(
    inputs=[input_upper, input_lower],
    gates=[CN, NC, CN],
)
swapcnots.diagram()

# Output
output_total = swapcnots.state(label="(ψ⊗φ)′")
output_upper = swapcnots.state(traces=[1], label="ψ′")
output_lower = swapcnots.state(traces=[0], label="φ′")
output_upper.kind = "pure"
output_lower.kind = "pure"
output_upper.simplify()
output_lower.simplify()

# Results
input_upper.print()
input_lower.print()
swapcnots.input().print()

output_upper.print()
output_lower.print()
output_total.print()
\end{minted}
\tcblowerspaced
\begin{minted}{python}
>>> swapcnots.diagram()
$\includegraphics[scale=1.25, trim=-0.02cm 0 0 -0.15cm]{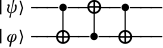}$
>>> input_upper.print()
|ψ⟩ = a|0⟩ + b|1⟩
>>> input_lower.print()
|φ⟩ = c|0⟩ + d|1⟩
>>> swapcnots.input().print()
|ψ⊗φ⟩ = a*c|0,0⟩ + a*d|0,1⟩ + b*c|1,0⟩ + b*d|1,1⟩
>>> output_upper.print()
|ψ′⟩⟨ψ′| = c*conjugate(c)|0⟩⟨0| + c*conjugate(d)|0⟩⟨1| + d*conjugate(c)|1⟩⟨0| + d*conjugate(d)|1⟩⟨1|
>>> output_lower.print()
|φ′⟩⟨φ′| = a*conjugate(a)|0⟩⟨0| + a*conjugate(b)|0⟩⟨1| + b*conjugate(a)|1⟩⟨0| + b*conjugate(b)|1⟩⟨1|
>>> output_total.print()
|(ψ⊗φ)′⟩ = a*c|0,0⟩ + b*c|0,1⟩ + a*d|1,0⟩ + b*d|1,1⟩
\end{minted}
\end{codetitled}

\subsubsection{Traces}\label{sec:circuits_traces}

Partial trace operations, representing the discarding of systems from the larger composite space, can be incorporated into any circuit by passing a list of the indices (corresponding to the systems which are to be traced over) to the \py{traces} argument (or property). This can be illustrated through the use of a simple SWAP circuit:\enlargethispage{\baselineskip}
\begin{codetitled}{Vector state SWAP with partial traces}{code:circuit_traces_swap}
\begin{minted}{python}
from qhronology.quantum.states import VectorState
from qhronology.quantum.gates import Swap
from qhronology.quantum.circuits import QuantumCircuit

# Input
input_upper = VectorState(
    spec=[("a", [0]), ("b", [1])],
    substitutions=[("a*conjugate(a) + b*conjugate(b)", 1)],
    label="ψ",
)
input_lower = VectorState(
    spec=[("c", [0]), ("d", [1])],
    substitutions=[("c*conjugate(c) + d*conjugate(d)", 1)],
    label="φ",
)

# Gate
SWAP = Swap(targets=[0, 1])

# Circuits
traced_none = QuantumCircuit(
    inputs=[input_upper, input_lower], gates=[SWAP]
)
traced_upper = QuantumCircuit(
    inputs=[input_upper, input_lower], gates=[SWAP], traces=[0]
)
traced_lower = QuantumCircuit(
    inputs=[input_upper, input_lower], gates=[SWAP], traces=[1]
)

traced_none.diagram()
traced_upper.diagram()
traced_lower.diagram()

# Output
output_total = traced_none.state(label="(ψ⊗φ)′")
output_lower = traced_upper.state(simplify=True, label="ψ′")
output_upper = traced_lower.state(simplify=True, label="φ′")
output_lower.kind = "pure"
output_upper.kind = "pure"

# Results
output_total.print(product=True)
output_lower.print(product=True)
output_upper.print(product=True)
\end{minted}
\tcblowerspaced
\begin{minted}{python}
>>> traced_none.diagram()
$\includegraphics[scale=1.25, trim=-0.02cm 0 0 -0.15cm]{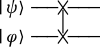}$
>>> traced_upper.diagram()
$\includegraphics[scale=1.25, trim=-0.02cm 0 0 -0.15cm]{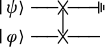}$
>>> traced_lower.diagram()
$\includegraphics[scale=1.25, trim=-0.02cm 0 0 -0.15cm]{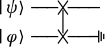}$
>>> output_total.print(product=True)
|(ψ⊗φ)′⟩ = a*c|0⟩⊗|0⟩ + b*c|0⟩⊗|1⟩ + a*d|1⟩⊗|0⟩ + b*d|1⟩⊗|1⟩
>>> output_lower.print(product=True)
|φ′⟩⟨φ′| = a*conjugate(a)|0⟩⟨0| + a*conjugate(b)|0⟩⟨1| + b*conjugate(a)|1⟩⟨0| + b*conjugate(b)|1⟩⟨1|
>>> output_upper.print(product=True)
|ψ′⟩⟨ψ′| = c*conjugate(c)|0⟩⟨0| + c*conjugate(d)|0⟩⟨1| + d*conjugate(c)|1⟩⟨0| + d*conjugate(d)|1⟩⟨1|
\end{minted}
\end{codetitled}
Note that any partial traces are performed in simultaneity with any specified postselections (given in \py{postselections}, see {\secn} \ref{sec:circuits_postselections}) at the termination (right-hand side) of the circuit, and so there must not be any overlap between the systems given in either of these arguments. Additionally, if fixed traces in the circuit itself are not desired, one can simply use the \py{traces} argument of the \py{state()} method, as described in {\secn} \ref{sec:circuits_outputs}.

\subsubsection{Postselections}\label{sec:circuits_postselections}

In Qhronology, the postselection of a circuit's output state is treated simply as a post-evolution operation at the termination of the circuit. In a \py{QuantumCircuit} object, postselection can be incorporated by passing a list of 2-tuples to the \py{postselections} argument, with each 2-tuple (representing an individual postselection) containing first a \py{QuantumState} instance (the postselection state itself) followed by a list of system indices which the postselection state is to target. Note that these indices must be consecutive, their number must match the number of systems spanned by the postselection state, and, most importantly, they cannot conflict with those specified in the circuit's partial traces (see {\secn} \ref{sec:circuits_traces}).

As an example, we can use Bell state preparation and postselection to transfer an arbitrary vector state between systems:\enlargethispage{2\baselineskip}
\begin{codetitled}{Bell postselection}{code:circuit_postselection}
\begin{minted}{python}
from qhronology.quantum.states import VectorState
from qhronology.quantum.circuits import QuantumCircuit

# Input
input_state = VectorState(
    spec=[("a", [0]), ("b", [1])],
    substitutions=[("a*conjugate(a) + b*conjugate(b)", 1)],
    label="ψ",
)
bell = VectorState(
    spec=[(1, [0, 0]), (1, [1, 1])],
    norm=False,
    label="Φ",
)

# Circuit
postselection = QuantumCircuit(
    inputs=[input_state, bell],
    postselections=[(bell, [0, 1])],
)
postselection.diagram(sep=(4, 1))

# Output
output_state = postselection.state(label="ψ′")

# Results
input_state.print()
output_state.print()
\end{minted}
\tcblowerspaced
\begin{minted}{python}
>>> postselection.diagram(sep=(4, 1))
$\includegraphics[scale=1.25, trim=-0.02cm 0 0 -0.15cm]{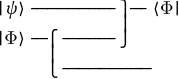}$
>>> input_state.print()
|ψ⟩ = a|0⟩ + b|1⟩
>>> output_state.print()
|ψ′⟩ = a|0⟩ + b|1⟩
\end{minted}
\end{codetitled}

\subsubsection{Measurement}\label{sec:circuits_measurement}

Quantum measurement on an instance of the \py{QuantumCircuit} class is performed on the circuit's total output state (after traces and postselections) and implemented in exactly the same manner as that of states described by the \py{QuantumState} class: {\via} the \py{measure()} method. This functionality is made available to \py{QuantumCircuit} purely as a way of obtaining measurement statistics (\py{statistics=True}) or the post-measurement state (as a \py{QuantumState} instance) (\py{statistics=False}) without needing to manually extract any output state(s) beforehand. The usage of this method is (mostly) unchanged compared to that of \py{QuantumState}, so please consult the relevant information regarding \py{measure()} in {\secn} \ref{sec:states_operations} for more detail, including a few helpful examples.

\enlargethispage{2\baselineskip}
\subsubsection{Visualization}\label{sec:circuits_visualization}

All instances of classes which have the \py{QuantumObject} class as an ancestor can be visualized by way of the \py{diagram()} method. This includes the \py{QuantumState} and \py{QuantumGate} classes (and their derivatives), in addition to the \py{QuantumCircuit} class. The shape, spacing, and style of the produced diagrams can be adjusted using the following five arguments:
\begin{itemize}
\item \py{pad}: A 2-tuple of non-negative integers specifying intra-gate padding.
\item \py{sep}: A 2-tuple of non-negative integers specifying inter-gate separation.
\item \py{uniform_spacing}: A Boolean value specifying whether to uniformly space the gates horizontally such that the midpoint of each is equidistant from those of its neighbours.
\item \py{force_separation}: A Boolean value specifying whether to force the horizontal gate separation to be exactly the value given in \py{sep} for all gates in the circuit. When not \py{False}, the effect provided by \py{uniform_spacing} is ignored.
\item \py{style}: A string specifying the style of the characters to be used to construct the diagram. Can be any of \py{"unicode"}, \py{"unicode_alt"}, or \py{"ascii"}.
\end{itemize}
Note that additional arguments are available to the \py{diagram()} method. To showcase some of the capabilities of Qhronology's visualization engine, we will use the quantum teleportation protocol example as defined in {\cdb} \ref{code:examples_teleportation}. After a small modification (in which the \py{Measurement} gates are combined using \py{GateInterleave}), the resulting \py{QuantumCircuit} instance (renamed here to ``\py{circuit}'') can be visualized with a few different configurations of values passed to the \py{diagram()} method:
\begin{code}
\begin{minted}{python}
>>> circuit.diagram(pad=(0, 0), sep=(1, 1), style="unicode")
$\includegraphics[scale=1.25, trim=-0.02cm 0 0 -0.15cm]{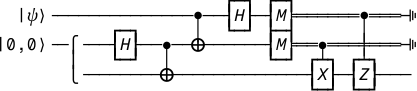}$
\end{minted}
\end{code}
\begin{code}
\begin{minted}{python}
>>> circuit.diagram(pad=(0, 0), sep=(1, 1), style="unicode_alt")
$\includegraphics[scale=1.25, trim=-0.02cm 0 0 -0.15cm]{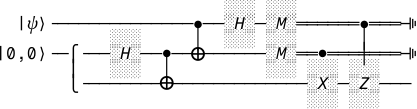}$
\end{minted}
\end{code}
\begin{code}
\begin{minted}{python}
>>> circuit.diagram(pad=(0, 0), sep=(1, 1), style="ascii")
$\includegraphics[scale=1.25, trim=-0.02cm 0 0 -0.15cm]{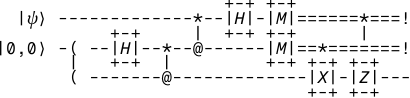}$
\end{minted}
\end{code}
\begin{code}
\begin{minted}{python}
>>> circuit.diagram(pad=(0, 0), sep=(1, 1), force_separation=True)
$\includegraphics[scale=1.25, trim=-0.02cm 0 0 -0.15cm]{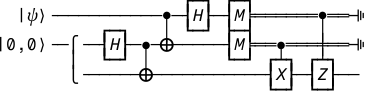}$
\end{minted}
\end{code}
\begin{code}
\begin{minted}{python}
>>> circuit.diagram(pad=(0, 0), sep=(1, 1), uniform_spacing=True)
$\includegraphics[scale=1.25, trim=-0.02cm 0 0 -0.15cm]{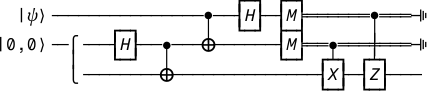}$
\end{minted}
\end{code}
\begin{code}
\begin{minted}{python}
>>> circuit.diagram(pad=(0, 0), sep=(2, 1), force_separation=True)
$\includegraphics[scale=1.25, trim=-0.02cm 0 0 -0.15cm]{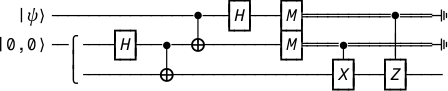}$
\end{minted}
\end{code}\enlargethispage{2\baselineskip}
\begin{code}
\begin{minted}{python}
>>> circuit.diagram(pad=(0, 0), sep=(0, 1), force_separation=True)
$\includegraphics[scale=1.25, trim=-0.02cm 0 0 -0.15cm]{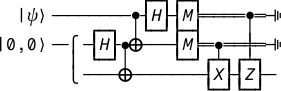}$
\end{minted}
\end{code}
\begin{code}
\begin{minted}{python}
>>> circuit.diagram(pad=(0, 0), sep=(1, 2), force_separation=True)
$\includegraphics[scale=1.25, trim=-0.02cm 0 0 -0.15cm]{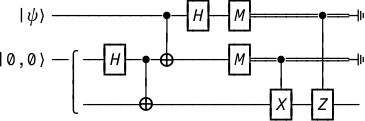}$
\end{minted}
\end{code}
\begin{code}
\begin{minted}{python}
>>> circuit.diagram(pad=(1, 0), sep=(1, 1), force_separation=True)
$\includegraphics[scale=1.25, trim=-0.02cm 0 0 -0.15cm]{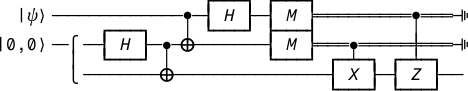}$
\end{minted}
\end{code}
\begin{code}
\begin{minted}{python}
>>> circuit.diagram(pad=(0, 1), sep=(1, 1), force_separation=True)
$\includegraphics[scale=1.25, trim=-0.02cm 0 0 -0.15cm]{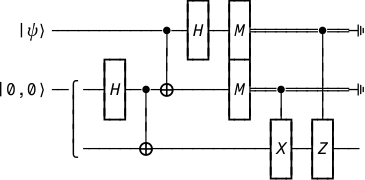}$
\end{minted}
\end{code}
Note that instead of printing the diagram, its multiline string can be returned by passing a value of \py{True} to the \py{diagram()} method's \py{return_string} argument.

\subsubsection{Examples}\label{sec:circuits_examples}

We can begin with a simple example: the quantum bit-flip. This circuit consists of a vector qubit transformed by a single NOT gate:\enlargethispage{-2\baselineskip}
\begin{codetitled}{Quantum bit-flip}{code:circuit_bitflip}
\begin{minted}{python}
from qhronology.quantum.states import VectorState
from qhronology.quantum.gates import Not
from qhronology.quantum.circuits import QuantumCircuit

# Input
input_state = VectorState(spec=[("a", [0]), ("b", [1])], label="ψ")

# Gate
NOT = Not()

# Circuit
bitflip = QuantumCircuit(inputs=[input_state], gates=[NOT])
bitflip.diagram()

# Output
output_state = bitflip.state(label="ψ′")

# Results
input_state.print()
output_state.print()
\end{minted}
\tcblowerspaced
\begin{minted}{python}
>>> bitflip.diagram()
$\includegraphics[scale=1.25, trim=-0.02cm 0 0 -0.15cm]{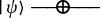}$
>>> input_state.print()
|ψ⟩ = a|0⟩ + b|1⟩
>>> output_state.print()
|ψ′⟩ = b|0⟩ + a|1⟩
\end{minted}
\end{codetitled}

For a more interesting example, we can apply a rotation gate and a diagonal gate to a $\ket{0}$ state, yielding an arbitrary qubit state parametrized by the angle $\theta$ and the phase $\phi$:\enlargethispage{\baselineskip}
\begin{codetitled}{Arbitrary state generation}{code:circuit_arbitrary}
\begin{minted}{python}
from qhronology.quantum.states import VectorState
from qhronology.quantum.gates import Rotation, Diagonal
from qhronology.quantum.circuits import QuantumCircuit

# Input
zero_state = VectorState(spec=[(1, [0])], label="0")

# Gates
R = Rotation(
    axis=1,
    angle="2*θ",
    symbols={"θ": {"real": True}},
    label="R(2θ)",
)
P = Diagonal(
    entries={1: "φ + pi/2"},
    exponentiation=True,
    symbols={"φ": {"real": True}},
    label="P(φ + π/2)",
)

# Circuit
generator = QuantumCircuit(inputs=[zero_state], gates=[R, P])
generator.diagram(force_separation=True)

# Output
arbitrary_state = generator.state(label="ψ")
arbitrary_state.simplify()

# Results
arbitrary_state.print()
\end{minted}
\tcblowerspaced
\begin{minted}{python}
>>> generator.diagram(force_separation=True)
$\includegraphics[scale=1.25, trim=-0.02cm 0 0 -0.15cm]{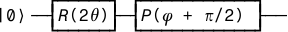}$
>>> arbitrary_state.print()
|ψ⟩ = cos(θ)|0⟩ + exp(I*φ)*sin(θ)|1⟩
\end{minted}
\end{codetitled}

For a general symbolic gate, unitarity can be imposed simply by passing the necessary relations to the gate's \py{substitutions} argument (along with the appropriate predicates in \py{symbols}). The example below demonstrates how this can be accomplished, with the unitarity of the resulting gate ($\op{U}$) verified ({\ie}, confirming that $\op{U}^\dagger \op{U} = \Identity$ holds true) by using the \py{QuantumCircuit} class to compute the (matrix) product of the unitary gate with its Hermitian conjugate:
\begin{codetitled}{Unitarity of a general symbolic gate}{code:circuit_unitarity}
\begin{minted}{python}
from qhronology.quantum.gates import QuantumGate
from qhronology.quantum.circuits import QuantumCircuit
from qhronology.mechanics.operations import dagger

import sympy as sp

# Construct unitary matrix, along with its substitutions and symbols
unitary = sp.MatrixSymbol("U", 2, 2).as_mutable()
substitutions = [
    ((Dagger(unitary) * unitary)[i, j], sp.eye(2)[i, j])
    for i in range(0, 2)
    for j in range(0, 2)
]
symbols = {
    unitary[i, j]: {"complex": True}
    for i in range(0, 2)
    for j in range(0, 2)
}

# Gates
U = QuantumGate(
    spec=unitary,
    symbols=symbols,
    substitutions=substitutions,
    label="U",
)
Ud = QuantumGate(
    spec=unitary,
    symbols=symbols,
    substitutions=substitutions,
    label="U^†",
    conjugate=True,
)

# Circuit
unitarity = QuantumCircuit(gates=[U, Ud])
unitarity.diagram(visible={"gates"})

# Output
print(repr(U))
print(repr(Ud))
print(repr(unitarity.gate(simplify=True)))
\end{minted}
\tcblowerspaced
\begin{minted}{python}
>>> unitarity.diagram(visible={"gates"})
$\includegraphics[scale=1.25, trim=-0.02cm 0 0 -0.15cm]{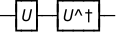}$
>>> U
Matrix([
[U[0, 0], U[0, 1]],
[U[1, 0], U[1, 1]]])
>>> Ud
Matrix([
[conjugate(U[0, 0]), conjugate(U[1, 0])],
[conjugate(U[0, 1]), conjugate(U[1, 1])]])
>>> unitarity.gate(simplify=True)
Matrix([
[1, 0],
[0, 1]])
\end{minted}
\end{codetitled}

With the help of Python's versatile data-manipulation abilities, the decomposition of the multipartite Fourier gate into Hadamard and (controlled) phase gates can easily be assembled:\enlargethispage{2\baselineskip}
\begin{codetitled}{Quantum Fourier transform decomposition}{code:fourier_transform}
\begin{minted}{python}
from qhronology.quantum.gates import Hadamard, Phase
from qhronology.quantum.circuits import QuantumCircuit

import sympy as sp

size = 4  # The number of qudits
dim = 2  # The dimensionality of the Fourier transform

# Gates
QFT = []
for i in range(0, size):
    count = size - i
    for j in range(0, count):
        if j == 0:
            QFT.append(
                Hadamard(
                    targets=[i],
                    num_systems=size,
                    dim=dim,
                )
            )
        else:
            QFT.append(
                Phase(
                    targets=[i],
                    controls=[i + j],
                    exponent=sp.Rational(1, dim**j),
                    num_systems=size,
                    dim=dim,
                    label=f"{dim**j}",
                    family="GATE",
                )
            )

# Circuit
fourier = QuantumCircuit(gates=QFT)
fourier.diagram(sep=(0, 1), visible={"gates"})

# Results
print(repr(fourier.gate()))
\end{minted}
\tcblowerspaced
\begin{minted}{python}
>>> fourier.diagram(sep=(0, 1), visible={"gates"})
$\includegraphics[scale=1.25, trim=-0.02cm 0 0 -0.15cm]{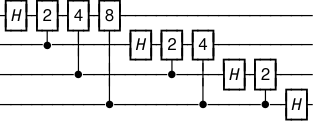}$
>>> fourier.gate()
# Too large to display here.
\end{minted}
\end{codetitled}

\subsection{Prescriptions}\label{sec:prescriptions}

Qhronology's novel functionality{\textemdash}the programmatic implementation of quantum models of CTCs{\textemdash}is facilitated chiefly by the \py{QuantumCTC} class:
\begin{code}
\begin{minted}{python}
>>> from qhronology.quantum.prescriptions import QuantumCTC
\end{minted}
\end{code}
This maintains an almost identical range of capabilities as the \py{QuantumCircuit} class (from which it is derived), differing only in the addition of the \py{systems_respecting} and \py{systems_violating} properties, which contain the CR and CV system indices, respectively. It is then the instances of the class which provide descriptions of specific interactions between the CR and CV systems.

By itself however, the \py{QuantumCTC} class does not possess the ability to compute CR and CV states{\textemdash}it merely serves as a base, upon which explicit implementations of prescriptions can be built (as subclasses). Qhronology provides two such subclasses, one for each of Deutsch's prescription (D-CTCs) and postselected teleportation (P-CTCs):
\begin{code}
\begin{minted}{python}
>>> from qhronology.quantum.prescriptions import DCTC, PCTC
\end{minted}
\end{code}
Closely related to the \py{systems_respecting} and \py{systems_violating} properties of the \py{QuantumCTC} base class are its subclasses' \py{state_respecting()} and \py{state_violating()} methods. Both of these operate identically to their \py{state()} peer, and simply return the respective CR and CV states (as \py{QuantumState} instances) according to the particular prescription to which the subclass corresponds.

Due to their close relationship, instances of the \py{QuantumCTC} class can be created directly from instances of \py{QuantumCircuit}. This is achieved by way of the \py{circuit} argument in the former's constructor, whereby all attributes of the given \py{QuantumCircuit} instance are copied (deeply) to the \py{QuantumCTC} instance during initialization. Importantly, this enables the various subclasses, such as \py{DCTC} and \py{PCTC}, to be instantiated using just a single \py{QuantumCTC} instance, thereby reducing duplication (of arguments) in cases where one wishes to study multiple prescriptions of the same CTC circuit. Various demonstrations of this functionality are included hereafter.

It is also important to be aware of a significant difference between the prescription subclasses, particularly concerning their CV states. In the absence of supplementary conditions, Deutsch's prescription naturally provides a multiplicity of solutions for the CV state (and therefore the CR state as well) for certain combinations of CR input states and interactions. The \py{DCTC} class therefore possesses the \py{free_symbol} property (defaulting to \py{"g"}), which holds the parameter(s) used to express a spectrum of (non-unique) solutions in cases where Qhronology's D-CTCs algorithm calculates such a plurality. When this occurs, setting the \py{maximum_entropy} property to \py{True} yields D-CTC solutions which correspond to the maximally entropic CV state. Naturally, as the CV solutions provided by the P-CTC prescription (through the implementation of the result in \cite{bishop_quantum_2025}) are always unique, the \py{PCTC} class does not have this property.

\subsubsection{Examples}\label{sec:prescriptions_examples}

Perhaps the simplest (and somewhat meaningful) interaction between the internal and external systems of a CTC would be a SWAP interaction. Using a general density matrix ($\op{\rho}$) as CR input, we find (as expected) that the CR and CV solution states for both prescriptions are identically the input:\enlargethispage{2\baselineskip}
\begin{codetitled}{SWAP interaction with a CTC}{code:prescription_swap}
\begin{minted}{python}
from qhronology.quantum.states import MixedState
from qhronology.quantum.gates import Swap, Pauli
from qhronology.quantum.prescriptions import QuantumCTC, DCTC, PCTC

import sympy as sp

# Input
rho = sp.MatrixSymbol("ρ", 2, 2).as_mutable()
input_state = MixedState(
    spec=rho,
    substitutions=[(rho[0, 0] + rho[1, 1], 1)],  # For normalization
    label="ρ",
    norm=1,
)

# Gate
S = Swap(targets=[0, 1], num_systems=2)
I = Pauli(index=0, targets=[0, 1], num_systems=2)

# CTC
SWAP_CTC = QuantumCTC(
    inputs=[input_state],
    gates=[S],
    systems_respecting=[0],
)
SWAP_CTC.diagram()

# Output
# D-CTCs
SWAP_DCTC = DCTC(circuit=SWAP_CTC)
SWAP_DCTC_CR = SWAP_DCTC.state_respecting(label="ρ_D")
SWAP_DCTC_CV = SWAP_DCTC.state_violating(label="τ_D")

# P-CTCs
SWAP_PCTC = PCTC(circuit=SWAP_CTC)
SWAP_PCTC_CR = SWAP_PCTC.state_respecting(label="ρ_P")
SWAP_PCTC_CV = SWAP_PCTC.state_violating(label="τ_P")

# Results
SWAP_DCTC_CR.print(simplify=True)
SWAP_DCTC_CV.print(simplify=True)
SWAP_PCTC_CR.print(simplify=True)
SWAP_PCTC_CV.print(simplify=True)
\end{minted}
\tcblowerspaced
\begin{minted}{python}
>>> SWAP_CTC.diagram()
$\includegraphics[scale=1.25, trim=-0.02cm 0 0 -0.15cm]{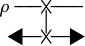}$
>>> SWAP_DCTC_CR.print(simplify=True)
ρ_D = ρ[0, 0]|0⟩⟨0| + ρ[0, 1]|0⟩⟨1| + ρ[1, 0]|1⟩⟨0| + ρ[1, 1]|1⟩⟨1|
>>> SWAP_DCTC_CV.print(simplify=True)
τ_D = ρ[0, 0]|0⟩⟨0| + ρ[0, 1]|0⟩⟨1| + ρ[1, 0]|1⟩⟨0| + ρ[1, 1]|1⟩⟨1|
>>> SWAP_PCTC_CR.print(simplify=True)
ρ_P = ρ[0, 0]|0⟩⟨0| + ρ[0, 1]|0⟩⟨1| + ρ[1, 0]|1⟩⟨0| + ρ[1, 1]|1⟩⟨1|
>>> SWAP_PCTC_CV.print(simplify=True)
τ_P = ρ[0, 0]|0⟩⟨0| + ρ[0, 1]|0⟩⟨1| + ρ[1, 0]|1⟩⟨0| + ρ[1, 1]|1⟩⟨1|
\end{minted}
\end{codetitled}
By following the flow of information in the circuit, it is easy to see how, under either prescription, the state of both systems at any given moment must always simply be the CR input state.

In the case of a CNOT interaction, where the CR system is the control, we obtain the following solutions:\enlargethispage{\baselineskip}
\begin{codetitled}{CNOT interaction with a CTC}{code:prescription_cnot}
\begin{minted}{python}
from qhronology.quantum.states import MixedState
from qhronology.quantum.gates import Not
from qhronology.quantum.circuits import QuantumCircuit
from qhronology.quantum.prescriptions import QuantumCTC, DCTC, PCTC

import sympy as sp

# Input
rho = sp.MatrixSymbol("ρ", 2, 2).as_mutable()
input_state = MixedState(
    spec=rho,
    substitutions=[(rho[0, 0] + rho[1, 1], 1)],  # For normalization
    label="ρ",
    norm=1,
)
input_state.simplify()

# Gate
CN = Not(targets=[0], controls=[1], num_systems=2)

# CTC
CNOT = QuantumCircuit(
    inputs=[input_state],
    gates=[CN],
)
CNOT_CTC = QuantumCTC(circuit=CNOT, systems_respecting=[1])
CNOT_CTC.diagram()

# Output
# D-CTCs
CNOT_DCTC = DCTC(circuit=CNOT_CTC)
CNOT_DCTC_CR = CNOT_DCTC.state_respecting(label="ρ_D")
CNOT_DCTC_CV = CNOT_DCTC.state_violating(label="τ_D")

# P-CTCs
CNOT_PCTC = PCTC(circuit=CNOT_CTC)
CNOT_PCTC_CR = CNOT_PCTC.state_respecting(norm=1, label="ρ_P")
CNOT_PCTC_CV = CNOT_PCTC.state_violating(label="τ_P")

# Results
CNOT_DCTC_CR.print(simplify=True)
CNOT_DCTC_CV.print(simplify=True)
CNOT_PCTC_CR.print(simplify=True)
CNOT_PCTC_CV.print(simplify=True)
\end{minted}
\tcblowerspaced
\begin{minted}{python}
>>> CNOT_CTC.diagram()
$\includegraphics[scale=1.25, trim=-0.02cm 0 0 -0.15cm]{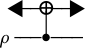}$
>>> CNOT_DCTC_CR.print(simplify=True)
ρ_D = ρ[0, 0]|0⟩⟨0| + 2*g*ρ[0, 1]|0⟩⟨1| + 2*g*ρ[1, 0]|1⟩⟨0| + ρ[1, 1]|1⟩⟨1|
>>> CNOT_DCTC_CV.print(simplify=True)
τ_D = 1/2|0⟩⟨0| + g|0⟩⟨1| + g|1⟩⟨0| + 1/2|1⟩⟨1|
>>> CNOT_PCTC_CR.print(simplify=True)
ρ_P = |0⟩⟨0|
>>> CNOT_PCTC_CV.print(simplify=True)
τ_P = 1/2|0⟩⟨0| + 1/2|1⟩⟨1|
\end{minted}
\end{codetitled}
Observe how the D-CTC CV solution describes a spectrum of states (parametrized by $g$), meaning that the CR solution is likewise non-unique. Alternatively, it is interesting to note how, regardless of the value of the input state, the P-CTC's CV solution is maximally mixed (thereby containing no information about the system), while its CR output has collapsed (and assumed the $\ket{0}$ pure state).

\newpage
\section{Examples}\label{sec:examples}

Presented here is a small collection of examples demonstrating some of the various uses of the package and much of the functionality that it offers. It is divided into two subsections: the first, \emph{Quantum algorithms, protocols, and procedures} ({\secn} \ref{sec:algorithms}), consists of standard processes and circuits that are significant to the fields of quantum computing and quantum information science, while the second, \emph{Quantum closed timelike curves and temporal paradoxes} ({\secn} \ref{sec:ctcs}), contains circuits which incorporate CTCs, particularly in the presence of physically interesting unitary interactions. It is useful to note that all of these examples (plus many more) are covered in more detail in the official documentation\footnote{\url{https://qhronology.org/text/examples/part-examples.html}} \cite{bishop_qhronology-documentation_2025}.

\subsection{Quantum algorithms, protocols, and procedures}\label{sec:algorithms}

\subsubsection{Quantum full adder}\label{sec:examples_adder_full}

A quantum full adder\footnote{\url{https://qhronology.org/text/examples/algorithms/adder_full.html}} producing the summation output state $\op{s}$ from augend $\ket{x}$, addend $\ket{y}$, and carry $\ket{c}$ input qubits:\enlargethispage{\baselineskip}
\begin{codetitled}{Quantum full adder}{code:examples_adder_full}
\begin{minted}{python}
from qhronology.quantum.states import VectorState
from qhronology.quantum.gates import Not
from qhronology.quantum.circuits import QuantumCircuit

import sympy as sp

# Input
augend_state = VectorState(
    spec=[("a", [0]), ("b", [1])],
    substitutions=[("a*conjugate(a) + b*conjugate(b)", 1)],
    label="x",
)
addend_state = VectorState(
    spec=[("u", [0]), ("v", [1])],
    substitutions=[("u*conjugate(u) + v*conjugate(v)", 1)],
    label="y",
)
carry_input_state = VectorState(spec=[(1, [1])], label="c")
zero_state = VectorState(spec=[(1, [0])], label="0")

# Gates
CCIN = Not(targets=[3], controls=[0, 1], num_systems=4)
CNII = Not(targets=[1], controls=[0], num_systems=4)
ICCN = Not(targets=[3], controls=[1, 2], num_systems=4)
ICNI = Not(targets=[2], controls=[1], num_systems=4)

# Circuit
adder = QuantumCircuit(
    inputs=[augend_state, addend_state, carry_input_state, zero_state],
    gates=[CCIN, CNII, ICCN, ICNI, CNII],
)
adder.diagram()

# Output
sum_state = adder.state(label="s", traces=[0, 1, 3])
carry_output_state = adder.state(label="c′", traces=[0, 1, 2])
carry_output_state.apply(sp.collect, arguments={"syms": ["b*conjugate(b)"]})

# Results
augend_state.print()
addend_state.print()
carry_input_state.print()
sum_state.print()
carry_output_state.print(simplify=True)
\end{minted}
\tcblowerspaced
\begin{minted}{python}
>>> adder.diagram()
$\includegraphics[scale=1.25, trim=-0.02cm 0 0 -0.15cm]{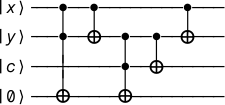}$
>>> augend_state.print()
|x⟩ = a|0⟩ + b|1⟩
>>> addend_state.print()
|y⟩ = u|0⟩ + v|1⟩
>>> carry_input_state.print()
|c⟩ = |1⟩
>>> sum_state.print()
s = (a*v*conjugate(a)*conjugate(v) + b*u*conjugate(b)*conjugate(u))|0⟩⟨0| + (a*u*conjugate(a)*conjugate(u) + b*v*conjugate(b)*conjugate(v))|1⟩⟨1|
>>> carry_output_state.print(simplify=True)
c′ = a*u*conjugate(a)*conjugate(u)|0⟩⟨0| + (a*v*conjugate(a)*conjugate(v) + b*conjugate(b))|1⟩⟨1|
\end{minted}
\end{codetitled}
Note that the official documentation contains numerous other examples of quantum arithmetic, including a ripple-carry adder\footnote{\url{https://qhronology.org/text/examples/algorithms/adder_ripple.html}}, carry-lookahead adder\footnote{\url{https://qhronology.org/text/examples/algorithms/adder_lookahead.html}}, and Fourier transform adder\footnote{\url{https://qhronology.org/text/examples/algorithms/adder_fourier.html}}.

\subsubsection{Quantum teleportation}\label{sec:examples_teleportation}

Quantum teleportation\footnote{\url{https://qhronology.org/text/examples/algorithms/teleportation.html}} of an arbitrary qubit $\ket{\psi} = a\ket{0} + b\ket{1}$:\enlargethispage{\baselineskip}
\begin{codetitled}{Quantum teleportation}{code:examples_teleportation}
\begin{minted}{python}
from qhronology.quantum.states import VectorState
from qhronology.quantum.gates import Hadamard, Not, Measurement, Pauli
from qhronology.quantum.circuits import QuantumCircuit
from qhronology.mechanics.matrices import ket

# Input
teleporting_state = VectorState(
    spec=[["a", "b"]],
    symbols={"a": {"complex": True}, "b": {"complex": True}},
    substitutions=[("a*conjugate(a) + b*conjugate(b)", 1)],
    label="ψ",
)
zero_state = VectorState(spec=[(1, [0, 0])], label="0,0")

# Gates
IHI = Hadamard(targets=[1], num_systems=3)
ICN = Not(targets=[2], controls=[1], num_systems=3)
CNI = Not(targets=[1], controls=[0], num_systems=3)
HII = Hadamard(targets=[0], num_systems=3)
IMI = Measurement(
    operators=[ket(0), ket(1)],
    targets=[1],
    num_systems=3,
)
MII = Measurement(
    operators=[ket(0), ket(1)],
    targets=[0],
    num_systems=3,
)
ICX = Pauli(index=1, targets=[2], controls=[1], num_systems=3)
CIZ = Pauli(index=3, targets=[2], controls=[0], num_systems=3)

# Circuit
teleporter = QuantumCircuit(
    inputs=[teleporting_state, zero_state],
    gates=[IHI, ICN, CNI, HII, IMI, MII, ICX, CIZ],
    traces=[0, 1],
)
teleporter.diagram(force_separation=True)

# Output
teleported_state = teleporter.state(label="ρ")

# Results
teleporting_state.print()
teleported_state.print()
print(teleporting_state.distance(teleported_state))
print(teleporting_state.fidelity(teleported_state))
\end{minted}
\tcblowerspaced
\begin{minted}{python}
>>> teleporter.diagram(force_separation=True)
$\includegraphics[scale=1.25, trim=-0.02cm 0 0 -0.15cm]{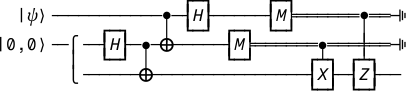}$
>>> teleporting_state.print()
|ψ⟩ = a|0⟩ + b|1⟩
>>> teleported_state.print()
ρ = a*conjugate(a)|0⟩⟨0| + a*conjugate(b)|0⟩⟨1| + b*conjugate(a)|1⟩⟨0| + b*conjugate(b)|1⟩⟨1|
>>> teleporting_state.distance(teleported_state)
0
>>> teleporting_state.fidelity(teleported_state)
1
\end{minted}
\end{codetitled}

\subsubsection{Deutsch-Jozsa algorithm}\label{sec:examples_deutsch_jozsa}\enlargethispage{2.5\baselineskip}

An implementation of the Deutsch-Jozsa algorithm\footnote{\vspace*{\baselineskip}\url{https://qhronology.org/text/examples/algorithms/deutsch_jozsa.html}}, in which a function $f$ (characterized by an oracle gate) is determined to be either constant or balanced on its input domain (\py{n = 4} qubits):
\begin{codetitled}{Deutsch-Jozsa algorithm}{code:examples_deutsch_jozsa}
\begin{minted}{python}
from qhronology.quantum.gates import Pauli, GateStack, Hadamard, QuantumGate
from qhronology.quantum.circuits import QuantumCircuit
from qhronology.mechanics.matrices import ket

import random
import numpy as np
from scipy.linalg import block_diag

n = 4  # The number of qubits in the address register

constant = None  # Whether the function is constant (True) or balanced (False)
constant = random.choice([True, False]) if constant is None else constant

class Oracle:
    def __init__(self, constant: bool, num_address: int):
        self.constant = constant
        self.num_address = num_address
        if self.constant is True:
            image = [random.choice([0, 1])] * 2**num_address
        else:
            image = [0] * (2**num_address // 2) + [1] * (2**num_address // 2)
        random.shuffle(image)
        self.image = image

    def f(self, x: int) -> int:
        return self.image[x]

    def operator(self) -> np.ndarray:
        blocks = [
            (1 - self.f(x)) * np.eye(2) + self.f(x) * np.eye(2)[::-1]
            for x in range(0, len(self.image))
        ]
        return np.array(block_diag(*blocks))

# Gates
IX = Pauli(index=1, targets=[n], num_systems=n + 1)
HH = GateStack(*[Hadamard()] * (n + 1))
HI = GateStack(*[Hadamard()] * n)

oracle = Oracle(constant=constant, num_address=n)
O = QuantumGate(
    spec=oracle.operator(),
    targets=list(range(0, n + 1)),
    num_systems=n + 1,
    label=" O ",
)

# Circuit
deutsch_jozsa = QuantumCircuit(
    gates=[IX, HH, O, HI],
    numerical=True,
    array=True,
)
deutsch_jozsa.diagram(pad=(1, 0), sep=(1, 2), force_separation=True)

# Measurement
probabilities = deutsch_jozsa.measure(
    operators=[ket([0] * n)],
    targets=list(range(0, n)),
    observable=False,
    statistics=True,
)
probability_zeroes = np.round(np.real(probabilities[0])).astype(int)

result_constant = True if probability_zeroes == 1 else False
result_function = "constant" if oracle.constant is True else "balanced"
result_algorithm = "constant" if result_constant is True else "balanced"

# Results
print(f"The function: {result_function}")
print(f"The Deutsch-Jozsa result: {result_algorithm}")
\end{minted}
\tcblowerspaced
\begin{minted}{python}
>>> deutsch_jozsa.diagram(pad=(1, 0), sep=(1, 2), force_separation=True)
$\includegraphics[scale=1.25, trim=-0.02cm 0 0 -0.15cm]{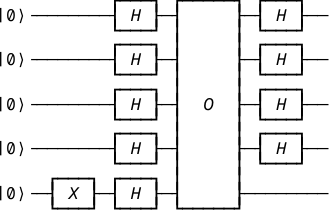}$
>>> print(f"The function: {result_function}")
The function: balanced
>>> print(f"The Deutsch-Jozsa result: {result_algorithm}")
The Deutsch-Jozsa result: balanced
\end{minted}
\end{codetitled}
Note that the example output here corresponds to the case where \py{constant} is set to \py{False}.

\subsubsection{Quantum phase estimation}\label{sec:examples_phase_estimation}

An implementation of the canonical quantum phase estimation algorithm\footnote{\url{https://qhronology.org/text/examples/algorithms/phase_estimation.html}}, in which the value of the phase (\py{phase = 0.64}) is estimated with a specific degree of precision (\py{precision = 4} bits):\enlargethispage{\baselineskip}
\begin{codetitled}{Quantum phase estimation}{code:examples_phase_estimation}
\begin{minted}{python}
from qhronology.quantum.gates import GateStack, Hadamard, Pauli, Phase, Fourier
from qhronology.quantum.circuits import QuantumCircuit
from qhronology.mechanics.matrices import encode, decode

import sympy as sp
import numpy as np

phase = 0.64  # The phase to be estimated
precision = 4  # The number of qubits used by the estimation

num_controls = precision
num_targets = 1
num_total = num_targets + num_controls

systems_controls = list(range(0, num_controls))
systems_targets = [m + num_controls for m in range(0, num_targets)]

# Gates
HX = GateStack(
    *[Hadamard()] * num_controls,
    Pauli(index=1),
)

unitaries = [
    Phase(
        phase=sp.exp(2 * sp.I * sp.pi * phase * 2**n),
        targets=systems_targets,
        controls=[systems_controls[n]],
        num_systems=num_total,
        label=f"U^{2**n}",
    )
    for n in range(0, num_controls)
]

IQFT = Fourier(
    targets=systems_controls,
    num_systems=num_total,
    composite=True,
    conjugate=True,
    label="QFT^†",
)

# Circuit
phase_estimator = QuantumCircuit(
    gates=[HX] + unitaries + [IQFT],
    numerical=True,
    array=True,
)

phase_estimator.diagram(pad=(1, 0), force_separation=True)

# Measurement
basis = [encode(k, num_controls) for k in range(0, 2**num_controls)]
probabilities = phase_estimator.measure(
    operators=basis,
    targets=systems_controls,
    observable=False,
    statistics=True,
)
probabilities = [np.real(probability) for probability in probabilities]

# Results
print(f"Input phase: {phase}")
expectation = 0
threshold = 0.001
for k, probability in enumerate(probabilities):
    value = decode(basis[k]) / 2**num_controls
    value = sp.N(value).round(precision)
    bitstring = encode(decode(basis[k]), num_controls, return_type=str)
    expectation += probability * value

    suffix = ""
    if probability == max(probabilities):
        suffix = " (most probable)"

    if probability >= threshold or probability == max(probabilities):
        probability = sp.N(probability).round(3)
        print(
            f"Bitstring={bitstring}, "
            + f"Probability={probability}, "
            + f"Phase={value}"
            + f"{suffix}"
        )

expectation = sp.N(expectation).round(precision)
print(f"Expectation (weighted average): {expectation}")
\end{minted}
\tcblowerspaced
\begin{minted}{python}
>>> phase_estimator.diagram(pad=(1, 0), force_separation=True)
$\includegraphics[scale=1.25, trim=-0.02cm 0 0 -0.15cm]{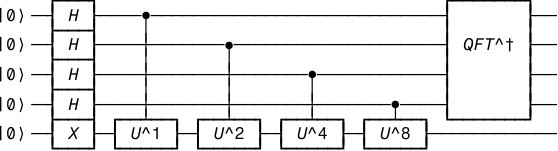}$
>>> print(f"Input phase: {phase}")
Input phase: 0.64
>>> # Cumulative output from the loop's print statement:
Bitstring=0000, Probability=0.002, Phase=0
Bitstring=0001, Probability=0.002, Phase=0.0625
Bitstring=0010, Probability=0.002, Phase=0.1250
Bitstring=0011, Probability=0.002, Phase=0.1875
Bitstring=0100, Probability=0.002, Phase=0.2500
Bitstring=0101, Probability=0.002, Phase=0.3125
Bitstring=0110, Probability=0.003, Phase=0.3750
Bitstring=0111, Probability=0.005, Phase=0.4375
Bitstring=1000, Probability=0.010, Phase=0.5000
Bitstring=1001, Probability=0.031, Phase=0.5625
Bitstring=1010, Probability=0.825, Phase=0.6250 (most probable)
Bitstring=1011, Probability=0.083, Phase=0.6875
Bitstring=1100, Probability=0.016, Phase=0.7500
Bitstring=1101, Probability=0.007, Phase=0.8125
Bitstring=1110, Probability=0.004, Phase=0.8750
Bitstring=1111, Probability=0.003, Phase=0.9375
>>> print(f"Expectation (weighted average): {expectation}")
Expectation (weighted average): 0.6245
\end{minted}
\end{codetitled}
See the official documentation for an implementation of Shor's factorization algorithm\footnote{\url{https://qhronology.org/text/examples/algorithms/shor.html}}, which leverages quantum phase estimation in order to factorize any given composite integer.

\subsubsection{Grover's algorithm}\label{sec:examples_grover}

An implementation of Grover's algorithm\footnote{\url{https://qhronology.org/text/examples/algorithms/grover.html}}, in which the marked value (\py{marked = 4}) is found in a list of $n = \lceil \log_{2}(N) \rceil$ (\py{N = 10}) items:
\begin{codetitled}{Grover's algorithm}{code:examples_grover}
\begin{minted}{python}
from qhronology.quantum.gates import GateStack, Hadamard, QuantumGate
from qhronology.quantum.circuits import QuantumCircuit
from qhronology.mechanics.matrices import ket, bra, encode, decode
from qhronology.mechanics.operations import densify

import math
import sympy as sp
import numpy as np

N = 10  # The size of the input space (domain)
marked = 4  # The value to find (should be smaller than N)

n = int(math.ceil(math.log(N, 2)))  # Encoding depth
iterations = int(math.pi * math.sqrt(2**n) / 4)

# Gates
H = GateStack(*[Hadamard()] * n)

# Construct the oracle from its definition
oracle = QuantumGate(
    spec=sp.eye(2**n) - 2 * densify(encode(marked, num_systems=n)),
    targets=list(range(0, n)),
    num_systems=n,
    label="O",
)

# Construct the Grover diffusion gate from its definition
diffusion = QuantumCircuit(
    gates=[H]
    + [
        QuantumGate(
            spec=2 * ket([0] * n) * bra([0] * n) - sp.eye(2**n),
            targets=list(range(0, n)),
        )
    ]
    + [H]
).gate(label="D")

# Construct the gate sequence of the Grover iterations
grover_iterations = [oracle, diffusion] * iterations

# Circuit
grover = QuantumCircuit(
    gates=[H] + grover_iterations,
    numerical=True,
    array=True,
)
grover.diagram(pad=(1, 0), sep=(1, 2), force_separation=True)

# Measurement
basis = [encode(k, n) for k in range(0, 2**n)]
probabilities = grover.measure(
    operators=basis,
    observable=False,
    statistics=True,
)
probabilities = [np.real(probability) for probability in probabilities]


# Results
print(f"Input size: {N}")
print(f"Marked value: {marked}")
expectation = 0
threshold = 0.001
for k, probability in enumerate(probabilities):
    value = decode(basis[k])
    bitstring = encode(value, n, return_type=str)
    expectation += probability * value

    suffix = ""
    if probability == max(probabilities):
        suffix = " (most probable)"

    if probability >= threshold or probability == max(probabilities):
        probability = sp.N(probability).round(3)
        print(
            f"Bitstring={bitstring}, "
            + f"Probability={probability}, "
            + f"Value={value}"
            + f"{suffix}"
        )

expectation = sp.N(expectation).round(3)
print(f"Expectation (weighted average): {expectation}")
\end{minted}
\tcblowerspaced
\begin{minted}{python}
>>> grover.diagram(pad=(1, 0), sep=(1, 2), force_separation=True)
$\includegraphics[scale=1.25, trim=-0.02cm 0 0 -0.15cm]{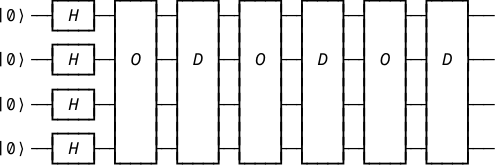}$
>>> print(f"Input size: {N}")
Input size: 10
>>> print(f"Marked value: {marked}")
Marked value: 4
>>> # Cumulative output from the loop's print statement:
Bitstring=0000, Probability=0.003, Value=0
Bitstring=0001, Probability=0.003, Value=1
Bitstring=0010, Probability=0.003, Value=2
Bitstring=0011, Probability=0.003, Value=3
Bitstring=0100, Probability=0.961, Value=4 (most probable)
Bitstring=0101, Probability=0.003, Value=5
Bitstring=0110, Probability=0.003, Value=6
Bitstring=0111, Probability=0.003, Value=7
Bitstring=1000, Probability=0.003, Value=8
Bitstring=1001, Probability=0.003, Value=9
Bitstring=1010, Probability=0.003, Value=10
Bitstring=1011, Probability=0.003, Value=11
Bitstring=1100, Probability=0.003, Value=12
Bitstring=1101, Probability=0.003, Value=13
Bitstring=1110, Probability=0.003, Value=14
Bitstring=1111, Probability=0.003, Value=15
>>> print(f"Expectation (weighted average): {expectation}")
Expectation (weighted average): 4.144
\end{minted}
\end{codetitled}

\subsection{Quantum closed timelike curves and temporal paradoxes}\label{sec:ctcs}

\subsubsection{Grandfather paradox}\label{sec:examples_grandfather}

A quantum version of the \emph{grandfather paradox}\footnote{\vspace*{\baselineskip}\url{https://qhronology.org/text/examples/ctcs/grandfather.html}} (based on \cite{lloyd_closed_2011}):\enlargethispage{2.5\baselineskip}
\begin{codetitled}{Quantum grandfather paradox}{code:examples_grandfather}
\begin{minted}{python}
from qhronology.quantum.states import MixedState
from qhronology.quantum.gates import Not, Swap
from qhronology.quantum.prescriptions import QuantumCTC, DCTC, PCTC

import sympy as sp

# Input
rho = sp.MatrixSymbol("ρ", 2, 2).as_mutable()
input_state = MixedState(
    spec=rho,
    substitutions=[(rho[0, 0], 1 - rho[1, 1])],
    label="ρ",
)

# Gates
NC = Not(targets=[0], controls=[1], num_systems=2)
S = Swap(targets=[0, 1], num_systems=2)

# CTC
grandfather = QuantumCTC(
    inputs=[input_state],
    gates=[NC, S],
    systems_respecting=[0],
)
grandfather.diagram()

# Output
# D-CTCs
grandfather_DCTC = DCTC(circuit=grandfather)
grandfather_DCTC_CR = grandfather_DCTC.state_respecting(norm=1, simplify=True, label="ρ_D")
grandfather_DCTC_CV = grandfather_DCTC.state_violating(norm=1, simplify=True, label="τ_D")

# P-CTCs
grandfather_PCTC = PCTC(circuit=grandfather)
grandfather_PCTC_CR = grandfather_PCTC.state_respecting(norm=1, simplify=True, label="ρ_P")
grandfather_PCTC_CV = grandfather_PCTC.state_violating(norm=1, simplify=True, label="τ_P")

# Results
grandfather_DCTC_CR.print()
grandfather_DCTC_CV.print()
grandfather_PCTC_CR.print()
grandfather_PCTC_CV.print()
\end{minted}
\tcblowerspaced
\begin{minted}{python}
>>> grandfather.diagram()
$\includegraphics[scale=1.25, trim=-0.02cm 0 0 -0.15cm]{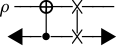}$
>>> grandfather_DCTC_CR.print()
ρ_D = 1/2|0⟩⟨0| + (ρ[0, 1] + ρ[1, 0])**2/2|0⟩⟨1| + (ρ[0, 1] + ρ[1, 0])**2/2|1⟩⟨0| + 1/2|1⟩⟨1|
>>> grandfather_DCTC_CV.print()
τ_D = 1/2|0⟩⟨0| + (ρ[0, 1]/2 + ρ[1, 0]/2)|0⟩⟨1| + (ρ[0, 1]/2 + ρ[1, 0]/2)|1⟩⟨0| + 1/2|1⟩⟨1|
>>> grandfather_PCTC_CR.print()
ρ_P = 1/2|0⟩⟨0| + 1/2|0⟩⟨1| + 1/2|1⟩⟨0| + 1/2|1⟩⟨1|
>>> grandfather_PCTC_CV.print()
τ_P = 1/2|0⟩⟨0| + (ρ[0, 1]/2 + ρ[1, 0]/2)|0⟩⟨1| + (ρ[0, 1]/2 + ρ[1, 0]/2)|1⟩⟨0| + 1/2|1⟩⟨1|
\end{minted}
\end{codetitled}

\subsubsection{Unproven-theorem paradox}\label{sec:examples_unproven}

A quantum version of the \emph{unproven-theorem paradox}\footnote{\url{https://qhronology.org/text/examples/ctcs/unproven.html}} \cite{deutsch_quantum_1991,lloyd_closed_2011,allen_treating_2014}:\enlargethispage{\baselineskip}
\begin{codetitled}{Quantum unproven-theorem paradox}{code:examples_unproven}
\begin{minted}{python}
from qhronology.quantum.states import VectorState
from qhronology.quantum.gates import Not, Swap
from qhronology.quantum.prescriptions import QuantumCTC, DCTC, PCTC

# Input
mathematician_state = VectorState(spec=[(1, [0])], label="0")
book_state = VectorState(spec=[(1, [0])], label="0")

# Gates
NIC = Not(targets=[0], controls=[2], num_systems=3)
CNI = Not(targets=[1], controls=[0], num_systems=3)
IS = Swap(targets=[1, 2], num_systems=3)

# CTC
unproven = QuantumCTC(
    inputs=[mathematician_state, book_state],
    gates=[NIC, CNI, IS],
    systems_respecting=[0, 1],
)
unproven.diagram()

# Output
# D-CTCs
unproven_DCTC = DCTC(circuit=unproven)
unproven_DCTC_CR = unproven_DCTC.state_respecting(label="ρ_D")
unproven_DCTC_CV = unproven_DCTC.state_violating(label="τ_D")

# P-CTCs
unproven_PCTC = PCTC(circuit=unproven)
unproven_PCTC_CR = unproven_PCTC.state_respecting(label="ψ_P")
unproven_PCTC_CV = unproven_PCTC.state_violating(label="τ_P")
unproven_PCTC_CR.normalize()

# Results
unproven_DCTC_CR.print()
unproven_DCTC_CV.print()
unproven_PCTC_CR.print()
unproven_PCTC_CV.print()
\end{minted}
\tcblowerspaced
\begin{minted}{python}
>>> unproven.diagram()
$\includegraphics[scale=1.25, trim=-0.02cm 0 0 -0.15cm]{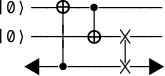}$
>>> unproven_DCTC_CR.print()
ρ_D = g|0,0⟩⟨0,0| + (1 - g)|1,1⟩⟨1,1|
>>> unproven_DCTC_CV.print()
τ_D = g|0⟩⟨0| + (1 - g)|1⟩⟨1|
>>> unproven_PCTC_CR.print()
|ψ_P⟩ = sqrt(2)/2|0,0⟩ + sqrt(2)/2|1,1⟩
>>> unproven_PCTC_CV.print()
τ_P = 1/2|0⟩⟨0| + 1/2|1⟩⟨1|
\end{minted}
\end{codetitled}

\newpage
\subsubsection{Billiard-ball paradox}\label{sec:examples_billiards}

A quantum version of the (simplified) \emph{billiard-ball paradox}\footnote{\vspace*{\baselineskip}\url{https://qhronology.org/text/examples/ctcs/billiards.html}} \cite{bishop_time-traveling_2021}:\enlargethispage{2.5\baselineskip}
\begin{codetitled}{Quantum billiard-ball paradox}{code:examples_billiards}
\begin{minted}{python}
from qhronology.quantum.states import VectorState
from qhronology.quantum.gates import QuantumGate, Swap, Diagonal
from qhronology.quantum.prescriptions import QuantumCTC, DCTC, PCTC
from qhronology.utilities.helpers import tensor_product

import sympy as sp

# Input
clock_state_unevolved = VectorState(
    spec=[[0, 1, 1]],
    dim=3,
    norm=1,
    label="ψ",
)
clock_state_evolved = VectorState(
    spec=[[0, 1, -1]],
    dim=3,
    norm=1,
    label="ψ",
)

# Gates
I = QuantumGate(
    spec=[[1, 0, 0], [0, 1, 0], [0, 0, 1]],
    targets=[0],
    num_systems=1,
    dim=3,
)
zero = QuantumGate(
    spec=[[1, 0, 0], [0, 0, 0], [0, 0, 0]],
    targets=[0],
    num_systems=1,
    dim=3,
)
IR = Diagonal(
    entries={2: "-pi*t"},
    exponentiation=True,
    symbols={"t": {"real": True, "positive": True}},
    targets=[1],
    num_systems=2,
    dim=3,
    label="R",
)

# Construct SWAP gate in clock subspace
S = Swap(targets=[0, 1], num_systems=2, dim=3)
S_matrix = S.matrix()
S = sp.Matrix(
    9, 9, lambda i, j:
    0 if (
        i % 3 == 0
        or j % 3 == 0
    )
    else S_matrix[i, j]
)

# Construct vacuum-excluding SWAP gate
S_vacuum = QuantumGate(
    spec=(
        tensor_product(I.matrix(), zero.matrix())
        + tensor_product(zero.matrix(), I.matrix())
        - tensor_product(zero.matrix(), zero.matrix())
        + S
    ),
    targets=[0, 1],
    num_systems=2,
    dim=3,
    family="SWAP",
)

# CTC
billiards = QuantumCTC(
    inputs=[clock_state_unevolved],
    gates=[S_vacuum, IR],
    systems_respecting=[0],
)
billiards.diagram()

# Output
# D-CTCs
billiards_DCTC = DCTC(circuit=billiards)
billiards_DCTC_CR = billiards_DCTC.state_respecting(label="ρ_D")
billiards_DCTC_CV = billiards_DCTC.state_violating(label="τ_D")
billiards_DCTC_CR.simplify()
billiards_DCTC_CR.apply(sp.factor)

# P-CTCs
billiards_PCTC = PCTC(circuit=billiards)
billiards_PCTC_CR = billiards_PCTC.state_respecting(label="ψ_P")
billiards_PCTC_CV = billiards_PCTC.state_violating(label="τ_P")
billiards_PCTC_CR.normalize()

# Results
billiards_DCTC_CR.print()
billiards_DCTC_CV.print()
billiards_PCTC_CR.print()
billiards_PCTC_CV.print()
\end{minted}
\tcblowerspaced
\begin{minted}{python}
>>> billiards.diagram()
$\includegraphics[scale=1.25, trim=-0.02cm 0 0 -0.15cm]{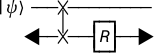}$
>>> billiards_DCTC_CR.print()
ρ_D = 1/2|1⟩⟨1| + -(g*exp(I*pi*t) - g - exp(I*pi*t))/2|1⟩⟨2| + (g*exp(I*pi*t) - g + 1)*exp(-I*pi*t)/2|2⟩⟨1| + 1/2|2⟩⟨2|
>>> billiards_DCTC_CV.print()
τ_D = g|0⟩⟨0| + (1/2 - g/2)|1⟩⟨1| + (1 - g)*exp(I*pi*t)/2|1⟩⟨2| + (1 - g)*exp(-I*pi*t)/2|2⟩⟨1| + (1/2 - g/2)|2⟩⟨2|
>>> billiards_PCTC_CR.print()
|ψ_P⟩ = sqrt(2)*sqrt(1/(cos(pi*t) + 3))|1⟩ + sqrt(2)*(1 + exp(-I*pi*t))*sqrt(1/(cos(pi*t) + 3))/2|2⟩
>>> billiards_PCTC_CV.print()
τ_P = 1/3|0⟩⟨0| + 1/3|1⟩⟨1| + exp(I*pi*t)/3|1⟩⟨2| + exp(-I*pi*t)/3|2⟩⟨1| + 1/3|2⟩⟨2|
\end{minted}
\end{codetitled}

\section{Performance}\label{sec:performance}

In this section, we embark upon an analysis of the performance of Qhronology's circuit simulation functionality. Before we commence however, it is useful to gain a better understanding of the package's inner workings, specifically its internal matrix machinery, which will be briefly discussed here.

In Qhronology, the conduit through which circuit simulations are facilitated is the \py{QuantumCircuit} class, which itself is composed of \py{QuantumState} and \py{QuantumGate} instances. When an output method is called on this class, such as \py{output()} or \py{state()}, the evolution of the initial state through the circuit (as dscribed by the states and gates) is simulated ({\via} the matrix mechanics of standard quantum theory), with the result (the output state) being returned as a SymPy matrix, NumPy array, or \py{QuantumState} instance. This functionality operates in two modes: numerical (NumPy backend) and non-numerical (SymPy backend), which are selected between using the \py{numerical} class property or method argument.

If any of a circuit's constituent objects contains a non-numerical type (such as a SymPy symbol), attempting to execute its simulation numerically results in the NumPy backend treating all array values as the \py{object} data type. This means that the advantages of NumPy's floating-point arithmetic ({\eg}, multithreaded array multiplication, efficient memory allocation in contiguous blocks) will be lost, resulting in significantly worse performance compared to that of purely numerical simulations. Note also that setting the \py{array} argument to \py{True} while \py{numerical} is similarly \py{True} can often increase performance, as fewer conversions between SymPy matrices and NumPy arrays may take place during the course of a simulation.

\subsection{Benchmarks}\label{sec:benchmarks}

The main facet of Qhronology which we are interested in benchmarking is the performance of its ability to simulate quantum circuits. By doing so, we will be able to characterize its time efficiency{\textemdash}how the execution times of its simulations scale with certain circuit properties{\textemdash}which we can then compare to the time complexity of its central algorithm: matrix multiplication. It is hoped that this analysis provides the user with a realistic idea of what they can expect to be able to accomplish with this component of the package.

The benchmarks presented in this paper aim to quantify how Qhronology's performance scales with two particular characteristics of any given circuit: the number of systems (\emph{circuit width}) and the number of gates (\emph{circuit depth}). Given that the vast majority of the modern applications of quantum mechanics concern only qubits (residing in 2-dimensional Hilbert spaces), the effect of dimensionality on the performance of Qhronology's simulations will not be examined here.

\newpage
The specific set of tests of which our benchmarking effort is comprised involves a varying number of identical equiprobabilistic qubit vector states $\bigl(\ket{0} + \ket{1}\bigr)/\sqrt{2}$ transformed by a varying number of identity gates $\Identity = \ket{0}\bra{0} + \ket{1}\bra{1}$. Choosing to use vector states rather than density matrices means that the calculations are performed within a vector space (not a matrix space), which substantially reduces the complexity (and therefore duration) of the computations. Similarly, the main incentive for using the identity gate is that its matrix representation is sparse (especially for larger compositions), corresponding to fewer operations per matrix multiplication and therefore considerably shorter tests.

It is also important to mention that neither the states nor the gates in these benchmarks possess any symbols, and so the data types of the elements in their matrix representations are purely numerical. This is because the incorporation of any non-numerical values (mainly SymPy symbols and expressions) into a circuit results in its simulation being significantly slower (primarily due to multiplication being a much more expensive operation when it involves such types). The specific format of our benchmarks was therefore chosen to be representative only of simple numerical simulations, with both the \py{numerical} and \py{array} arguments set to \py{True} in the tests. More complex benchmarks, such as those involving symbols, higher dimensions, dense gates, or matrix states, are radically slower, and so would not be feasible to perform on a personal computer. Note that the specification of the particular hardware used in the benchmarking is summarized in \linebreak {\tbl} \ref{tbl:benchmarking_system}.

\begin{center}
\vspace*{0.25cm}
\renewcommand{\arraystretch}{1.4}
\begin{NiceTabular}{*{3}{l}}[corners,hlines]
\Block[c, fill=lightblue, respect-arraystretch]{2-1}{\textbf{\textsf{Hardware}}} & CPU & AMD Ryzen 7 5800X3D \textsf{@} 4.552 GHz \\
 & RAM & 64 GB ECC CL22 DDR4-3200 \\
\Block[c, fill=lightblue, respect-arraystretch]{4-1}{\textbf{\textsf{Software}}} & Kernel & Linux 7.1.5 \\
 & Operating system & NixOS 26.11 (x86\_64) \\
 & Python & CPython 3.14.6 \\
 & Packages & \Block[l, respect-arraystretch]{}{Qhronology 1.1.1 \\ SymPy 1.14.0 \\ NumPy 2.5.1} \\
\end{NiceTabular}
\captionof{table}{Specification of the system used in the performance tests.}\label{tbl:benchmarking_system}
\end{center}

\begin{figure}[H]
\vspace*{-0.5cm}
\centering
\begin{subfigure}{1.0\textwidth}
    \hspace*{0.45cm}
    \includegraphics[scale=1.69]{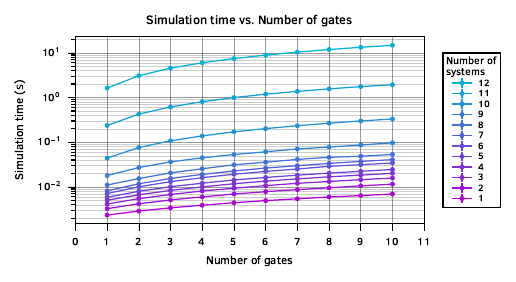}
    \vspace*{-0.45cm}
    \caption{}
    \label{fig:benchmark_gates_log}
\end{subfigure}
\begin{subfigure}{1.0\textwidth}
    \vspace*{0.2cm}
    \hspace*{0.45cm}
    \includegraphics[scale=1.69]{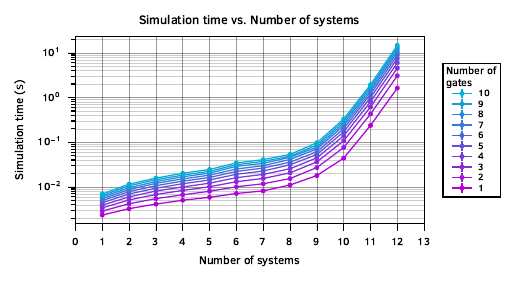}
    \vspace*{-0.45cm}
    \caption{}
    \label{fig:benchmark_systems_log}
    \vspace*{-0.1cm}
\end{subfigure}
\caption{The results of the benchmarks, plotted in two different ways. Note that even though only 5 samples were taken for each data point, the error bars are vanishingly small for all of them. It should also be noted that, when plotted on linearly (not logarithmically) scaled axes, the data in {\subfig} \ref{fig:benchmark_gates_log} can be observed to follow a virtually perfect linear trend, indicating that the execution time of Qhronology's simulations scale linearly with the number of gates (circuit depth), as one might expect.}
\label{fig:benchmark}
\end{figure}

Performing the benchmarking simulations (as outlined above) with the system described in {\tbl} \ref{tbl:benchmarking_system} yielded the results visualized in {\fig} \ref{fig:benchmark}. The absence of significant variability in this dataset reinforces our belief in its validity, meaning that any conclusions we draw from it, particularly those regarding Qhronology's performance, can be considered to be reasonably representative of the program's bulk behaviour and the algorithms from which it is built. This is supported in particular by the uniformity of the data in {\subfig} \ref{fig:benchmark_systems_log}: in the context of the vertical axis's logarithmic scale, the series are evenly spaced and remain consistently parallel for the entire benchmark. Accordingly, our estimation of its so-called ``effective'' time complexity should be credible, at least in regards to how the performance of the simulations scales with the number of systems (albeit only for \linebreak small numbers).

The main observation of these results can be seen clearly in {\subfig} \ref{fig:benchmark_systems_log}. With the number of systems represented by $N$, all series appear to follow a fairly linear trend on the logarithmic scale in the region $N \geq 6$. This suggests that the trend here is exponential in nature (with respect to $N$), and so potentially corresponds to an underlying exponential time complexity, {\ie}, $O(b^{N})$ (with $b > 1$). Note that convex shape of the curves for $N \leq 6$ also hints at such behaviour: a vertically shifted exponential function ({\eg}, $b^N + k$) plotted on a logarithmic scale necessarily possesses convex curvature (for $k > 0$, since $\log_{b}(b^N + k) \neq N + \log_{b}(k)$).

\subsubsection{Characterization of simulation scaling behaviour from benchmark data}

From the pervious discussion, our operating assumption is that the overarching trend we observe in our results can be described as being exponential with respect to $N$. With this in mind, we can use the recorded data in an attempt to precisely quantify the behaviour of Qhronology's quantum circuit simulations. This can be accomplished {\via} non-linear regression analysis, with our initial exponential model being simply:
\begin{equation}
    f_{\mathsf{E}}(N) = a b^{N} + u. \label{eq:regression_exponential}
\end{equation}
Performing regression analysis with a function such as this involves the determination of the values of its parameters ({\ie}, $a$, $b$, $u$) with which it best fits the data. This is typically accomplished by minimizing the sum of squared residuals between the data and the model, yielding a trendline that follows the data as closely as the particular model allows.

\begin{figure}[b!]
    \centering
    \vspace*{-0.25cm}
    \includegraphics[scale=1.69]{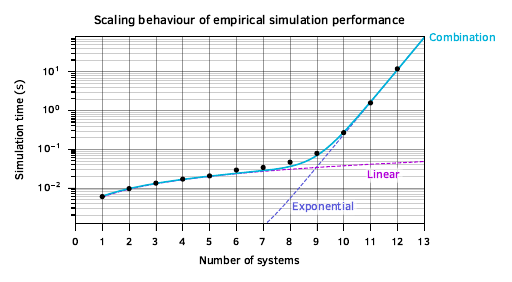}
    \vspace*{-0.25cm}
    \caption[Scaling behaviour of empirical simulation performance]{\label{fig:benchmark_systems_complexity_log}The scaling behaviour of Qhronology's quantum circuit simulation algorithm can be characterized by determining the trend of the data as a function of the number of systems involved in the simulation. By combining ({\via} summation) two different models (exponential and linear functions), we obtain a single model that appears to accurately describe the non-linear scaling of the program. Note that while the series we analyze here corresponds to a gate depth of 8, all others exhibit exactly the same behaviour (as evident in {\fig} \ref{fig:benchmark_systems_log}), and so the conclusions obtained are universal. Note also that the specific values of the regression parameters obtained in these analyses are unimportant, as they would differ between test systems.}
\end{figure}

The line of our exponential model (\ref{eq:regression_exponential}), fitted to a single series of the dataset, is depicted in {\fig} \ref{fig:benchmark_systems_complexity_log}. It is easy to see that the purely exponential model does not capture the curvature of the series exactly, especially in the region $2 \leq N \leq 5$. This suggests that the simulations exhibit more sophisticated behaviour, to which we can attempt to provide a better fit using a more complicated model (which also appears in {\fig} \ref{fig:benchmark_systems_complexity_log}).

To formulate an improved model, a good place to start is to focus on how the data lies above the exponential line in region of small $N$. Though the benchmark's method of data collection endeavoured to measure the duration of only those of Qhronology's calculations that are exclusively part of the circuit simulations (and not any other processes, like memory allocation and compilation, which are not strictly part of the matrix machinery), it is inevitable that sources of extraneous computation time contributed to the final measurements. If we assume, of these instances of so-called ``overhead'', that each takes roughly a constant amount of time to perform, and that their number is proportional to the number of systems $N$ ({\ie}, a simulation possessing more systems entails proportionally more overhead), then their contribution to the benchmark data would be linear in $N$. For small $N$, this would non-insignificantly raise the simulation execution time, similar to what we observe in the data. Alternatively, for large $N$, the relatively long duration of the matrix-specific calculations would dominate additions to the simulation time accrued from any incidental computations, meaning that the linear contribution from systematic overhead would be negligible.

Accordingly, by first defining a linear model,
\begin{equation}
    f_{\mathsf{L}}(N) = c N + v, \label{eq:regression_linear}
\end{equation}
with two parameters ($c$, $v$), we can construct a combination model by summing the exponential (\ref{eq:regression_exponential}) and linear (\ref{eq:regression_linear}) models. This yields the function
\begin{equation}
    f_{\mathsf{C}}(N) \equiv f_{\mathsf{E}}(N) + f_{\mathsf{L}}(N) = a b^{N} + c N + d, \label{eq:regression_combination}
\end{equation}
where the respective shift parameters $u$ and $v$ of the exponential and linear models have been amalgamated into a single parameter $d \equiv u + v$. The fitted line of this model, obtained from regression analysis, appears alongside its constituent models in {\fig} \ref{fig:benchmark_systems_complexity_log}.

\newpage
Evidently, this model fits the data reasonably well, and so we come to the conclusion that, at least for the specific simulations tested, Qhronology's performance (that is, the execution time of its simulations) scales exponentially with the number of systems $N$ (with slight linear corrections, presumably due to execution overhead). Of course, whether or not this model provides an accurate description of the program's true behaviour is of course something which regression analysis itself cannot answer.

\subsubsection{Comparison of the empirical behaviour with algorithmic time complexity}

An algorithm's time complexity typically describes the upper bound of the asymptotic behaviour of its idealized execution time, that is, how the its performance is predicted to behave in the limit (tending to infinity) of some scaling variable (usually the size $n$ of its input). Essentially, time complexity is an entirely theoretical notion, meaning that attempting to estimate it empirically is a fundamentally flawed pursuit. To be specific, given a set of performance data (such as that in {\fig} \ref{fig:benchmark}), it is impossible to ascertain whether the scaling variable $n$ is large enough for the limit to be apparent, especially when the data describes only a narrow region of relatively small values of $n$. This problem is exacerbated by the fact that actually executing an algorithm on physical hardware incurs all kinds of performance penalties associated with realistic ({\ie}, unidealistic) computers, such as memory limitations (latency, bandwidth constraints), CPU architecture (pipeline stalls from branch prediction, scheduling inefficiencies), operating system overhead (context switching, interrupts), cache misses, and the behaviours of the specific versions, implementations, and compilations of all software in the entire stack. While effort can certainly be made to minimize sources of performance variability in the course of benchmarking, only so much can be done.

Qhronology's core functionality is based upon the linear algebra of discrete-variable quantum mechanics, and so the matrix multiplication involved with such machinery forms an integral part of the package. As this operation is both relatively expensive and constitutes the majority of computations involved with any simulation in Qhronology, it often contributes the most to the execution time. Consequently, it is reasonable to expect that the performance of Qhronology's simulations are within the neighbourhood of that of matrix multiplication.

Given two $n \times n$ matrices, the time complexity of their multiplication is, according to the na\"{i}ve algorithm, $O(n^3)$, under the standard assumption that the exactly $n^3$ scalar multiplication operations and $n^3 - n^2$ scalar addition operations required all execute in constant time, {\ie}, $O(1)$. Similarly, the time complexity of multiplying one such matrices with an $n \times 1$ vector is $O(n^2)$, due to the procedure necessitating $n^2$ multiplication and $n(n-1)$ addition operations. The shared characteristic of these algorithms is that they are all in the polynomial complexity class $\mathsf{P}$, that is, their time complexities have the form $n^{O(1)} \cong O(n^p)$ for some $p > 0$.

In Qhronology's $d$-dimensional quantum mechanics, states on $N$ systems are represented by either $d^N \times d^N$ matrices (for density matrices) or $d^N \times 1$ vectors (for vector states), while operators which act on these take the form of $d^N \times d^N$ matrices, as standard in quantum theory. This means that the relationship between a quantum circuit's matrix size $n$ and number of systems $N$ can be expressed as $n = d^N$. Accordingly, the time complexity of multiplication involving such matrices and vectors in terms of $N$ is simply $O(n^p) = O(d^{pN})$, with $p$ depending on the kind of product being computed ({\eg}, matrix-matrix, matrix-vector, matrix-matrix-matrix, {\etc}). The main observation of this result is that it describes exponential time complexity in the number of systems $N$. The fact that the empirical data exhibits similarly exponential behaviour (with respect to the number of systems) is therefore not surprising, and suggests that Qhronology's simulations perform precisely how one might expect (at least with respect to how their execution times scale with $N$).

\section{Conclusion}\label{sec:conclusion}

Of the many areas of research that lie within the intersection of foundational quantum mechanics and relativity, the quantum physics of time travel is one of the most fascinating. The various quantum models of CTCs, which collectively form a critical component of the numerous research efforts into this topic, represent some of the most meaningful efforts to unify time travel and the standard laws of physics. Contributing to this endeavour, Qhronology's programmatic implementation of prominent quantum prescriptions of antichronological time travel constitutes perhaps the first major attempt at facilitating the systematic study of CTCs in a computational, simulative capacity. By providing a simple yet powerful framework by which these prescriptions can be analyzed near-limitlessly within the context of specific quantum interactions, the project hopes to make the research field more accessible to both experienced and inexperienced academics alike.

In this paper, we formally introduced Qhronology, discussing its architecture, design philosophy, usage, and performance. The plethora of examples provided, along with tutorial-style instruction on some of its various features, are intended to enable the keen user to quickly gain a good understanding of Qhronology's interface and capabilities. While detail on a large amount of advanced functionality has been omitted for the sake of brevity, the project's official documentation \cite{bishop_qhronology-documentation_2025} naturally contains exhaustive information on every component by the package. Additionally, one may find a brief but complete treatise on the theory upon which the project is built, including the mathematical foundations of quantum mechanics, quantum circuitry, CTCs, and the physical theories of time travel. The interested user is therefore encouraged to explore Qhronology's official documentation to find out exactly what the package has to offer.

\printbibliography
\addcontentsline{toc}{section}{References}

\end{document}